\pdfoutput=1
\RequirePackage{ifpdf}
\ifpdf 
\documentclass[pdftex]{sigma}
\else
\documentclass{sigma}
\fi
   		
\usepackage{mathtools}
\usepackage{xfrac}
\usepackage{tikz}
\newcommand{\ket}[1]{\left| #1 \right>}
\newcommand{\bra}[1]{\left< #1 \right|}
\newcommand{\braket}[2]{\left< #1 \vphantom{#2} \right|
 \left. #2 \vphantom{#1} \right>}

\renewcommand{\Re}{\operatorname{Re}}
\renewcommand{\Im}{\operatorname{Im}}
\newcommand{\rmi}{\mathrm{i}}
\newcommand{\rmd}{\mathrm{d}}
\newcommand{\eps}{\varepsilon}
\newcommand{\rme}{\mathrm{e}}
\newcommand{\tr}{\mathrm{tr}}
\newcommand{\Log}{\mathrm{Log}}

\numberwithin{equation}{section}

\newtheorem{Theorem}{Theorem}[section]
\newtheorem*{Theorem*}{Theorem}
\newtheorem{Corollary}[Theorem]{Corollary}
\newtheorem{Lemma}[Theorem]{Lemma}

 { \theoremstyle{definition}

\newtheorem{Example}[Theorem]{Example}
\newtheorem{Remark}[Theorem]{Remark} }

\begin{document}
\allowdisplaybreaks

\renewcommand{\thefootnote}{}

\newcommand{\arXivNumber}{2105.13359}

\renewcommand{\PaperNumber}{098}

\FirstPageHeading

\ShortArticleName{Exact correlations in topological quantum chains}

\ArticleName{Exact Correlations in Topological Quantum Chains\footnote{This paper is a~contribution to the Special Issue on Evolution Equations, Exactly Solvable Models and Random Matrices in honor of Alexander Its' 70th birthday. The~full collection is available at \href{https://www.emis.de/journals/SIGMA/Its.html}{https://www.emis.de/journals/SIGMA/Its.html}}}

\Author{Nick G. JONES~$^{\rm ab}$ and Ruben VERRESEN~$^{\rm cd}$}

\AuthorNameForHeading{N.G.~Jones and R.~Verresen}

\Address{$^{\rm a)}$~St John's College, University of Oxford, UK}
\Address{$^{\rm b)}$Mathematical Institute, University of Oxford, UK}
\EmailD{\href{mailto:nick.jones@maths.ox.ac.uk}{nick.jones@maths.ox.ac.uk}}

\Address{$^{\rm c)}$~Department of Physics, Harvard University, Cambridge, MA 02138, USA}
\EmailD{\href{mailto:rubenverresen@g.harvard.edu}{rubenverresen@g.harvard.edu}}
\Address{$^{\rm d)}$~Department of Physics, Massachusetts Institute of Technology, Cambridge, MA 02139, USA}

\ArticleDates{Received March 06, 2023, in final form November 27, 2023; Published online December 15, 2023}

\Abstract{Although free-fermion systems are considered exactly solvable, they generically do not admit closed expressions for nonlocal quantities such as topological string correlations or entanglement measures. We derive closed expressions for such quantities for a dense subclass of certain classes of topological fermionic wires (classes BDI and AIII). Our results also apply to spin chains called generalised cluster models. While there is a bijection between general models in these classes and Laurent polynomials, restricting to polynomials with degenerate zeros leads to a plethora of exact results: (1) we derive closed expressions for the string correlation functions---the order parameters for the topological phases in these classes; (2) we obtain an exact formula for the characteristic polynomial of the correlation matrix, giving insight into ground state entanglement; (3) the latter implies that the ground state can be described by a matrix product state (MPS) with a finite bond dimension in the thermodynamic limit---an independent and explicit construction for the BDI class is given in a concurrent work [\textit{Phys. Rev. Res.} \textbf{3} (2021), 033265, 26~pages, arXiv:2105.12143]; (4) for BDI models with even integer topological invariant, all non-zero eigenvalues of the transfer matrix are identified as products of zeros and inverse zeros of the aforementioned polynomial. General models in these classes can be obtained by taking limits of the models we analyse, giving a further application of our results. To the best of our knowledge, these results constitute the first application of Day's formula and Gorodetsky's formula for Toeplitz determinants to many-body quantum physics.}


\Classification{82B10; 81V74}

 \begin{flushright}
 \begin{minipage}{60mm}
 \textit{Dedicated to Alexander Its\\ on the occasion of his 70th birthday}
 \end{minipage}
 \end{flushright}

\renewcommand{\thefootnote}{\arabic{footnote}}
\setcounter{footnote}{0}

\section{Introduction}

Modelling certain many-body quantum systems in terms of non-interacting fermions has proved to be unreasonably effective. Historically forming the basis of Fermi liquid theory \cite{Sachdev01}, it has in recent times led to the discovery of topological insulators and superconductors \cite{Bernevig06,Fu07,Hasan10,Kane05,Kitaev09,Moore07,Roy09,Ryu10,Schnyder08}. Free-fermion chains---dual to famous spin models such as the two-dimensional classical Ising model and one-dimensional quantum XY model---have also been fertile ground for mathematical innovations. Whilst correlation functions of local fermionic operators are simple, non-local string correlation functions---essential for calculating certain local spin correlations as well as understanding the topological nature of fermionic systems---required and indeed stimulated significant developments in the theory of the asymptotics of Toeplitz determinants (determinants of matrices that are constant along each diagonal) \cite{Deift2013}.

Considering only the two-dimensional classical Ising model as an example, there are a vast number of remarkable exact results for spin correlations \cite{McCoy2010}. However, it is unusual to find simple closed formulae for these correlators. Typically, in analysing string correlations or the entanglement properties of fermionic systems, one finds closed expressions for dominant terms in the \emph{asymptotic expansion} for, respectively, large string length \cite{Barouch1971,Jones2019} or large subsystem size (for the entanglement entropy) \cite{Its2007,Jin2004,Keating2004}. A rare and famous exception\footnote{Another exceptional case, less relevant to this work, is the spin-spin correlator `along the diagonal' at $T=T_c$ in the 2D classical Ising model (equivalently, at the phase transition in the 1D quantum transverse-field Ising model). This can be expressed in an exact closed form as a ratio of Barnes G-functions \cite{Deift2013}.} that does admit simple closed formulae for arbitrary distances is the `disorder line' in the XY spin chain---this is dual to a nearest-neighbour free-fermion model and there are two symmetry-broken product state ground states \cite{Barouch1971,Chung01,Franchini,Its2005,Its2007,Kurmann82,Mueller85,Taylor83}.

In the present work, we consider Hamiltonians for one-dimensional classes of spinless fermions with phases classified by a topological winding number (commonly referred to as the BDI and AIII classes \cite{Altland97,Kitaev09,Motrunich01,Ryu10}). The BDI class consists of superconducting time-reversal invariant spinless fermions, while the AIII class consists of charge-conserving spinless fermions with a sublattice symmetry; these classes are non-interacting instances of symmetry-protected topological (SPT) phases \cite{Chen11,Fidkowski11,Pollmann10,Schuch11,Senthil15}. We show that for a simply characterised subclass of both the BDI and AIII classes, we can find closed expressions for certain physically relevant string correlation functions for models within this subclass. Moreover, we give an exact formula for the characteristic polynomial of the correlation matrix---intimately related to entanglement properties of the system---where this formula depends only parametrically on the size of the matrix. We further show that our subclass of interest is dense in the whole class, meaning that any Hamiltonian in the BDI or AIII class can be obtained as the limit of a sequence of Hamiltonians studied in this work.

The above results allow us to prove the existence of an exact matrix product state (MPS) representation \cite{Cirac2020,Pollmann2017} of the ground state for this dense subclass.
An MPS is one of the simplest tensor networks for describing many-body wavefunctions and has proven to be a valuable concept both analytically (e.g., in the discovery of fixed-point SPT states \cite{Affleck87}) and as a numerical ansatz (underlying the density-matrix renormalization group (DMRG)) \cite{White92,White93}. In a concurrent work~\cite{Jones21}, we give an explicit construction for the MPS ground state in this subclass of the BDI class (up to a measure zero set of exceptional cases), where this subclass is referred to as the MPS skeleton. Particular examples of models in this subclass were previously studied in \cite{Smith19,Wolf06}. While the explicit MPS representation is a valuable tool (for example, they could serve as initial states for DMRG or dynamic quenches in non-integrable models \cite{Jones21}), it is not straightforward to extract closed analytic expressions for the string correlations and entanglement spectrum, motivating the analysis in this work. Moreover, we prove the existence of an MPS for a subclass of class AIII, where the construction in \cite{Jones21} does not directly apply. It is interesting that the techniques used here, from Toeplitz determinant theory, are rather different to those used in the explicit construction, which is based on Witten's conjugation argument for frustration-free models \cite{Jones21,Wouters2020}. Both are in turn different to approaches to free-fermion MPS based on Gaussian states \cite{Kraus10,Schuch08}. One can think of the MPS ground state as two $\chi \times \chi$ matrices, where $\chi$ is the bond dimension of the MPS. As well as proving that $\chi < \infty$ (even for infinite system sizes) in both classes, for the BDI class we find a rigorous lower bound on $\chi$. The explicit construction in~\cite{Jones21} gives an upper bound on $\chi$ and in certain cases this upper bound coincides with the lower bound given here, proving that this is the optimal bond dimension in these cases.

We now expand on the above claims. As we review in detail below, any (finite-range) Hamiltonian in theses classes is equivalent to a Laurent polynomial $f(z) = \sum_\kappa t_\kappa z^\kappa$, (where the coupling constants $t_\kappa \in \mathbb{R}$ for BDI and $t_\kappa \in \mathbb{C}$ for AIII). This Laurent polynomial is characterised by the degree of the pole (or zero) at $z=0$ and a finite number of zeros away from $z=0$. If no zero is on the unit circle, then the corresponding Hamiltonian is gapped and the ground state phase diagram is labelled by a winding number \cite{Jones2019,Verresen2018}. Physically, a non-zero winding number is topologically non-trivial, as evidenced---for instance---by topologically-protected zero-energy edge modes \cite{Balabanov21,Kitaev2001,Verresen2018}.
Each gapped phase has a string order parameter \cite{Jones2019,Perez2008}. These are non-local string correlation functions defined as a ground state expectation value of a product of an extensive number ($\Theta(N)$)
of fermionic operators. In the phase with winding number $\omega$, the corresponding string order parameter is non-zero in the limit of large $N$, while the string correlation functions that are order parameters for other phases decay with $N$. The subclass of models that we consider in this work is the case where every zero of $f(z)$ away from $z=0$ has even multiplicity. An exact formula for all string correlation functions for each model in this subclass constitutes our first main result.

For free-fermion systems, the entanglement spectrum of a subsystem is simply calculated from the eigenvalues of the correlation matrix---a matrix with elements consisting of two-point fermionic correlators \cite{Peschel2003,Peschel_2009,Vidal2003}. From these eigenvalues one can easily calculate the entanglement entropy. Our second main result gives a method for finding a closed formula for the characteristic polynomial of the correlation matrix for a subsystem of size $N$, with explicit $N$ dependence. As mentioned, this result allows us to prove the existence of an exact MPS representation of the ground state. We also use this to give the characteristic polynomial in a series of examples.

Given an MPS, a useful construction in analysing correlations is the \emph{MPS-transfer matrix}. Our results on string correlation functions allow us to derive properties of this transfer matrix, without needing the MPS itself. We focus on the BDI class; then, given a property that is generically satisfied in our class of models, we use the transfer matrix to give a lower bound on the bond dimension of any MPS representation of the ground state. The proof of this lower bound is through identifying eigenvalues of the MPS-transfer matrix that appear in correlation functions. As mentioned, in certain cases this lower bound coincides with an upper bound derived in \cite{Jones21}, giving the optimal bond dimension of the exact MPS ground state. In these cases, given the upper bound, we can find the entire spectrum of the transfer matrix---the eigenvalues are products of zeros and inverse zeros of $f(z)$ (multiplied by a sign that we determine). We also show how our results constrain the eigenvectors of the transfer matrix.

The class of models that we consider is a class of exceptional cases in the full class of BDI or AIII Hamiltonians. However, in this work we show that any (finite-range) model in these classes can arise as a limit of Hamiltonians considered in this paper. We remark that as we take the limit, we allow longer and longer range couplings. We illustrate how taking the limit of the string order parameter recovers the result for general models obtained in \cite{Jones2019}. The concurrent work \cite{Jones21} discusses a concrete example, where the transverse-field Ising chain is approximated by a sequence of models in this subclass.

The key object determining ground state correlations for these models is the symbol, or generating function, $\sqrt{f(z)/\overline{f}(1/z)}$ (note $\overline{f}(z) = \sum_\kappa \overline{t_\kappa} z^\kappa$). In particular, Fourier coefficients of this function give the fermionic two-point function, and other correlations follow from this through Wick's theorem. In the general case, this is a multivalued function that we can analyse with branch cuts in the complex plane. For example, we can calculate asymptotic Fourier coefficients using the Darboux principle \cite{Dingle}. In our case, since every zero has even multiplicity, we have a rational function. This allows us to find a closed form for the Fourier coefficients (this observation has been used for identifying the correlation length of Gaussian MPS in \cite{Schuch06}). Moreover, our result on string correlation functions is an application of Day's formula for Toeplitz determinants with rational symbol\footnote{This means that the matrix elements are Fourier coefficients of a rational function, in this case related to $\sqrt{f(z)/\overline{f}(1/z)}$. We review this terminology and the relevant theory in Section \ref{sec:toeplitz}.} \cite{Day} and our result on the correlation matrix is an application of Gorodetsky's formula for block Toeplitz determinants with rational symbol \cite{Bottcher,Gorodetsky1981,Gorodetsky}. Day's formula arises from a reduction of the determinant of a rationally generated $N\times N$ Toeplitz matrix, to a product of determinants of matrices of fixed size (independent of $N$). Evaluating the determinant leads to a relatively simple closed formula in terms of the zeros and poles of the generating function. Toeplitz determinants have long been associated with problems in statistical physics (see, for example, the review \cite{Deift2013}), and Toeplitz determinants with rational symbols have previously appeared in problems in statistical mechanics and quantum many body physics \cite{Basor1994,Forrester1992,Forrester1992selberg}. However, to the best of our knowledge, this is the first application of Day's formula and Gorodetsky's formula in this context.

We note that similar conditions on particular generating functions being rational have previously appeared in works on Gaussian MPS---Gaussian states with finite bond dimension \cite{Cirac2020,Kraus10,Schuch08}. One characterisation of these Gaussian MPS is that the correlation matrix in Fourier space is a rational function of $\rme^{\rmi k_j}$, for momenta $k_1,\dots,k_d$ (this result applies in any spatial dimension $d$). Hence, in the models we consider, the correlation matrix having a rational symbol is a necessary condition that we have an MPS ground state. Our results show that in the one-dimensional BDI and AIII classes this is also sufficient.

The paper is organised as follows. In Section \ref{sec:model}, we introduce the BDI class, and the corresponding subclass of interest. We then introduce the necessary theory of Toeplitz determinants in Section \ref{sec:toeplitz}, including Day's formula and Gorodetsky's formula. For each gapped phase of the BDI class, we have a string order parameter that can be viewed as a correlation function of two string operators. Section \ref{sec:stringresults} gives our results for the value of the correlation function of all such string operators in any model in the given subclass. Section \ref{sec:corrmatrixresults} then gives our results for the characteristic polynomial of the correlation matrix. We give examples of how to use this to find the eigenvalues of the correlation matrix. In Section \ref{sec:AIII}, we introduce the AIII class and explain how the main results of Sections \ref{sec:stringresults} and~\ref{sec:corrmatrixresults} apply to a subclass of these models.
Following these results and corresponding remarks, we give the proofs. In Section \ref{sec:string} (Section \ref{sec:corrmatrix}), we prove the results of Section \ref{sec:stringresults} (Section \ref{sec:corrmatrixresults}).
Finally, we include two applications of our results.
In Section \ref{sec:transfer}, we discuss how our results allow us to understand properties of the ground state MPS-transfer matrix in class BDI. Then
Section \ref{sec:generic} shows how to obtain general models in both classes as limits of models studied in this work; we moreover demonstrate how this can be used to obtain the formula for the order parameter in general BDI models.

\section{The BDI class}\label{sec:model}
Consider a periodic one-dimensional chain with $L$-sites, where for each site we have a spinless fermionic degree of freedom $c^\dagger_n$ (for $n=1,\dots, L$) that satisfies $\{ c_n, c_m \}= 0$ and $\big\{c_n^\dagger ,c_m \big\} = \delta_{nm}$. We will analyse this model in the thermodynamic limit $L\rightarrow \infty$. Define the Majorana operators \begin{align*}
\gamma_n = c^\dagger_n + c_n, \qquad \tilde \gamma_n = \rmi \bigl(c^\dagger_n -c_n\bigr),
\end{align*}
which satisfy $\{ \gamma_n,\gamma_m \} = \{ \tilde \gamma_n, \tilde \gamma_m \} = 2 \delta_{nm}$ and $\{ \gamma_n, \tilde \gamma_m \} = 0$.
The class of BDI Hamiltonians is defined as the vector space of all Hamiltonians that are quadratic in fermionic operators and that are symmetric under the anti-unitary involution $T \gamma_n T = \gamma_n$ and $T \tilde \gamma_n T = -\tilde \gamma_n$ \cite{Verresen2018}.
The most general translation-invariant one-dimensional BDI Hamiltonian is of the form
\begin{align}
H = \frac{\rmi}{2}\sum_\kappa \sum_{n \in \mathrm{sites}}t_\kappa \tilde \gamma_n \gamma_{n+\kappa}, \qquad t_\kappa \in \mathbb{R}. \label{HBDI}
\end{align}
This model was first analysed in \cite{Suzuki71}. We will work in settings where $t_\kappa$ is only non-zero for a~finite number of $\kappa \in \mathbb Z$ (physically, this corresponds to local Hamiltonians).
Using the Jordan--Wigner transformation, the above Hamiltonian is equivalent to a spin-\sfrac{1}{2} chain---details are given in Appendix \ref{app:spin}.
 As discussed above, it is helpful \cite{Jones2019,Verresen2018} to introduce the complex function
\begin{align}
f(z) = \sum_\kappa t_\kappa z^\kappa. \label{fz}
\end{align}
The Hamiltonian \eqref{HBDI} is diagonalised as follows. For each momentum $k \in [0,2\pi)$, $\big\lvert f\bigl(\rme^{\rmi k}\bigr)\big\rvert$ is the one-particle energy of the mode labelled by $k$, while the mode itself is given by
\begin{align}
\eta_k = \frac{1}{2}\left(1 +\frac{ {f\bigl(\rme^{\rmi k}\bigr)}}{\big|f\bigl(\rme^{\rmi k}\bigr)\big|} \right)c^\dagger_{k}+ \frac{1}{2}\left(1 -\frac{ {f\bigl(\rme^{\rmi k}\bigr)}}{\big|f\bigl(\rme^{\rmi k}\bigr)\big|} \right)c_{-k},\label{modes}
\end{align}
where $c_{k}$ is the Fourier transform of $c_n$.
The eigenstates of the many-body Hamiltonian then corresponding to filling these modes. Other properties of the model follow from knowing the zeros of this function. For a finite-range Hamiltonian, $f(z)$ is a Laurent polynomial: asymptotic correlations for that case were analysed in \cite{Jones2019} and asymptotics of the characteristic polynomial of the correlation matrix and the entanglement entropy were analysed for gapped Hamiltonians in \cite{Its2008} and for certain\footnote{Subject to the restriction that $f(z) = f(1/z)$---physically this means an additional symmetry under spatial inversion.} gapless Hamiltonians in \cite{Keating2004}.

If the Hamiltonian corresponding to \eqref{fz} is gapped, then the ratio appearing in \eqref{modes} satisfies
\begin{align}
\frac{ {f(z)}}{|f(z)|}=\sqrt{\frac{f(z)}{f(1/z)}}. \label{sqrtfz}
\end{align}
In this paper, we are interested in continuous families of `exceptional cases' for this model. In particular, we restrict to the case where \eqref{sqrtfz} is a rational function. Let us consider two of the simplest examples. First, the case $f(z) = \frac{1}{z}(z-a)^2$ for $a\in\mathbb{R}$, which corresponds to the fermionization (Jordan--Wigner transformation) of the disorder line of the XY model \cite{Barouch1971,Chung01,Franchini,Its2005,Its2007,Kurmann82,Mueller85,Taylor83}:
\begin{align*}
H ={}& \frac{\rmi}{2} \sum_{n \in \mathrm{sites}}\bigl( \tilde\gamma_n \gamma_{n-1}-2a \tilde\gamma_n \gamma_{n}+a^2\tilde\gamma_n \gamma_{n+1}\bigr) \\
&\xleftrightarrow{\textrm{J--W}} -\frac{1}{2} \sum_{n \in \mathrm{sites}}\bigl( Y_nY_{n+1}+2a Z_n+a^2X_nX_{n+1}\bigr).
\end{align*}
Second, the case $f(z) = (z-a)^2$ which is the fermionization of a cluster-Ising chain \cite{Jones21,Smith19,Wolf06}:
\begin{align*}
H ={}&\frac{\rmi}{2} \sum_{n \in \mathrm{sites}}\bigl( \tilde\gamma_n \gamma_{n}-2a \tilde\gamma_n \gamma_{n+1}+a^2\tilde\gamma_n \gamma_{n+2}\bigr) \\ &\xleftrightarrow{\textrm{J--W}} \frac{1}{2} \sum_{n \in \mathrm{sites}}\bigl( Z_n +2a X_nX_{n+1} -a^2X_nZ_{n+1}X_{n+2}\bigr).
\end{align*}

In general the condition that \eqref{sqrtfz} is a rational function means that we can take\footnote{For completeness, let $h(z)$ denote any Laurent polynomial that satisfies $h(z)=h(1/z)$, has no zeros on the unit circle and has positive constant term. Then we can multiply \eqref{fzcanon} by $h(z)$ without changing the ground state---see the discussion in \cite{Jones21}, as well as comments in Remark \ref{remark:nongeneric}.} every zero to have even multiplicity, i.e., $f(z)$ has the following form:
\begin{gather}
f(z) = \sigma \frac{1}{z^{n_P}} \prod_{j=1}^{n_z} \left(z-z_j\right)^2\prod_{k=1}^{n_Z} \left(z-Z_k\right)^2, \qquad \sigma \in \{\pm 1\}, \label{fzcanon}
\end{gather}
where $n_P\in \mathbb{Z}$ (an integer shift affecting the interaction range of the Hamiltonian \eqref{HBDI}) and we have that $0<\lvert z_j \rvert <1 <\lvert Z_k \rvert$. (In Section~\ref{sec:generic}, we will see that general models in the BDI class can be obtained from limits of these exceptional cases.) We have implicitly normalised $f(z)$, or equivalently the Hamiltonian, since this overall normalisation does not affect the ground state. Note that since $t_\alpha\in\mathbb{R}$, zeros of $f(z)$ are~either real, or come in complex-conjugate pairs. Given this form, we will assume the generic case, that $\big\{z_{j_1}^{\vphantom {-1}},z_{j_2}^{-1},Z_{k_1}^{\vphantom {-1}},Z_{k_2}^{-1}\big\}_{j_1,j_2=1,\dots, n_z; k_1,k_2=1,\dots, n_Z\vphantom{\big|}}$ are~pairwise distinct.
Assuming this generic case is mainly for ease of presentation. In fact, our results on the correlation matrix (implying the existence of an exact MPS ground state) can be applied directly to the non-generic case, while other results can be generalised to non-generic cases by taking an appropriate limit, this is discussed in Remark \ref{remark:nongeneric}. It is also helpful to index each set of zeros by proximity to the unit circle, i.e., $\lvert Z_1\rvert \leq\lvert Z_2\rvert \leq \dots \leq \lvert Z_{n_Z}\rvert$ and $\lvert z_1\rvert \geq \lvert z_2 \rvert \geq \dots \geq \lvert z_{n_z}\rvert$.
Unless explicitly mentioned, we fix $\sigma=1$ (see Remark \ref{remark:nongeneric} for the case~$\sigma=-1$). Since this model has no zeros on the unit circle, it is gapped\footnote{Our analysis can be applied to gapless models where the multiplicity of zeros on the unit circle is even. As discussed in Remark \ref{remark:nongeneric} and \cite{Jones21} these models have the same ground state as a gapped model in our class.} and the gapped phase diagram is labelled by a winding number, given by $\omega =2n_z-n_P$.

Ground state correlations are determined, using Wick's theorem, by the fermionic two-point correlation function \cite{Lieb61}. This correlator is calculated in Appendix \ref{app:fourier} for the class of models given by \eqref{fzcanon}, but the closed form is not required for our main results.

The fermionic string operators that form the string order parameter for the phase $\omega = \alpha$ are given by\footnote{The phase factor ensures that $\mathcal{O}_\alpha=\mathcal{O}_\alpha^\dagger$. Note that there is an erroneous absolute value in the corresponding term in \cite{Jones2019}. Other formulae in that paper are unchanged given the correct definition \eqref{Odef}.} \cite{Jones2019}:
\begin{align}\label{Odef}
\mathcal{O}_\alpha(n)=\begin{cases} \displaystyle  \rmi^{\lfloor \alpha/2 \rfloor} \left(\prod_{m=1}^{n-1} \rmi\tilde\gamma_m\gamma_m\right)\gamma_{n}\gamma_{n+1}\cdots\gamma_{n+\alpha-1}, \quad& \alpha>0, \\
\displaystyle  (-\rmi)^{\lfloor \alpha/2 \rfloor}\left(\prod_{m=1}^{n-1} \rmi\tilde\gamma_m\gamma_m\right) (-\rmi\tilde\gamma_{n})\cdots(-\rmi\tilde\gamma_{n+\lvert\alpha\rvert-1}), \quad &\alpha<0,\\
\displaystyle \prod_{m=1}^{n-1} \rmi\tilde\gamma_m\gamma_m, \quad &\alpha=0.\end{cases}
\end{align}
The (string) order parameter in the phase $\omega$ is $\lim_{N\rightarrow\infty}\lvert\langle\mathcal{O}_\omega(1)\mathcal{O}_\omega(N+1)\rangle \rvert ,$ where angle brackets denote the ground state expectation value. In this phase,
 \[
\lim_{N\rightarrow\infty}\lvert\langle\mathcal{O}_\alpha(1)\mathcal{O}_\alpha(N+1)\rangle \rvert = 0
\]
 for~$\alpha \neq \omega$.

In the dual spin-\sfrac{1}{2} chain the physics is different, and the $\mathcal{O}_\alpha(n)$ are local operators for odd $\alpha$ and non-local for even $\alpha$. Hence, for the spin chain, we have a non-zero string order parameter given by $\lim_{N\rightarrow\infty}\lvert\langle\mathcal{O}_\omega(1)\mathcal{O}_\omega(N+1)\rangle \rvert$ for even $\omega$, while we have a (spin-flip) symmetry-breaking order parameter $\left(\lim_{N\rightarrow\infty}\lvert\langle\mathcal{O}_\omega(1)\mathcal{O}_\omega(N+1)\rangle \rvert\right)^{1/2}$ for odd $\omega$. The nature of the symmetry-breaking and (interacting) SPT order in the spin-chain is discussed in \cite{Jones2019}.

\section{Toeplitz determinants}\label{sec:toeplitz}
The theory of Toeplitz determinants is central to our analysis, and we review key notation and theory in this section.

An $N\times N$ Toeplitz matrix takes the following form:
\begin{align}
(T_{mn}) = ({t}_{m-n}) = \left(\begin{matrix}{t}_0 & {t}_{-1} & {t}_{-2} & \hdots & {t}_{-(N-1)} \\{t}_{1} & {t}_0 & {t}_{-1} & \hdots &{t}_{-(N-2)} \\{t}_2 & {t}_{1} & {t}_0 & \hdots&{t}_{-(N-3)} \\\vdots & \vdots & \vdots & \ddots&\vdots\\ {t}_{N-1} &{t}_{N-2} & {t}_{N-3} & \hdots & {t}_0 \end{matrix}\right).\label{toeplitzform}
\end{align}
We say the Toeplitz matrix has (scalar) symbol $t(z)$ if the entries of the Toeplitz matrix are the Fourier coefficients of the function $t(z)$, i.e.,
\begin{align}
t_{n} = \frac{1}{2\pi} \int^{2 \pi}_{0}t\bigl(\rme^{\rmi k}\bigr) \rme^{- \rmi n k} \rmd k. \label{fourier}
\end{align}
Given this symbol, we can generate an $N \times N$ Toeplitz matrix for any $N$. We define $D_N[t(z)]$ to be the determinant of the $N\times N$ Toeplitz matrix generated by the symbol $t(z)$. Toeplitz determinants have a rich theory; see, for example, the review \cite{Deift2013} and references therein. We will also need to consider block Toeplitz matrices---generalising the above to symbols that are matrix valued functions. Our cases of interest have a $2\times2$ matrix symbol $\Phi(z)$, where the determinant $D_N[\Phi(z)]$ will be the determinant of the $2N \times 2N$ matrix with form as in \eqref{toeplitzform}, with $2\times 2$ matrix Fourier coefficients defined as in \eqref{fourier}.

Now, for translation-invariant free-fermion chains in class BDI, we have that the correlator of string operators is the following Toeplitz determinant:
\begin{align}
(-1)^{N(\alpha-1)}\langle\mathcal{O}_\alpha(1) \mathcal{O}_\alpha(N+1)\rangle &= \det \left( \frac{1}{2\pi} \int_0^{2\pi} \frac{ {f\bigl(\rme^{\rmi k}\bigr)}}{\big\lvert f\bigl(\rme^{\rmi k}\bigr)\big\rvert}\rme^{-\rmi\alpha k} \rme^{-\rmi(m-n) k}\rmd k\right)_{m,n=1}^{N}\nonumber\\\label{detcorr}
&= D_N\left[ \sqrt{\frac{f(z)}{f(1/z)}}z^{-\alpha}\right]= D_N\left[ \sqrt{\frac{f(1/z)}{f(z)}}z^{\alpha}\right],\end{align}
\big(the final equality is $\det M = \det M^{\rm T}$\big). Since the non-trivial fermionic two-point function is given by
\begin{align*}
\langle -\rmi\tilde\gamma_n \gamma_m \rangle = \int_0^{2\pi} \frac{ {f\bigl(\rme^{\rmi k}\bigr)}}{\big\lvert f\bigl(\rme^{\rmi k}\bigr)\big\rvert} \rme^{-\rmi(m-n) k}\rmd k,
\end{align*} the formula \eqref{detcorr} follows from Wick's theorem---a derivation is given in \cite{Jones2019}. At $z=1$, $f(z)$ is real and we choose the branch of the square-root in \eqref{detcorr} so that $\sqrt{f(z)/f(1/z)}$ is equal to the sign of $f(1)$. For $f(z)$ defined in \eqref{fzcanon}, the scalar symbol can be simplified to a rational function. The Fourier coefficients that give the matrix entries are evaluated in Appendix \ref{app:fourier}, while Day's formula gives us an expression for the determinant in this case \cite{Bottcher,Bottcher2006,Day}. Day's formula may also be applied to evaluate string correlations in class AIII, introduced in Section \ref{sec:AIII}.
\begin{Theorem}[Day 1975]\label{thm:Day}
Consider symbols of the form:
\begin{align}
t(z)= \rho \frac{\prod_{j=1}^{s} \left(z-\tau_j\right)}{\prod_{j=1}^{q} \left(1-z/\gamma_j\right)\prod_{j=1}^{p} \left(z-\delta_j\right)}\label{daycanon}
\end{align}
with $p\geq 1$, $q \geq 1$, $s \geq p+q$, $|\delta_j|<1$, $|\gamma_j|>1$ and $\{\tau_1, \dots, \tau_s\}$ pairwise distinct. Then we have
\begin{align*}
D_N[t(z)] = \sum_M C_M^{\vphantom N} r_M^N, \qquad N\geq 1,
\end{align*}
where the sum is taken over all subsets of $M \subseteq \{1, \dots,s\}$ such that $\lvert M \rvert = p$ and $C_M$, $r_M$ are defined as follows. For each $M$ define $M^c = \{1, \dots, s\} \setminus M$, and also define $P=\{ 1, \dots, p\}$ and $Q= \{ 1, \dots, q\}$. Then
\begin{align*}
r_M&= (-1)^{s-p} \rho \prod_{k \in M^c} \tau_k, \\
C_M &= \prod_{k \in M^c,m\in P} (\tau_k - \delta_m) \prod_{l \in Q, j\in M} (\gamma_l - \tau_j) \prod_{l \in Q, m\in P} (\gamma_l - \delta_m)^{-1} \prod_{k \in M^c,j\in M} (\tau_k - \tau_j)^{-1} .
\end{align*}
\end{Theorem}
The condition that the $\tau_j$ are pairwise distinct is necessary because differences $\tau_k - \tau_j$ appear in the denominator of $C_M$. However, we will want to study cases where the $\tau_j$ are degenerate, this is important since the factor $z^k = (z-0)^k$ appears in our analysis of correlators. As pointed out in \cite{Bottcher2006}, one can still use Day's formula by taking an appropriate limit. In particular, let us write $z^k = \lim_{\eps\rightarrow 0} \prod_{j=1}^k (z-\eps x_j)$ for $x_j$ pairwise distinct. For finite $\eps$ we do not have degenerate zeros and can apply Day's formula. Moreover, the Fourier coefficients, and hence the matrix elements of the Toeplitz determinant will depend continuously on $\eps$. Indeed, the Fourier coefficients of $\prod_{j=1}^k (z-\eps x_j)t(z) $ will just be sums over shifted Fourier coefficients of $t(z)$. Since we are taking a finite Toeplitz determinant, we can then take the limit of Day's formula as $\eps \rightarrow 0$ and this will be the determinant we are interested in.

We are also interested in the eigenvalues of the correlation matrix. For translation-invariant free-fermion chains in class BDI, the correlation matrix for a subsystem of size $N$ is a $2N\times 2N$ block Toeplitz matrix generated by the symbol $\Phi(z,0)$ \cite{Its2008}, where
 \begin{align}
\Phi(z,\lambda) = \left(\begin{matrix}\rmi \lambda & \displaystyle\sqrt{\frac{f(z)}{f(1/z)}} \\ \displaystyle -\sqrt{\frac{f(1/z)}{f(z)}} & \rmi \lambda \end{matrix}\right). \label{blocksymbol}
\end{align}
 By evaluating the determinant generated by this symbol for general $\lambda$, we can find the eigenvalues of the correlation matrix. Note that changing the branch of the square-root takes $\Phi(z,\lambda)\rightarrow-\Phi(z,-\lambda)$; since the eigenvalues come in $\pm \lambda$ pairs, they are independent of this branch choice. Moreover, notice that $\Phi(1/z,\lambda)=-\Phi(z,-\lambda)$. Again, since the eigenvalues come in $\pm \lambda$ pairs this means that this determinant is unchanged under $f(z) \rightarrow f(1/z)$. Note that following \cite{Basor2019} for $\lambda=0$, we have
 \begin{align}
D_N[\Phi(z,0)]&=(-1)^N D_N\left[\sqrt{{f(z)}/{f(1/z)}}\right]D_N\left[-\sqrt{{f(1/z)}/{f(z)}}\right]\nonumber\\&=(\langle \mathcal{O}_0(1) \mathcal{O}_0(N+1) \rangle)^2.\label{eq:detAN2}
 \end{align}
In our analysis of the entanglement spectrum in class AIII, a closely related block Toeplitz matrix arises, the symbol is given in Appendix \ref{app:AIIIproof}.

 As before, in the models that we consider these symbols simplify to a matrix where each element is a rational function. For a certain class of such rational matrix functions, we have Gorodetsky's formula \cite{Bottcher,Gorodetsky1981,Gorodetsky}. The general statement is too long to include here, but let us consider a simplified version sufficient for illustration:
\begin{Theorem}[Gorodetsky 1981]\label{thm:Gorodetsky}
Suppose $a(z) = \sum_{j=0}^s a_j z^j $ is a matrix polynomial, where $a_j \in \mathbb{C}^{r \times r}$ and suppose that $a_s$ is invertible. For symbols of the form
\begin{align*}
\Phi(z)= \frac{1}{\prod_{j=1}^{p} (1-z/\gamma_j)\prod_{j=1}^{q} (z-\delta_j)} a(z),
\end{align*}
where $|\delta_j|<1$ and $|\gamma_j|>1$, and subject to some simple conditions involving $p$, $q$, $s$ and the matrix $a_0$, then \begin{align*}
D_N[\Phi(z)] =(\det{a_s})^N \frac{\det{\mathcal{M}[N,\Phi(z)]}}{\det{\mathcal{M}[0,\Phi(z)]}}, \qquad N\geq 1,
\end{align*}
where $\mathcal{M}[n,\Phi(z)]$ is an $rs \times rs$ matrix such that $\det{\mathcal{M}[0,\Phi(z)]}\neq 0$, and that depends parametrically on $n$.
\end{Theorem}
The construction of $\mathcal{M}[n,\Phi(z)]$ can be found in \cite{Bottcher} and requires the \emph{Smith canonical form} of the matrix polynomial $a(z)$ \cite{Bottcher,Gohberg}. In particular, given $a(z),$ there exist $r \times r$ matrix polynomials~$y(z)$ and $w(z)$ with non-vanishing and $z$-independent determinants, such that $y(z) a(z) w(z)$ is a diagonal matrix $d(z)$. This matrix $d(z)$ has entries of the form $d_{kk}(z) = \prod_{j=1}^R (z-t_j)^{m_{jk}}$, where the $t_j$ are the $R$ zeros of $\det(a(z))$, and $0\leq m_{j1} \leq \dots \leq m_{jr}$ . The integers $m_{jk}$ are such that $\sum_k m_{jk} =m_j$, the multiplicity of the zero $t_j$ of $\det(a(z))$. $\mathcal{M}[n,\Phi(z)]$ requires both the entries of $y(z)$ and the $m_{jk}$. When we apply this theorem to our problem, $r=2$, $m_{j1} =0$ and~$m_{j2}=2$. We give the explicit construction of $\mathcal{M}[n,\Phi(z)]$ for this special case.

\section{Results: correlations of string operators in BDI}\label{sec:stringresults}
Our first result concerns the behaviour of the correlation functions $\langle\mathcal{O}_\alpha(1)\mathcal{O}_\alpha(N+1)\rangle$ for finite~$N$. The case with $n_z=0$ and $n_Z=0$ is trivial: in that case $f(z) = 1/z^{n_P}$ and the string correlators satisfy
$\langle\mathcal{O}_\alpha(1) \mathcal{O}_\alpha(N+1)\rangle = \delta_{\alpha,-n_P}(-1)^{N(n_P-1)}$.
For all other cases, we have an exact formula in terms of the zeros of $f(z)$. In fact, the correlators are zero for $N$ sufficiently large, apart from the cases $\omega - n_z\leq \alpha \leq \omega +n_Z$. For these cases, the formula is a linear combination of terms $r_M^N$, where each $r_M$ is a product of zeros and inverse zeros. The precise statement is quite involved, so let us first give as an example the case $f(z) = (z-a)^2(z-b)^2$, where either $a,b \in \mathbb{R}$ or $a =\overline{b}$.
\begin{Example}\label{correxample}
Let $f(z) = (z-a)^2(z-b)^2$ where $\lvert a\rvert<1$ and $\lvert b\rvert>1$. Since $n_z =n_Z= 1$, we are in the phase $\omega =2$, and the only correlators that are not exactly zero (for sufficiently large~$N$) are those with $1\leq \alpha \leq 3$. We have that
\begin{gather}
\langle\mathcal{O}_1(1) \mathcal{O}_1(N+1)\rangle = (-1)^N \frac{\bigl(b^2-1\bigr) (a b-1) }{a b^2 (b-a)}a^{N}, \qquad N\geq 1,\nonumber\\
\langle\mathcal{O}_2(1) \mathcal{O}_2(N+1)\rangle = (-1)^N \bigg(\underbrace{\frac{\bigl(1-a^2\bigr) \bigl(1-1/b^{2}\bigr) }{(1-a/b)^2}}_{\textrm{Order parameter}} + \frac{(1-ab)^2}{(b-a)^2}\left(\frac{a}{b}\right)^N \bigg),\qquad N\geq 1, \nonumber\\
\langle\mathcal{O}_3(1) \mathcal{O}_3(N+1)\rangle = (-1)^N \frac{\bigl(1-a^2\bigr) (1-a b) b}{(b-a)}b^{-N}, \qquad N\geq 1.\label{exampleO2}
\end{gather}
Let $f(z) = (z-a)^2(z-b)^2$ where $\lvert a\rvert<1$ and $\lvert b\rvert<1$. Since $n_z = 2$, $n_Z= 0$, we are in the phase~$\omega =4$, and the only correlators that are not exactly zero (for sufficiently large $N$) are those with $2\leq \alpha \leq 4$. We have that
\begin{gather}
\langle\mathcal{O}_2(1) \mathcal{O}_2(N+1)\rangle = (-1)^N \left(ab\right)^{N}, \qquad N\geq 1, \label{O2inside}\\
\langle\mathcal{O}_3(1) \mathcal{O}_3(N+1)\rangle = (-1)^N \frac{1}{b-a}\nonumber\\
\phantom{\langle\mathcal{O}_3(1) \mathcal{O}_3(N+1)\rangle =}{}\times\Big(b (1-a^2)(1-ab)b^N -a (1-b^2)(1-ab)a^N \Big), \qquad N\geq 1, \label{O3inside}\\
\langle\mathcal{O}_4(1) \mathcal{O}_4(N+1)\rangle = (-1)^N \underbrace{{\bigl(1-a^2\bigr) \bigl(1-b^2\bigr) (1-ab)^2}}_{\textrm{order parameter}} , \qquad N\geq 2. \label{O4inside}
\end{gather}
Let $f(z) = (z-a)^2(z-b)^2$ where $\lvert a\rvert>1$ and $\lvert b\rvert>1$. Since $n_z = 0$, $n_Z= 2$, we are in the phase~$\omega =0$, and the only non-zero correlators (for sufficiently large $N$) are those with~${0\leq \alpha \leq 2}$. We have that $\langle\mathcal{O}_0(1) \mathcal{O}_0(N+1)\rangle $ is given by \eqref{O4inside}, $\langle\mathcal{O}_1(1) \mathcal{O}_1(N+1)\rangle$ is given by \eqref{O3inside} and~$\langle\mathcal{O}_2(1) \mathcal{O}_2(N+1)\rangle$ is given by \eqref{O2inside} under the replacements $a\rightarrow1/a$, $b\rightarrow 1/b$.
\end{Example}
The correlations take the form of a sum of terms where each term is a coefficient that is a~rational function of zeros and inverse zeros of $f(z)$, multiplied by $r_M^N$, where $r_M$ is a product of zeros from inside the unit circle and inverse zeros from outside unit the circle. This term $r_M$ is labelled by a set $M$ that tells you which zeros and inverse zeros contribute.
The order parameter term appears for the particular set where $r_M=1$. Informally, for values of $\alpha$ outside the range of $\alpha$ given, we `run out' of zeros to include in the product, and so the correlator is zero.
Notice that the formulae given are symmetric functions of zeros inside the unit circle and, separately, the zeros outside the unit circle. Furthermore, we are able to define the non-trivial correlators for $\lvert a\rvert,\lvert b\rvert>1$ in terms of the non-trivial correlators for $\lvert a\rvert,\lvert b\rvert<1$, under the transformation $a\rightarrow1/a$, $b\rightarrow 1/b$. Moreover, for $\lvert a\rvert <1$ and $\lvert b\rvert>1$, the formula for $\langle\mathcal{O}_1(1) \mathcal{O}_1(N+1)\rangle$ is identical to the formula $\langle\mathcal{O}_3(1) \mathcal{O}_3(N+1)\rangle$ under the replacements $a\rightarrow 1/b$ and $b\rightarrow 1/a$. These relationships apply more generally, due to an identity between the $ \langle\mathcal{O}_\alpha(1) \mathcal{O}_\alpha(N+1)\rangle$ in the chain corresponding to $f(z)$ and $ \langle\mathcal{O}_{-\alpha}(1) \mathcal{O}_{-\alpha}(N+1)\rangle$ in the chain corresponding to $f(1/z)$. We will see in the proof that depending on the sign of $n_z+n_Z-n_P-\alpha$ the calculation uses either~$f(z)$ or $f(1/z)$ (this inverts the zeros), giving the two cases in the following general result:

\begin{Theorem}\label{thm:correlators}
Consider a chain with a non-trivial $f(z)$ as defined in \eqref{fzcanon}, assuming the generic case, and for each $\alpha \in \mathbb{Z}$ define $N_{\alpha} = \max\{\lvert n_z+n_Z-n_P-\alpha \rvert,1\}$. Then
\begin{enumerate}\itemsep=0pt
\item[$(1)$] for $\alpha < \omega - n_z$ and $\alpha > \omega +n_Z$
\begin{align*}
\langle\mathcal{O}_\alpha(1) \mathcal{O}_\alpha(N+1)\rangle &= 0, \qquad N\geq N_{\alpha};
\end{align*}
\item[$(2)$] for $\omega - n_z\leq \alpha \leq \omega +n_Z$ \begin{align}
\langle\mathcal{O}_\alpha(1) \mathcal{O}_\alpha(N+1)\rangle &= (-1)^{N(n_P+1)}\sum_{M} C_M^{\vphantom N} r_M^N, \qquad N\geq N_{\alpha},\label{eq:thm1}
\end{align}
where for $n_z>0$ and $n_Z>0$:\begin{itemize}\itemsep=0pt \item for $\omega - n_z\leq \alpha\leq \omega +n_Z-n_z$ the sum is over sets $M$ that label sets of zeros, with the non-zero constants $C_M$ and $r_M$ defined in Case $1$ below.
\item for $\omega +n_Z-n_z \leq \alpha\leq \omega +n_Z$ the sum is over sets $M$ that label sets of inverse zeros, with the non-zero constants $C_M$ and $r_M$ defined in Case $2$ below.\end{itemize}
If either $n_z=0$ or $n_Z=0$:
\begin{itemize}\itemsep=0pt
\item if $n_z =0$, and $\alpha<\omega+n_Z$, then put $n_z=1$, $\alpha \rightarrow \alpha+2$ and take the limit $z_1 \rightarrow 0$ with Case $1$. If $n_z=0$ and $\alpha = \omega+n_Z$ then the correlator is given by $(-1)^{N(\omega-1)} \prod_{k=1}^{n_Z} Z_k^{-N}$.
\item if $n_Z =0$ and $\alpha>\omega-n_z$, then put $n_Z=1$, and take the limit $Z_1 \rightarrow \infty$ with Case $2$. If $n_Z =0$ and $\alpha=\omega-n_z$, then the correlator is given by $(-1)^{N(\omega-1)} \prod_{j=1}^{n_z} z_j^N$.
\end{itemize}
\end{enumerate}

In both cases, in the formula \eqref{eq:thm1}, the sum is over all subsets $M \subseteq \{1,\dots,n_z+n_Z\}$ of some fixed size $\lvert M \rvert$. For each such $M$, we also define:
$
 M^c = \{1,\dots,n_z+n_Z\}\setminus M $ and further define $M_z = \{1,\dots, n_z\}$ and $M_Z= \{1,\dots, n_Z\}$. Then we have

Case $1$:
Define $\tau_j = z_j$ for $1\leq j \leq n_z$ and $\tau_{n_z+j}=Z_j$ for $1\leq j \leq n_Z$.
Then the sum in~\eqref{eq:thm1} is over all subsets $M$ of size $\lvert M \rvert =\alpha+n_P-n_z$, and $C_M$ and $r_M$ are defined by
\begin{align*}
&r_M= \frac{\prod_{k \in M^c} \tau_k} {\prod_{j=1}^{n_Z}Z_j},\\
&C_M=\prod_{k \in M^c,m\in M_Z} \bigl(\tau_k - Z^{-1}_m\bigr) \prod_{l \in M_z, j\in M} \bigl(z_l^{-1} - \tau_j\bigr) \prod_{l\in M_z}z_l^{-(n_z+n_Z -n_P-\alpha)} \\
&\phantom{C_M=}{}\times\prod_{l \in M_z, m\in M_Z} \bigl(z_l^{-1} - Z^{-1}_m\bigr)^{-1} \prod_{k \in M^c,j\in M} (\tau_k - \tau_j)^{-1} \prod_{k\in M^c}\tau_k^{-(n_z+n_Z -n_P-\alpha)} .
\end{align*}

Case $2$:
Define $\tau_j = z_j^{-1}$ for $1\leq j \leq n_z$ and $\tau_{n_z+j}=Z_j^{-1}$ for $1\leq j \leq n_Z$.
Then the sum in~\eqref{eq:thm1} is over all subsets $M$ of size $\lvert M \rvert =2n_z+n_Z-n_P-\alpha$, and $C_M$ and $r_M$ are defined by\begin{align*}
&r_M= \prod_{j=1}^{n_z}z_j{\prod_{k \in M^c} \tau_k} ,\nonumber\\
&C_M= \prod_{k \in M^c,m\in M_z} (\tau_k - z_m) \prod_{l \in M_Z, j\in M} (Z_l - \tau_j)\prod_{l \in M_Z} Z_l^{n_P+\alpha-n_z-n_Z}\nonumber\\
&\phantom{C_M=}{}\times \prod_{l \in M_Z, m\in M_z} (Z_l - z_m)^{-1} \prod_{k \in M^c,j\in M} (\tau_k - \tau_j)^{-1} \prod_{k \in M^c}\tau_k^{-(n_P+\alpha-n_z-n_Z)}. \end{align*}
Note that in both Case $1$ and Case $2$, we can have $\lvert M \rvert =0$ or $\lvert M^c \rvert =0$. For $\lvert M \rvert =0$, we define double products with one variable in $M$ that appear in the above formulae, such as~$\prod_{l \in M_z, j\in M} \bigl(z_l^{-1} - \tau_j\bigr)$, to be equal to one. Similarly double products with one variable in ${\lvert M^c \rvert =0}$ are equal to one. Note also that $\alpha=\omega +n_Z-n_z$ can be evaluated using both Case~$1$ and Case~$2$.
\end{Theorem}
\begin{Remark}\label{remark1}
Theorem \ref{thm:correlators} is proved by considering a particular limiting case of Day's formula for Toeplitz determinants (see Theorem \ref{thm:Day}). The limit in question arises by writing ${z^k = \lim_{\eps\rightarrow 0} \prod_{j=1}^k (z-\eps x_j)}$, for $x_j$ pairwise distinct, so that we can apply Day's formula. The limit of the final formula simplifies for $N\geq N_{\alpha}$ leading to the result above. However, for any $N$ such that $1\leq N<N_{\alpha}$ the steps outlined in the proof can be used to evaluate the correlator. \end{Remark}
\begin{Remark}One evaluation of this limit for $N<N_{\alpha}$ is the following: for $n_z\neq0$ and $n_Z=0$ we have that for any $\nu>0$:
\begin{align*}
 \langle \mathcal{O}_{\omega+\nu}(1) \mathcal{O}_{\omega+\nu}(N+1) \rangle &=\begin{cases} \displaystyle (-1)^{n_z(\omega-1)}\biggl(\prod_{j=1}^{n_z} z_j^\nu\biggr) \prod_{j_1=1}^{n_z} \prod_{j_2=1}^{n_z} (1-z_{j_1}z_{j_2}), & N= n_z, \\
0, & N>n_z.\end{cases}
\end{align*}
This follows from the results of \cite[Section 7]{Hartwig1969}. Note that in this case, Theorem \ref{thm:correlators} gives that this correlator is zero for $N\geq n_z + \nu$. There is an analogous formula for the case $n_Z\neq 0$ and~$n_z = 0$.
\end{Remark}
\begin{Remark}\label{remark:orderparameter}
For $\alpha = \omega$, there is always one choice of $M$ where $r_M=1$. For this choice of~$M$,~$C_M$ is the order parameter. We can evaluate this $C_M$ using Theorem \ref{thm:correlators} to reach
\begin{align}
\lim_{N\rightarrow\infty}\lvert\langle\mathcal{O}_\omega(1)\mathcal{O}_\omega(N+1)\rangle \rvert = \frac{ \prod_{j_1,j_2=1}^{n_z} (1-z_{j_1} z_{j_2})
	\prod_{k_1,k_2=1}^{n_Z} \bigl(1-\frac{1}{Z_{k_1} Z_{k_2}}\bigr)}{\prod_{j=1}^{n_z}\prod_{k=1}^{n_Z} \bigl(1-\frac{z_j}{ Z_k}\bigr)^2}. \label{eq:orderparameter}
\end{align}
This formula also holds in the non-generic case---see Remark \ref{remark:nongeneric}. Note that this result, following from Day's formula, agrees with the general formula for the order parameter given in \cite{Jones2019} as applied to our case. The general formula is proved using Szeg\H{o}'s theorem \cite{Szego}. In fact, by taking a~limit of models, one can use \eqref{eq:orderparameter} to recover the result of \cite{Jones2019} for the general case (see Section~\ref{sec:generic}).\looseness=-1

\end{Remark}
\begin{Remark}\label{remark:simplecorrelator}
For $\alpha = \omega +n_Z$ and $\alpha = \omega-n_z$, we see that there is only one choice of $M$, and hence the correlator has a particularly simple form.
Indeed, one has
\begin{gather*}
\langle\mathcal{O}_{\omega +n_Z}(1) \mathcal{O}_{\omega +n_Z}(N+1)\rangle = (-1)^{N(n_P+1)} C_M \biggl(\prod_{k=1}^{n_Z} Z_k \biggr)^{-N},\qquad N \geq \max\{n_z,1\}, \\
\langle\mathcal{O}_{\omega-n_z}(1) \mathcal{O}_{ \omega-n_z}(N+1)\rangle = (-1)^{N(n_P+1)} C_M \biggl(\prod_{j=1}^{n_z} z_j \biggr)^N,\qquad N \geq \max\{n_Z,1\}.
\end{gather*}
For $n_z=0$ or $n_Z=0$, these formulae apply and this means that the order parameter reaches its limiting value for $N\geq N_{\alpha}$ with no correction term. This property was emphasised for the case $f(z) = (z-a)^2$ in~\cite{Smith19}. While the result for $n_z=0$ or $n_Z=0$ follows from our analysis, it may also be derived as an application of the results of \cite[Section 7]{Hartwig1969}.
\end{Remark}
\begin{Remark}\label{remark:asymptotics}
The large $N$, asymptotics for each correlator can be deduced easily from Theorem~\ref{thm:correlators}. To find the dominant term, one chooses the subset $M$ such that the corresponding~$r_M$ is as large as possible. Note that choosing $M$ of size $m$ is equivalent to choosing $M^c$ of size~$n_z+n_Z-m$. We maximise $r_M$ by choosing the $n_z+n_Z-m$ largest zeros (or inverse zeros) to make up $M^c$. Concretely, for a correlator in Case~1 we choose $M^c = \{Z_{n_Z},Z_{n_Z-1}, \dots \}$ of the appropriate size. Since all zeros are either real or come in complex-conjugate pairs, this is not necessarily a unique maximum. If we find a dominant $r_M$ as described and it contains a complex zero but not its conjugate, then there will be a corresponding $M'$ with the zero replaced by its conjugate such that $\lvert r_M \rvert = \lvert r_M' \rvert$. In general, we could have more than two zeros of the same absolute value. The dominant asymptotic term will then be a sum over contributions from all~$M$ with the maximal~$r_M$. The result is in agreement with \cite[Theorem~3]{Bultheel} (where it is assumed that there is no degeneracy of the smallest zero that appears in~$r_M$ and so we have a~single dominant contribution).

The second most dominant term will be where we take the dominant $M^c$, remove the smallest zero that appears, and replace it by the largest zero that did not feature in the dominant term (up to accounting for zeros of the same absolute value as above). In general, by ordering the size of $\prod_{k \in M^c}\tau_k$, one can find as many terms in the large $N$ asymptotics as desired.
\end{Remark}
We can define the correlation length, $\xi_\alpha$ for the operators $\mathcal{O}_\alpha$ with $\alpha \neq \omega$ by
\begin{align*}
\langle\mathcal{O}_{\alpha}(1) \mathcal{O}_{\alpha}(N+1)\rangle = O^*\bigl(\rme^{-N/\xi_\alpha}\bigr)
\end{align*} as $N\rightarrow\infty$. Following \cite{Schuch06}, we write $f(x)=O^*(g(x))$ if $g(x)$ is a tight upper bound for $f(x)$; i.e., $f(x)=O(g(x))$ but $f(x)\neq o(g(x))$. This allows for oscillatory cases where $\xi$ gives the exponentially decaying envelope, but $f(x)$ may vanish for some values of $x$.
Theorem \ref{thm:correlators} and Remark \ref{remark:asymptotics} lead to:
\begin{Corollary}\label{corollary:xi}
Consider a chain with a non-trivial $f(z)$ as defined in \eqref{fzcanon}. Then the correlation length is given by
\begin{align*}
\xi_\alpha^{-1} = \begin{cases}
\displaystyle\sum_{j=1}^{\omega-\alpha} \log\bigl\lvert z_j^{-1}\bigr\rvert, & \omega-n_z\leq\alpha<\omega,\\
\displaystyle\sum_{j=1}^{\alpha-\omega} \log\vert Z_j\rvert, & \omega<\alpha\leq\omega+n_Z,\\
\infty, & \alpha<\omega-n_z\quad \mathrm{or} \quad \alpha>\omega+n_Z.
 \end{cases}
\end{align*}
\end{Corollary}
\begin{Remark}
In \cite{Jones2019}, an expression for the large $N$ asymptotics of each correlator is given. The models we consider here arise as limiting cases of models in that paper, so it is interesting to compare our results. We firstly point out that the method used in \cite{Jones2019} for evaluating the asymptotics requires the evaluation of the determinant of an $\lvert\omega-\alpha\rvert\times \lvert\omega-\alpha\rvert$ matrix, where the matrix elements are, up to some error terms, the asymptotically large Fourier coefficients of functions $l(z)$ and $m(z)$ that are defined in Appendix \ref{app:asymptotics} \cite{Hartwig1969}. Analysing this determinant for the cases studied here is non-trivial, except when $\lvert\omega-\alpha\rvert=1$. Firstly, let us suppose that~$n_Z \geq 1$ and~$n_z \geq 1$. In Appendix~\ref{app:asymptotics}, we show that $l_N =O^*\bigl(\lvert Z_1\rvert ^{-N}\bigr)$ and $m_N =O^*\bigl(\lvert z_1\rvert ^{N}\bigr)$. Then we have that for $\omega-\alpha=1$, the correlator behaves like $m_N$ and for $\omega-\alpha=-1$ the correlator behaves like $l_N$. This agrees with our results here. For example, suppose the order parameter can be calculated using Case 1. Then for $\alpha = \omega$ the dominant term comes from $M^c = \{Z_{n_Z},Z_{n_Z-1}, \dots, Z_1\}$. For $\alpha = \omega-1$, $\lvert M \rvert$ is one smaller, so $\lvert M^c \rvert$ is one larger and so dominant terms come from $ M^c = \{Z_{n_Z},Z_{n_Z-1}, \dots, Z_1,z_j\}$ where $\lvert z_j\rvert =z_1$---each of these sets has $\lvert r_M\rvert =\lvert z_1\rvert$, and hence the asymptotic behaviour $O^*\bigl(\lvert r_M\rvert^N\bigr)$ agrees with $m_N$. We give an explicit example in Appendix \ref{app:asymptotics}, showing that the terms match exactly.

An important difference with the results of \cite{Jones2019} is that, in the more general class considered there, \emph{generically in that class} the correlation length of each string operator is given by ${\xi_\alpha^{-1} = \xi^{-1}\lvert \omega-\alpha \rvert}$; where the basic correlation length $\xi$ is given by \begin{align*} \xi^{-1} = \min\big\{\log\big\lvert z_1^{-1}\big\rvert,\log\lvert Z_1\rvert\big\}. \end{align*}
This means that the decay of these correlators depends only on the zero closest to the unit circle. By comparing this result with Corollary \ref{corollary:xi}, for $\lvert \omega-\alpha \rvert>1$ we see that generically in the class of models given by \eqref{fzcanon} we have faster decay than in the generic class of models considered in~\cite{Jones2019}.
\end{Remark}
\begin{Remark}
Recall that the one-particle energies of our model are given by $\big\lvert f\bigl(\rme^{\rmi k}\bigr)\big\rvert$. Allowing for a non-zero temperature, $T$, the following equation appears in studying zeros of the partition function \cite{Timonin21,Tong06} (often one considers a large but finite system, so that $k$ is quantised):
\begin{align}
\big\lvert f\bigl(\rme^{\rmi k}\bigr)\big\rvert = \rmi(2n+1)\pi T, \qquad n\in\mathbb{Z}. \label{eq:partitionfunction}
\end{align}
For example, the Fisher zeros correspond to solutions of this equation\footnote{More generally, for a system of size $L$, Fisher zeros are solutions of $Z(L,\beta)=0$. In our case, we have that $Z(L,\beta)=\prod_{k_m}\bigl(1+\rme^{-\beta \lvert f(k_m)\rvert}\bigr)$ (where we have fixed the ground state to have zero energy), leading to \eqref{eq:partitionfunction}.} in the complex $\beta$ plane for $\beta = 1/T$ \cite{Fisher65}. If we instead take this equation and complexify $z=\rme^{\rmi k}$, we see that at zero temperature the zeros of $f(z)$ and zeros of $f(1/z)$ coincide with the set of zeros of \eqref{eq:partitionfunction}; this method is used to study disorder lines in \cite{Timonin21}. Note that to analytically continue the left-hand side of \eqref{eq:partitionfunction} we can write $\lvert f(z) \rvert = \sqrt{f(z)f(1/z)}$ on the unit circle.
The subclass of models considered in this paper then corresponds to degeneracy of solutions to~\eqref{eq:partitionfunction}---it would be of interest to understand how our results generalise to the finite-temperature case.
\end{Remark}
\begin{Remark}\label{remark:emptiness}
Another important correlator is the emptiness formation probability \cite{Abanov2002,Franchini2005,Korepin1997,Korepin1994,Shiroishi2001}, given by
\begin{align*}
P(N)=\left\langle \prod_{j=1}^N c^{\vphantom \dagger}_j c^\dagger_j \right \rangle .
\end{align*}
This is a string correlator for complex fermions, rather than the Majorana fermions considered above, and is the probability, in the ground state, that $N$ consecutive fermionic sites are unoccupied (equivalently, in the spin chain picture, that there are $N$ consecutive up-spins). A closed formula for $P(N)$ was found for the disorder line in the XY model in \cite{Franchini2005}, using the simple formula for the ground state in that case. We remark that closed formulae for $P(N)$ can be found for the class of models considered in this paper, since this correlator may also be evaluated using Day's formula. We give an example below, and in Section \ref{sec:emptiness} outline the general approach.
\end{Remark}
\begin{Example}\label{example:emptiness}
Let $f(z) = z^{-2} (z-a)^2(z-b)^2$, with $\lvert a\rvert<1$ and $\lvert b\rvert>1$. Then
\begin{align*}
P(N) = \left(\frac{b^{-1}-a}{2}\right)^N\left(\frac{(1-a)(1+b)}{2(b-a)}+ (-1)^N\frac{(1+a)(1-b)}{2(a-b)} \right), \qquad N\geq 1.
\end{align*}
This is derived using Day's formula in Section \ref{sec:emptiness}.
\end{Example}
\section{Results: correlation matrix in BDI}\label{sec:corrmatrixresults}
Let $a_{2n-1} = \gamma_n$ and $a_{2n} =\tilde\gamma_n$. The correlation matrix for a subsystem of size $N$ is defined by
\begin{align*}
A_N= \left( \rmi \langle a_j a_k \rangle - \rmi\delta_{jk}\right)_{j,k=1,\dots, 2N} .
\end{align*}
The matrix $A_N$ has $2N$ imaginary eigenvalues, $\{\pm\rmi\nu_1,\dots, \pm\rmi\nu_N\}$. $A_N$ always has an even number of eigenvalues that are equal to zero. If there are $2m$ such eigenvalues then $\nu_1,\dots, \nu_m=0$ and the remaining $\nu_j$ are non-zero. With a slight abuse of terminology to simplify our discussion, we will call the $N$ non-negative imaginary parts $\{\nu_1,\dots, \nu_N\}$ `the eigenvalues' of the correlation matrix (these eigenvalues are zeros of the characteristic polynomial $\det(\rmi\lambda-A_N)$). Let $\rho_N$ be the density matrix of a subsystem of our chain of size $N$, the eigenvalues of $\rho_N$ can be computed from the eigenvalues of the correlation matrix. Indeed, the following formula (given in, for example, \cite{Vidal2003}) allows us to write the eigenvalues of the reduced density matrix in terms of the eigenvalues of the correlation matrix:
\begin{align*}
\lambda_{\zeta_1\dots \zeta_N} =\prod_{j=1}^N \frac{1+(-1)^{\zeta_i}\nu_j}{2}, \qquad \zeta_j=0,1.
\end{align*}
We go over all values $\zeta_i$, so have $2^N$ eigenvalues of $\rho_N$ from the $N$ eigenvalues of the correlation matrix $\{\nu_1,\dots, \nu_N\}$. That the $\lambda$ are eigenvalues of a density matrix means that each $\nu_j \leq 1$. An eigenvalue $\nu_j=1$ we call trivial. Moreover, the von Neumann entanglement entropy ($S(N)=-\tr(\rho_N\log{\rho_N})$), or more generally the R\'enyi entropy $\bigl(S_\alpha(N)=\frac{1}{1-\alpha}\log\tr(\rho_N^\alpha)\bigr)$, has a simple expression in terms of the non-trivial eigenvalues $\nu_j$ \cite{Franchini,Its2009,Its2008,Keating2004,Vidal2003}.
\par
In our class of interest, there are only a finite number of non-trivial eigenvalues as $N\rightarrow \infty$. For $f(z)=z^k$, then \[\det(\rmi\lambda-A_N)=\bigl(-\lambda^{2}\bigr)^{\lvert k\rvert}\bigl(1-\lambda^2\bigr)^{N-\lvert k\rvert}.\] In all other cases, the claim follows from:
\begin{Theorem}\label{thm:corrmatrix}
Consider a chain with a non-trivial $f(z)$ as defined in \eqref{fzcanon} such that $n_P=n_z+n_Z$. We allow the non-generic case where zeros may coincide, but assume that $z_j\neq Z_k^{-1}$ for any $j$, $k$. Denote the correlation matrix on a subsystem of size $N$ by $A_N$. Then
\begin{align*}
\det(\rmi \lambda-A_N)= \bigl(1- \lambda^2\bigr)^N \left(\frac{\prod_{j=1}^{n_z} z_j^2}{\prod_{k=1}^{n_Z} Z_k^2}\right)^N \frac{\det{M(N,\lambda)}}{\det{M(0,\lambda)}}, \end{align*}
where $M(n,\lambda)$ is a $4(n_z+n_Z)\times 4(n_z+n_Z)$ matrix that can be determined from the zeros of~$f(z)$ and the Smith canonical form of a known $2\times 2$ matrix polynomial $($defined in \eqref{a11}--\eqref{a21}$)$. Moreover, there exist $d', d'' \in \mathbb{N}$ $($independent of $n)$ such that
\begin{align*}
\det{M(n,\lambda)} = \mu \left(\frac{\prod_{j=1}^{n_z} z_j^2}{\prod_{k=1}^{n_Z} Z_k^2}\right)^{-n} \prod_{j_1=1}^{d'}\bigl(\lambda^2-\tilde \nu_{j_1}(n)^2\bigr)\prod_{j_2=1}^{d''}(\lambda-c_{j_2}). \end{align*}
The $\pm\tilde{\nu}_j$ and $c_j$ are zeros of $\det{M(n,\lambda)}$ as a function of $\lambda$. We differentiate between the zeros that depend on $n$, $\pm\tilde\nu_j(n)$, and those that do not, $c_j$. $($Along with the overall constant $\mu$, the~$c_j$ drop out of the ratio of interest.$)$ The $n$-dependent zeros satisfy $\tilde \nu_j(0)=1$, $0\leq\tilde \nu_j(n)\leq 1$ for~$n\geq 1$ and $\tilde \nu_j(n)<1$ for some $n>0$.

For any $f(z)$ as defined in \eqref{fzcanon} such that $n_P \neq n_z+n_Z$, the characteristic polynomial of the correlation matrix can be found by taking a limit of a related case where $n_P = n_z+n_Z$ that can be evaluated using the above. Details of this limit are discussed in Remark {\rm \ref{limitremark}}.
\end{Theorem}
As we explain below in Remark \ref{remark:nongeneric}, the assumption that $f(z)$ cannot contain mutually inverse zeros can be trivially accounted for. Then we have

\begin{Corollary}\label{corollary:characteristicpolynomial}
Given any $f(z)$ as defined in \eqref{fzcanon}, with no assumptions of a generic case, Theorem {\rm \ref{thm:corrmatrix}} leads to the following expression for the characteristic polynomial of the correlation matrix:
\begin{align}
\det(\rmi\lambda-A_N)= \bigl(1- \lambda^2\bigr)^{N-d} \times \prod_{j=1}^{d}\bigl(\tilde \nu_j(N)^2-\lambda^2\bigr),\label{characteristicpolynomial}
\end{align}
where $d\in\mathbb{N}$ is independent of $N$.
\end{Corollary}

By computing $\det{M(N,\lambda)}$ we can determine $d$, the number of non-trivial eigenvalues of the correlation matrix. This is illustrated in Examples \ref{examplecorrmatrix} and \ref{examplecorrmatrix2}.

\begin{Example} \label{examplecorrmatrix} Let $f(z) = \frac{1}{z}(z-b)^2$ with $\lvert b\rvert <1$. This corresponds to the aforementioned disorder line of the XY model \cite{Chung01}. We can evaluate the terms appearing in Theorem \ref{thm:corrmatrix} as follows. Using the definition of $M(N,\lambda)$ given in Section \ref{sec:corrmatrix} and the Smith canonical form for this example given in Appendix \ref{app:smith}, we have
\begin{align*}
&M(N,\lambda)=\bigl(b^2-1\bigr)\left(
\begin{matrix}
 \lambda\bigl(1 -b^2\bigr) & 0 & 0 & 0 \\
 \lambda \bigl(\lambda ^2-3\bigr)b & \rmi \lambda ^2 & \lambda b^N & 0 \\
 0 & 0 & 0 & -\rmi \lambda ^2\bigl(b^2-1\bigr)b^{-(N+2)} \\
 0 & -\rmi \lambda ^2 & -\lambda ^3 b^{-N} & -\rmi \lambda ^2 b^{-(N+1)} \bigl(\bigl(b^2-1\bigr) N-2\bigr) \\
\end{matrix}
\right),\\
&\det{M(N,\lambda)}=\bigl(-b^{-2}\bigl(b^2-1\bigr)^{6} \lambda^{6}\bigr) b^{-2N}\bigl(\lambda^2 -b^{2N}\bigr),\\
&\frac{\det{M(N,\lambda)}}{\det{M(0,\lambda)}}=\frac{b^{-2N} \bigl(\lambda^2 -b^{2N}\bigr)}{\lambda^2-1},
\qquad \det(\rmi\lambda-A_N)= \bigl(1- \lambda^2\bigr)^{N-1}\bigl(b^{2N}-\lambda^2\bigr). \end{align*}
This means that there is one non-trivial eigenvalue of the correlation matrix for a subsystem of size $N\colon \nu_1=\lvert b\rvert^{N}$. As $N\rightarrow \infty$ the limit gives one zero eigenvalue. This agrees with taking the limit as we approach the disorder line in the XY model---there the infinitely many correlation matrix eigenvalues in the limit $N\rightarrow \infty$ are known for all values of the couplings (away from the disorder line these models are outside the class analysed here) \cite{Franchini2,Its2005,Its2007}.

For $f(z) = \frac{1}{z}(z-b)^2$ with $\lvert b\rvert >1$, there is one non-trivial eigenvalue for a subsystem of size $N\colon\nu_1=\lvert b\rvert^{-N}$. This follows from the result for $\lvert b\rvert <1$ and Remark \ref{limitremark} below.
\end{Example}
\begin{Example} \label{examplecorrmatrix2} Let $f(z) = \frac{1}{z^2}(z-a)^2(z-b)^2$. First consider the case $\lvert a\rvert <1$ and $\lvert b\rvert <1$. We can evaluate the terms appearing in Theorem \ref{thm:corrmatrix} (see also Appendix \ref{app:smith}) to give
\begin{align}
\frac{\det{M(N,\lambda)}}{\det{M(0,\lambda)}}={}&\frac{a^{-2N} b^{-2N}}{(\lambda^2-1)^2}\Bigg(\lambda^4 + \lambda^2 \frac{a^N b^N}{(a-b)^2}\nonumber\\
&\times\left( 2(1-a^2)(1-b^2)-(1-a b)^2\left(\left(\frac{a}{b}\right)^N +\left(\frac{b}{a}\right)^N\right)\right) +a^{2N}b^{2N}\Bigg) \label{example3eq}.
\end{align}
Thus we have two non-trivial eigenvalues of the correlation matrix for a subsystem of size $N$, given by the zeros of \eqref{example3eq}. As $N\rightarrow \infty$ the coefficients of $\lambda^2$ and $\lambda^0$ go to zero, so in this limit we have two non-trivial eigenvalues equal to zero.

Now consider the case $\lvert a\rvert <1$ and $\lvert b\rvert >1$. We can evaluate the terms appearing in Theorem~\ref{thm:corrmatrix} to give
\begin{align}
\frac{\det{M(N,\lambda)}}{\det{M(0,\lambda)}}={}&\frac{a^{-2N} b^{2N}}{(\lambda^2-1)^2}\Bigg(\lambda^4+\frac{ \lambda^2 }{(b-a)^2}\bigl( 2\bigl(1-a^2\bigr)\bigl(1-b^2\bigr)-(1-a b)^2\bigl( a^{2N}+b^{-2N}\bigr)\bigr)\nonumber\\
&+\left( \frac{a^Nb^{-N}(1-ab)^2-\bigl(1-a^2\bigr)\bigl(1-b^2\bigr)}{(b-a)^2}\right)^2\Bigg).\label{example3eq2}
\end{align}
Taking the limit $N\rightarrow \infty$ gives us that the two non-trivial eigenvalues of the correlation matrix are degenerate and given by
\begin{align*}
\nu_1^2=\nu_2^2 = {\frac{\bigl(1-a^2\bigr)\bigl(1-1/b^2\bigr)}{(1-a/b)^2}}.
\end{align*}
Note that these limiting eigenvalues are equal to each other and to the order parameter, given in \eqref{exampleO2}---these points are discussed in Remarks \ref{remark:symmetry} and \ref{remark:orderparameter2}.

The final case $\lvert a\rvert >1$ and $\lvert b\rvert >1$ is equivalent to the first case, where we make the replacement $a\rightarrow 1/a$ and $b \rightarrow 1/b$ in all formulae. We explain this in general terms in Remark \ref{limitremark}.
\end{Example}
\begin{Remark}\label{MPSremark}
Corollary \ref{corollary:characteristicpolynomial} tells us that as we take a subsystem of size $N\rightarrow \infty$, there are a finite number of non-zero eigenvalues of the reduced density matrix. This in turn directly implies that the ground state for $f(z)$ of the form \eqref{fzcanon} can be written as an MPS of fixed bond dimension. Except for a measure zero set of cases, an MPS is explicitly constructed in \cite{Jones21} using different methods. The results of that paper give us an upper bound on $d$, the number of eigenvalues that depend on $n$ in \eqref{characteristicpolynomial}. The bond dimension, $\chi$, of the optimal MPS representation of the ground state is related to $d$ by $2^d \leq \chi^2$. The upper bound is determined by the range of the Hamiltonian: $d \leq 2\log_2(\chi) \leq 2\lceil \mathrm{range}(H)/2\rceil$, where $\mathrm{range}(H)$ is the largest value of $\lvert \alpha \rvert$ such that $t_\alpha\neq0$. This holds in Examples \ref{examplecorrmatrix} and \ref{examplecorrmatrix2}. It would be interesting to derive this bound in general using the methods of this work.
\end{Remark}
\begin{Remark} The correlation matrix is of block Toeplitz form, defined in Section \ref{sec:toeplitz}.
The Szeg\H{o}--Widom theorem for block Toeplitz determinants \cite{Widom1974} tell us that as $N \rightarrow \infty$ \begin{align*}
\det(\rmi\lambda-A_N) = \bigl(1- \lambda^2\bigr)^N E_W(1+o(1)),\end{align*}
where $E_W$, Widom's constant, may be zero (see, for example, \cite{Basor2019}).
This theorem is discussed in the context of the correlation matrix of a quantum chain in \cite{Its2009,Its2008}.

Applying Theorem \ref{thm:corrmatrix} allows us to derive formulae for $E_W$ in our case. For example, for $f(z)=\frac{1}{z^2}(z-a)^2(z-b)^2$ with $\lvert a \rvert <1$ and $\lvert b\rvert >1$, by taking the limit of \eqref{example3eq2} we have that $E_W$ is given by
\begin{align*}
E_W=\frac{1}{\bigl(1-\lambda^2\bigr)^2}\left(\lambda^2-\frac{\bigl(1-a^2\bigr)\bigl(1-1/b^2\bigr)}{(1-a/b)^2}\right)^2.
\end{align*}
We prove Theorem \ref{thm:corrmatrix} using Gorodetsky's formula (see Theorem \ref{thm:Gorodetsky}). Note that \cite{Bottcher88} gives a detailed discussion of asymptotics of determinants of block Toeplitz matrices that can be analysed using Gorodetsky's formula. As with our Theorem \ref{thm:corrmatrix}, to use these results to derive the non-trivial part of the asymptotics, $E_W$, we require an expression for the matrices in the Smith canonical form of the relevant matrix polynomial.

The models we consider are limiting cases of those studied in \cite{Its2008}. Another approach to finding the asymptotics would be to take a limit of the results of that paper.
\end{Remark}
\begin{Remark}\label{limitremark}
Given $f(z)$ defined in \eqref{fzcanon}, and suppose that $n_P=n_z+n_Z-k$ for $k> 0$. We can write this as
\begin{align*}
f(z) &= \frac{z^k}{z^{n_z+n_Z}}\prod_{j=1}^{n_z} \left(z-z_j\right)^2\prod_{k'=1}^{n_Z} \left(z-Z_{k'}\right)^2=\lim_{\eps \rightarrow 0}\frac{1}{z^{n_z+n_Z+k}} \prod_{j=1}^{n_z+k} \left(z-z_j\right)^2\prod_{k'=1}^{n_Z} \left(z-Z_{k'}\right)^2,
\end{align*}
where $z_j = \eps x_j$ for $n_z+1 \leq j \leq n_z+k$. Before taking the limit $\eps \rightarrow 0$, we can define $n'_z=n_z+k$ and then apply Theorem \ref{thm:corrmatrix} with $n_P=n'_z+n_Z$. Since we are evaluating the finite determinant $\det(\rmi\lambda-A_N)$, taking the limit $\eps \rightarrow 0$ of this result gives us the characteristic polynomial of the correlation matrix for cases with $n_P=n_z+n_Z-k$.

We also have that the eigenvalues of the correlation matrix for the system corresponding to $f(z)$ are identical to the eigenvalues of the correlation matrix for the system corresponding to $f(1/z)$. This follows from the block Toeplitz form defined in Section \ref{sec:toeplitz}. Moreover, for the purposes of calculating the eigenvalues of the correlation matrix, this transformation amounts to replacing all zeros by their inverse, and replacing $n_P \rightarrow 2(n_z+n_Z)-n_P$.
This allows us to analyse $n_P=n_z+n_z+k$ for $k>0$ as follows:
\begin{align*}
f(1/z) &= \frac{z^k}{z^{n_z+n_Z}}\prod_{j=1}^{n_z} \left(z-1/z_j\right)^2\prod_{k'=1}^{n_Z} \left(z-1/Z_{k'}\right)^2.
\end{align*}
Using the same reasoning as above, we then apply Theorem \ref{thm:corrmatrix} with $n_P= n'_z+n'_Z$, where $n'_z=n_Z+k$ and $n'_Z=n_z$, and then take the limit where $k$ zeros inside the unit circle go to zero.
\end{Remark}
\begin{Example}
We can use Remark \ref{limitremark} and Example \ref{examplecorrmatrix2} to find the correlation matrix eigenvalues for $f(z) = (z-b)^2$. For $\lvert b \rvert <1$ we take the limit $a\rightarrow 0$ in \eqref{example3eq}. For $\lvert b \rvert >1$, we take the limit $a\rightarrow 0$ in \eqref{example3eq2}. Moreover, the correlation matrix eigenvalues for $f(z) =\frac{1}{z^2} (z-b)^2$ can be found by replacing $b \rightarrow 1/b$ appropriately in the above results.
\end{Example}
\begin{Remark}\label{remark:symmetry}
In Example \ref{examplecorrmatrix2}, on taking the subsystem size $N \rightarrow \infty$, we had two non-trivial eigenvalues such that $\nu_1=\nu_2$. Such a symmetry, where we have a twofold degeneracy of non-trivial correlation matrix eigenvalues, was observed for the infinitely many non-trivial $\nu_m$ in the XY model in \cite{Its2005,Its2007}. In fact in that case $\nu_1$ is not degenerate, but for all $m>1$ we have $\nu_{2m}=\nu_{2m+1}$ in the limit. This has a straightforward physical origin, as pointed out in \cite{Peschel2004}.
Let us consider the spin chain picture, with the model given in Appendix \ref{app:spin}. If the winding number is even, there is no symmetry breaking, and the correlation length is thus finite. Hence, for a~large enough block, the two edges will decouple. Moreover, there is a symmetry between the two edges.\footnote{For example, note that bond-centered inversion swaps the two edges and maps $f(z) \to f(1/z)$; an on-site rotation over an angle $\pi$ also maps $f(z) \to f(1/z)$. In combination, we thus have a symmetry of the model which exchanges the two edges of the block.} All correlation eigenvalues (in the limit $N \rightarrow \infty$) must thus come in degenerate pairs, corresponding to the correlations associated to each edge and the exterior of the block. For odd winding numbers, the ground state spontaneously breaks spin-flip symmetry. Hence, a~symmetry-preserving ground state can be seen as a macroscopic `cat state' superposition of two ground states which obey the aforementioned doubled correlation matrix eigenvalues, resulting in one additional, non-degenerate correlation eigenvalue. This explains the observations in Example~\ref{examplecorrmatrix2} and \cite{Its2005,Its2007}. On these physical grounds, we conjecture this property to hold much more generally, although we do not have a proof of this even for the class considered in this paper.\looseness=-1
\end{Remark}
\begin{Remark}\label{remark:orderparameter2}
We have the following relationship between the correlation matrix and a string correlation function that may be evaluated with Theorem \ref{thm:correlators}:
\begin{align}
\det(A_N) = \prod_{j=1}^N \nu_j^2= (\langle \mathcal{O}_0(1) \mathcal{O}_0(N+1) \rangle)^2.\label{eq:detAN}
\end{align}
This follows from the block Toeplitz structure, see also \eqref{eq:detAN2}. This relationship holds for all $N$, and one can check, for example,
that the product of the zeros of \eqref{example3eq2} is equal to the square of the right-hand side\footnote{Example \ref{correxample} has $n_P=0$, while Example \ref{examplecorrmatrix2} has $n_P=2$, so this is the relevant correlator.} of \eqref{exampleO2}. Moreover, suppose that we have two non-trivial correlation eigenvalues, and assume the symmetry argued in Remark \ref{remark:symmetry}, so that $\nu_1=\nu_2$ in the limit~$N\rightarrow \infty$. Then
it follows immediately from \eqref{eq:detAN} that the limiting correlation eigenvalues are equal to the order parameter, as observed in Example \ref{examplecorrmatrix2}. Equation \eqref{eq:detAN} also shows that, in the limit $N\rightarrow \infty$, the correlation matrix has at least one zero eigenvalue if and only if the winding number is non-zero. Equivalently, we have a degeneracy of the entanglement spectrum if and only if the trivial string order parameter $\mathcal O_0(N)$ does not have long-range order, consistent with the physical fingerprints associated to non-trivial SPT phases \cite{Pollmann10}.
\end{Remark}
\begin{Remark}\label{remark:nongeneric}
Throughout we have assumed that $f(z)$ defined in \eqref{fzcanon} has a positive overall sign. For our purposes, if the overall sign is negative, the correlators $\langle\mathcal{O}_\alpha(1) \mathcal{O}_\alpha(N+1)\rangle $ are multiplied by $(-1)^N$---this is shown in Section \ref{sec:string}. Moreover, the correlation matrix eigenvalues are unchanged---see Section \ref{sec:toeplitz}. However, at the level of the Hamiltonian the sign-change can cause frustration, and in certain cases care must be taken with boundary conditions and the thermodynamic limit before evaluating physical quantities \cite{Maric2020b,Maric2020c,Maric2020}.

We define the generic case for \eqref{fzcanon} to be where $\big\{z_{j_1}^{\vphantom {-1}},z_{j_2}^{-1},Z_{k_1}^{\vphantom {-1}},Z_{k_2}^{-1}\big\}_{j_1,j_2=1,\dots, n_z; k_1,k_2=1,\dots, n_Z\vphantom{\big|}}$ are pairwise distinct. If we allow mutually inverse zeros, i.e., some $z_j = Z_k^{-1}$, then it turns out that the ground state is equivalent to a related generic model: we simply remove all of these mutually inverse pairs of zeros and shift $n_P$. This is also the case when we allow zeros on the unit circle of even multiplicity. A full analysis of these cases can be found in the paper \cite{Jones21}. We can further relax the condition that the zeros are pairwise distinct by writing, say, $z_j=z_{j'}+\eps$ and taking the limit $\eps\rightarrow 0$, as in Remarks \ref{remark1} and \ref{limitremark}.

Let us consider more carefully the case of degenerate zeros with $z_j=z_{j'}+\eps$, or $Z_j = Z_j+\eps $ in Theorem \ref{thm:correlators}. An example would be given by the correlator $\langle\mathcal{O}_3(1)\mathcal{O}_3(N+1)\rangle$ given in \eqref{O3inside}, where we set $b=a+\eps$. Then for $\eps\rightarrow 0$,
\begin{align}
\langle\mathcal{O}_3(1)\mathcal{O}_3(N+1)\rangle = (-1)^N \bigl(1-a^2\bigr)\bigl(\bigl(1-a^2\bigr)N+1+a^2\bigr)a^N (1+o(1)). \label{eq:nongenericcorrelator}
\end{align}
This limit is non-trivial in that the denominator of \eqref{O3inside} is order $\eps$---the constant term in the numerator cancels to give a finite limit. This illustrates the general case. Suppose $M$ contains~$z_j$ but not $z_j+\eps$, then the term $\prod_{k \in M^c,j\in M} (\tau_k - \tau_j)^{-1}$ that appears in $C_M$ diverges as $\eps \rightarrow 0$. However, there will be a corresponding set $M'$ that is the same as $M$ except that it contains~$z_j+\eps$ in place of $z_j$. Then $C_{M'}$ also diverges, but we have that $C_M^{\vphantom N}r_M^N+C_{M'}^{\vphantom N}r_{M'}^N$ is finite. This discussion further generalises to multiple nearly degenerate zeros. This cancellation must occur in general since Day's formula is an exact formula for the determinant, and by considering the matrix elements this limit is well behaved (see the discussion in Sections~\ref{sec:toeplitz} and~\ref{sec:string}).\looseness=1

This analysis allows us to deduce that the formula for the order parameter given in Remark~\ref{remark:orderparameter} applies also for the case of degenerate zeros. In fact, this formula holds in all non-generic cases. Recall that if we have mutually inverse zeros, this model has the same ground state as the model with those zeros removed (and with a shift of $n_P$ that fixes the winding number) \cite{Jones21}. By removing the zeros, we are in a case that we have already analysed and so can evaluate the order parameter with \eqref{eq:orderparameter}. However, this result agrees with keeping the zeros and evaluating~\eqref{eq:orderparameter}, any terms involving zeros that are mutually inverse will cancel.\looseness=1
\end{Remark}
\section{Results: correlations in AIII} \label{sec:AIII}
\subsection{The AIII class}
We now consider a different class of models on a one-dimensional fermionic chain. Let us take $2L$ spinless fermionic degrees of freedom $c^\dagger_{A,n}$ and $c^\dagger_{B,n}$ for $n=1,\dots, L$. The class of AIII Hamiltonians~\cite{Altland97,Ryu10} are charge-conserving and are symmetric under the anti-unitary involution~$T$ such that $Tc^\dagger_{A,n} T=c_{A,n}$ and $T c^\dagger_{B,n}T=-c_{B,n}$ (physically $T$ is the sublattice symmetry). The class of models which are translation-invariant with respect to the two-site unit cell is given by
\begin{align}
H = \sum_{\kappa,n} \tau_\kappa^{\vphantom \dagger} c^\dagger_{B,n}c^{\vphantom \dagger}_{A,n+\kappa} + \overline{\tau}_\kappa^{\vphantom \dagger} c^\dagger_{A,n+\kappa}c^{\vphantom\dagger}_{B,n}. \label{H_AIII}
\end{align}
We can diagonalise $H=\sum_k \lvert f\bigl(\rme^{\rmi k}\bigr)\rvert\bigl(\eta^\dagger_{+,k}\eta^{\vphantom\dagger}_{+,k} - \eta^\dagger_{-,k}\eta^{\vphantom\dagger}_{-,k}\bigr)$ with bands given\footnote{The branch of the square-root is chosen so that this is equal to $f(z)/\lvert f(z)\rvert$ for $z$ on the unit circle.} by
\begin{align}
\eta_{\pm,k}^\dagger = \frac{1}{\sqrt{2}}\left(c^\dagger_{A,k} \pm \sqrt{\frac{f(z)}{\overline{f}(1/z)}}c^\dagger_{B,k}\right), \label{AIIImode}
\end{align}
where $f(z) =\sum_{\kappa} \tau_\kappa z^\kappa$ and $\overline{f}(z) =\sum_{\kappa} \overline{\tau}_\kappa z^\kappa$. As in BDI, the gapped phase diagram is characterised by $\omega$, the winding number of $f\bigl(\rme^{\rmi k}\bigr)$. Since $f(z)$ is a Laurent polynomial, we have that~$\omega = N_z-N_P$ where $N_z$ is the number of zeros inside the unit circle, and $N_P$ is the degree of the pole at zero. Examples in this class include the Su--Schrieffer--Heeger model \cite{Su1979}. Non-trivial ground state correlations may be easily derived from \eqref{AIIImode} as
\begin{align}
\overline{\big\langle c^\dagger_{B,n-\alpha}c^{\vphantom\dagger}_{A,n} \big\rangle}= \big\langle c^{ \dagger}_{A,n}c^{\vphantom\dagger}_{B,n-\alpha} \big\rangle = -\frac{1}{4\pi\rmi}\int\sqrt{\frac{f(z)}{\overline{f}(1/z)}}z^{-\alpha-1} \rmd z \label{AIIItwopoint}
\end{align}
and $\big\langle c^\dagger_{A,n}c^{\vphantom \dagger}_{A,n-\alpha}\big\rangle=\big\langle c^\dagger_{B,n}c^{\vphantom \dagger}_{B,n-\alpha}\big\rangle=\frac{1}{2}\delta_{\alpha,0}$. For the subclass of models we focus on, we give these two-point functions in Appendix \ref{app:fourier}.

A natural change of variables is defined by
\begin{align*}
c_{A,n}=\frac{1}{2}(-\gamma_{2n} + \rmi\gamma_{2n-1} ), \qquad
c_{B,n}=\frac{1}{2}(\tilde\gamma_{2n-1} + \rmi\tilde\gamma_{2n} ),
\end{align*}
where the $\gamma_n$, $\tilde\gamma_n$ are Majorana operators as introduced in Section \ref{sec:model}. The Hamiltonian \eqref{H_AIII} becomes:
\begin{align}
H = \frac{\rmi}{2} \sum_{\alpha,n}\bigl( \Re(\tau_\alpha) \tilde{\gamma}_n \gamma_{n+2\alpha} + \Im(\tau_\alpha) (-\tilde{\gamma}_{2n-1} \gamma_{2n+2\alpha}+\tilde{\gamma}_{2n}\gamma_{2n+2\alpha-1})\bigr)\label{H_AIII_majorana}.
\end{align}
The sublattice symmetry $T$ acts as $T \gamma_n T =\gamma_n$, $T\tilde\gamma_nT=-\tilde \gamma_n$, and we see that \eqref{H_AIII_majorana} is a BDI Majorana chain, with two-site unit cell. If $\tau_\alpha \in \mathbb{R}$ then the model is actually translation-invariant, and is a special case of \eqref{HBDI}, while for complex $\tau_\alpha$ this class of models is distinct.
\subsection{Correlations of string operators}
The string order parameters in this class will, as before, be given by string correlators of fermionic parity strings with certain end-point operators
\begin{align}
\tilde{\mathcal{O}}_0(n) &= \exp\left(\sum_{m=1}^{n-1}\rmi\pi\bigl(c^\dagger_{A,m}c^{\vphantom \dagger}_{A,m}+c^\dagger_{B,m}c^{\vphantom \dagger}_{B,m}\bigr)\right) =(-1)^{n-1}\prod_{m=1}^{n-1} \rmi\tilde{\gamma}_{2m-1}\gamma_{2m-1}\rmi\tilde{\gamma}_{2m}\gamma_{2m},\nonumber\\
\tilde{\mathcal{O}}_\alpha(n) &=\exp\left(\sum_{m=1}^{n-1}\rmi\pi\bigl(c^\dagger_{A,m}c^{\vphantom \dagger}_{A,m}+c^\dagger_{B,m}c^{\vphantom \dagger}_{B,m}\bigr)\right)\prod_{j=n}^{n+\alpha-1}\bigl(1-2c^\dagger_{A,j}c^{\vphantom \dagger}_{A,j}\bigr)\nonumber\\& =(-1)^{n-1}\Bigg(\prod_{m=1}^{n-1} \rmi\tilde{\gamma}_{2m-1}\gamma_{2m-1}\rmi\tilde{\gamma}_{2m}\gamma_{2m} \Bigg)\prod_{j=n}^{n+\alpha-1}(-\rmi\gamma_{2j-1}\gamma_{2j})\qquad \textrm{for~}\alpha>0,\nonumber\\
\tilde{\mathcal{O}}_{\alpha}(n) &= \exp\left(\sum_{m=1}^{n-1}\rmi\pi\bigl(c^\dagger_{A,m}c^{\vphantom \dagger}_{A,m}+c^\dagger_{B,m}c^{\vphantom \dagger}_{B,m}\bigr)\right)\prod_{j=n}^{n+\lvert\alpha\rvert-1}\bigl(1-2c^\dagger_{B,j}c^{\vphantom \dagger}_{B,j}\bigr)\nonumber\\&=(-1)^{n-1}\Bigg(\prod_{m=1}^{n-1} \rmi\tilde{\gamma}_{2m-1}\gamma_{2m-1}\rmi\tilde{\gamma}_{2m}\gamma_{2m} \Bigg)\prod_{j=n}^{n+\lvert\alpha\rvert-1}(-\rmi\tilde\gamma_{2j-1}\tilde\gamma_{2j})\qquad \textrm{for~} \alpha<0. \label{stringcorrelatorsAIII}
\end{align}
String order parameters are usually defined for interacting SPT phases \cite{Perez2008,Pollmann12}, our case is different since we have a free-fermion system---see the discussion in \cite{Jones2019}. In Theorem \ref{thm:AIII} below, we prove that these string correlators are order parameters for the different phases; i.e., $\tilde{\mathcal O}_\alpha$ has long-range order if and only if $\alpha = \omega$, the latter being the topological invariant defined above. Special cases of these fermionic string order parameters\footnote{For example, the correlator of $\bigl(\tilde{\mathcal{O}}_{-1}(n)-\tilde{\mathcal{O}}_{1}(n)\bigr)/2$ is used to detect a phase with two Majorana edge modes. This would correspond to winding number $\pm1$ in our model, consistent with $\tilde{\mathcal{O}}_{\pm1}(n)$ being the order parameter for those phases.} appear in \cite{Bahri14}. We can evaluate the ground-state correlation $\big\langle \tilde{\mathcal{O}}_\alpha(1)\tilde{\mathcal{O}}_\alpha(N+1)\big\rangle$ using Wick's theorem \cite{Barouch1971,Lieb61} and the two-point correlators given above. We show in Appendix \ref{app:AIIIproof} that these correlators are equal to the square of the absolute value of a Toeplitz determinant. This determinant can be evaluated using the same techniques as in the BDI case. We now justify that these are indeed the AIII analogues of the string order parameters in the BDI class---this follows by generalising the analysis of \cite{Jones2019} to this class of models. We will not pursue this fully, but give the following result:
\begin{Theorem}\label{thm:AIII}
Consider a general model in the AIII class. This corresponds to \begin{align}
  f(z) = \rme^{\rmi \theta} \frac{1}{z^{N_P}} \prod_{j=1}^{N_z} (z-z_j)\prod_{k=1}^{N_Z} (z-Z_k),\label{fz_genericAIII}
\end{align}
where $\lvert z_j\rvert<1$, $\lvert Z_k\rvert>1$ and $\theta\in [0,2\pi)$. The winding number is given by $\omega=N_z-N_P$ and the string correlators satisfy
\begin{align*}
\lim_{N\rightarrow\infty}\big\langle\tilde{\mathcal{O}}_\alpha(1)\tilde{\mathcal{O}}_\alpha(N+1)\big\rangle = \mathrm{const} \times \delta_{\alpha,\omega}.
\end{align*}
The non-zero constant is the value of the order parameter, given by
\begin{align*}
\lim_{N\rightarrow\infty}\big\langle\tilde{\mathcal{O}}_\omega(1)\tilde{\mathcal{O}}_\omega(N+1)\big\rangle = \Bigg \vert\frac{ \prod_{j_1,j_2=1}^{N_z} (1-z_{j_1} \overline{z}_{j_2})
	\prod_{k_1,k_2=1}^{N_Z} \bigl(1-Z_{k_1}^{-1} \overline{Z}_{k_2}^{-1}\bigr)}{\prod_{j=1}^{N_z}\prod_{k=1}^{N_Z} \Big\lvert 1-\dfrac{z_j}{ Z_k}\Big\rvert^2}\Bigg \vert^{1/2}.
\end{align*}
\end{Theorem}
We outline the proof, using Szeg\H{o}'s theorem \cite{Deift2011,Szego} and results of Hartwig and Fisher \cite{Hartwig1969}, in Appendix \ref{app:AIIIproof}.

For models with $f(z)$ of the form:
\begin{align}
f(z) = \rme^{\rmi\theta} \frac{1}{z^{n_P}} \prod_{j=1}^{n_z} \left(z-z_j\right)^2\prod_{k=1}^{n_Z} \left(z-Z_k\right)^2, \label{fzcanonAIII}
\end{align}we can proceed using the methods of this paper, leading to an analogue of Theorem \ref{thm:correlators}. The generic case here is that \smash{$\big\{z_{j_1}^{\vphantom {-1}},\overline{z}_{j_2}^{-1},Z_{k_1}^{\vphantom {-1}},\overline{Z}_{k_2}^{-1}\big\}_{j_1,j_2=1,\dots, n_z; k_1,k_2=1,\dots, n_Z}$} are pairwise distinct.

\begin{Theorem}[AIII restatement of Theorem \ref{thm:correlators}]\label{thm:correlatorsAIII}
Consider a chain with a non-trivial $f(z)$ as defined in \eqref{fzcanonAIII}, assuming the generic case, and for each $\alpha \in \mathbb{Z}$ define $N_{\alpha} = \max\{\lvert n_z+n_Z-n_P-\alpha \rvert,1\}$. Then
\begin{enumerate}\itemsep=0pt
\item[$(1)$] for $\alpha < \omega - n_z$ and $\alpha > \omega +n_Z$
\begin{align*}
\big\langle\tilde{\mathcal{O}}_\alpha(1) \tilde{\mathcal{O}}_\alpha(N+1)\big\rangle &= 0, \qquad N\geq N_{\alpha};
\end{align*}
\item[$(2)$] for $\omega - n_z\leq \alpha \leq \omega +n_Z$ \begin{align}
\big\langle\tilde{\mathcal{O}}_\alpha(1) \tilde{\mathcal{O}}_\alpha(N+1)\big\rangle &= \Big\lvert \sum_{M} C_M^{\vphantom N} r_M^N \Big\rvert^2, \qquad N\geq N_{\alpha},\label{eq:thm1b}
\end{align}
where for $n_z>0$ and $n_Z>0$:\begin{itemize}\itemsep=0pt \item for $\omega - n_z\leq \alpha\leq \omega +n_Z-n_z$ the sum is over sets $M$ that label sets of zeros, with the non-zero constants $C_M$ and $r_M$ defined in Case $1$ below.
\item for $\omega +n_Z-n_z \leq \alpha\leq \omega +n_Z$ the sum is over sets $M$ that label sets of inverse zeros, with the non-zero constants $C_M$ and $r_M$ defined in Case $2$ below.\end{itemize}
If either $n_z=0$ or $n_Z=0$:\begin{itemize}\itemsep=0pt
\item if $n_z =0$, and $\alpha<\omega+n_Z$, then put $n_z=1$, $\alpha \rightarrow \alpha+2$ and take the limit $z_1 \rightarrow 0$ with Case $1$. If $n_z=0$ and $\alpha = \omega+n_Z$ then the correlator is given by $ \prod_{k=1}^{n_Z} \lvert Z_k\rvert^{-2N}$.
\item if $n_Z =0$ and $\alpha>\omega-n_z$, then put $n_Z=1$, and take the limit $Z_1 \rightarrow \infty$ with Case $2$. If $n_Z =0$ and $\alpha=\omega-n_z$, then the correlator is given by $\prod_{j=1}^{n_z} \lvert z_j\rvert^{2N}$.
\end{itemize}
\end{enumerate}

In both cases, in the formula \eqref{eq:thm1b}, the sum is over all subsets $M \subseteq \{1,\dots,n_z+n_Z\}$ of some fixed size $\lvert M \rvert$. For each such $M$, we also define:
\begin{align*}
 M^c &= \{1,\dots,n_z+n_Z\}\setminus M \end{align*} and further define $M_z = \{1,\dots, n_z\}$ and $M_Z= \{1,\dots, n_Z\}$. Then we have

Case $1$:
Define $\tau_j = z_j$ for $1\leq j \leq n_z$ and $\tau_{n_z+j}=Z_j$ for $1\leq j \leq n_Z$.
Then the sum in~\eqref{eq:thm1b} is over all subsets $M$ of size $\lvert M \rvert =\alpha+n_P-n_z$, and $C_M$ and $r_M$ are defined by
\begin{align*}
&r_M= \frac{\prod_{k \in M^c} \tau_k} {\prod_{j=1}^{n_Z}\overline{Z}_j},\nonumber\\\nonumber
&C_M= \prod_{k \in M^c,m\in M_Z} \bigl(\tau_k - \overline{Z}^{-1}_m\bigr) \prod_{l \in M_z, j\in M} \bigl(\overline{z}_l^{-1} - \tau_j\bigr) \prod_{l\in M_z}z_l^{-(n_z+n_Z -n_P-\alpha)} \\
&\phantom{C_M=}{}\times\prod_{l \in M_z, m\in M_Z} \bigl(\overline{z}_l^{-1} - \overline{Z}^{-1}_m\bigr)^{-1} \prod_{k \in M^c,j\in M} (\tau_k - \tau_j)^{-1} \prod_{k\in M^c}\tau_k^{-(n_z+n_Z -n_P-\alpha)}.
\end{align*}

Case $2$:
Define $\tau_j = z_j^{-1}$ for $1\leq j \leq n_z$ and $\tau_{n_z+j}=Z_j^{-1}$ for $1\leq j \leq n_Z$.
Then the sum in~\eqref{eq:thm1b} is over all subsets $M$ of size $\lvert M \rvert =2n_z+n_Z-n_P-\alpha$, and $C_M$ and $r_M$ are defined by
\begin{align*}
&r_M= \prod_{j=1}^{n_z}z_j\prod_{k=1}^{n_Z}\frac{Z_k}{\overline{Z}_k}{\prod_{k \in M^c} \tau_k} ,\nonumber\\
&C_M= \prod_{k \in M^c,m\in M_z} (\tau_k - \overline{z}_m) \prod_{l \in M_Z, j\in M} \bigl(\overline{Z}_l - \tau_j\bigr)\prod_{l \in M_Z} \overline{Z}_l^{n_P+\alpha-n_z-n_Z}\nonumber\\
&\phantom{C_M=}{}\times \prod_{l \in M_Z, m\in M_z} \bigl(\overline{Z}_l - \overline{z}_m\bigr)^{-1} \prod_{k \in M^c,j\in M} (\tau_k - \tau_j)^{-1} \prod_{k \in M^c}\tau_k^{-(n_P+\alpha-n_z-n_Z)}.
\end{align*}
\end{Theorem}
Much of the discussion in Section \ref{sec:stringresults} applies analogously here. We will simply give one further example.
\begin{Example}
Let $f(z) = (z-a)^2(z-b)^2$ where $\lvert a\rvert<1$ and $\lvert b\rvert>1$. Since $n_z =n_Z= 1$, we are in the phase $\omega =2$, and the correlator $\big\langle\tilde{\mathcal{O}}_2(1) \tilde{\mathcal{O}}_2(N+1)\big\rangle$ is given by
\begin{align*}
\big\langle\tilde{\mathcal{O}}_2(1) \tilde{\mathcal{O}}_2(N+1)\big\rangle={}&\Bigg\vert \frac{\bigl(1-\lvert a\rvert^2\bigr) \bigl(\lvert b\rvert^{2}-1\bigr) }{\lvert b-a\rvert^2}\left(\frac{b}{\overline{b}}\right)^N + \frac{\lvert 1-b\overline{a}\rvert^2}{\lvert b-a\rvert ^2}\left(\frac{a}{\overline{b}}\right)^N \Bigg\vert^2 \\
={}&\underbrace{\left(\frac{\bigl(1-\lvert a\rvert^2\bigr) \bigl(\lvert b\rvert^{2}-1\bigr) }{\lvert b-a\rvert^2}\right)^2}_{\textrm{order parameter}}+\frac{\lvert 1-b\overline{a}\rvert^4}{\lvert b-a\rvert ^4}\left(\frac{\lvert a\rvert^2}{\lvert b \rvert^2}\right)^N\\
& +\frac{\bigl(1-\lvert a\rvert^2\bigr) \bigl(\lvert b\rvert^{2}-1\bigr)\lvert 1-b\overline{a}\rvert^2 }{\lvert b-a\rvert^4}\left(\left(\frac{a}{b}\right)^N +\left(\frac{\overline{a}}{\overline{b}}\right)^N \right), \qquad N\geq 1.
\end{align*}
\end{Example}
Note that for $a,b\in\mathbb{R}$ this example reduces to a translation-invariant BDI model. Given the form of \eqref{H_AIII_majorana}, we see that we have two decoupled BDI chains, each with $f(z) =(z-a)^2(z-b)^2 $. The correlator $\langle{\mathcal{O}}_2(1) {\mathcal{O}}_2(N+1)\rangle$ can be calculated for each of these decoupled chains using Example \ref{correxample} and we see that in that case $\langle{\mathcal{O}}_2(1) {\mathcal{O}}_2(N+1)\rangle^2=\big\langle\tilde{\mathcal{O}}_2(1) \tilde{\mathcal{O}}_2(N+1)\big\rangle $.

\subsection{Correlation matrix}
For the AIII class, the correlation matrix for a subsystem containing $N$ two-site unit cells is defined by
\begin{align*}
C_N= \bigl(\big\langle b^\dagger_j b^{\vphantom{\dagger}}_k \big\rangle\bigr)_{j,k=1,\dots, 2N},
\end{align*}
where we define $b_{2j-1}=c_{A,j}$ and $b_{2j}=c_{B,j}$. Following the discussion in \cite{Peschel_2009}, to find the eigenvalues of the reduced density matrix or, equivalently, the entanglement spectrum, we want the eigenvalues of the matrix $\hat{A}_N= 1-2C_N$. The eigenvalues, $\lambda$, of $\hat{A}$ are real and satisfy~${-1\leq \lambda \leq 1}$. The values $\lambda =\pm 1$ are trivial eigenvalues that do not correspond to entanglement. Our main result for class AIII is that the proof of Theorem \ref{thm:corrmatrix} carries over, leading to:
\begin{Theorem}[AIII statement of Theorem \ref{thm:corrmatrix}]\label{thm:corrmatrixAIII}
Consider a Hamiltonian of the form \eqref{H_AIII} with a non-trivial $f(z)$ of the form \eqref{fzcanonAIII} such that $n_P=n_z+n_Z$. Recall that in this class zeros need not come in conjugate pairs, and we allow the non-generic case where zeros may coincide, but assume without loss of generality\footnote{As in Remark \ref{remark:nongeneric}, if there were such zeros they can be removed without affecting the ground state.} that $z_j\neq \overline{Z}_k^{-1}$ for any $j$, $k$. Denote the correlation matrix on a subsystem of size $N$ by $C_N$ and define $\hat{A}_N=1-2C_N$. Then
\begin{align*}
\det\bigl(\lambda-\hat{A}_N\bigr)= \bigl(\lambda^2-1\bigr)^N \left(\frac{\prod_{j=1}^{n_z} \overline{z}_j^2\prod_{j=1}^{n_Z} \overline{Z}_j^2}{\bigl(\prod_{j=1}^{n_Z} \lvert Z_j\rvert^2\bigr)^2}\right)^N \frac{\det{M(N,\lambda)}}{\det{M(0,\lambda)}}, \end{align*}
where $M(n,\lambda)$ is a $4(n_z+n_Z)\times 4(n_z+n_Z)$ matrix that can be determined from the zeros of~$f(z)$ and the Smith canonical form of a known $2\times 2$ matrix polynomial $($defined in \eqref{a11AIII}--\eqref{a21AIII}$)$. Moreover, there exist $d', d'' \in \mathbb{N}$ $($independent of $n)$ such that
\begin{align*}
\det{M(n,\lambda)} = \mu \left(\frac{\prod_{j=1}^{n_z} \overline{z}_j^2\prod_{k=1}^{n_Z} \overline{Z}_k^2}{\bigl(\prod_{k'=1}^{n_Z} \lvert Z_{k'}\rvert^2\bigr)^2}\right)^{-n} \prod_{j_1=1}^{d'}\bigl(\lambda^2-\tilde \nu_{j_1}(n)^2\bigr)\prod_{j_2=1}^{d''}(\lambda-c_{j_2}) , \end{align*}
where the constants $\mu$ and $c_j$ are independent of $n$ and $\tilde \nu_j(n)$ satisfies $\tilde \nu_j(0)=1$, $0\leq\tilde \nu_j(n)\leq 1$ for $n\geq 1$ and $\tilde \nu_j(n)<1$ for some $n>0$.

For any $f(z)$ as defined in \eqref{fzcanonAIII} such that $n_P \neq n_z+n_Z$, then the characteristic polynomial of the correlation matrix can be found by taking a limit of a related case where $n_P = n_z+n_Z$ that can be evaluated using the above. Details of this limit are discussed in Remark {\rm \ref{limitremark}}---in particular, this discussion does not rely on zeros of $f(z)$ appearing in complex-conjugate pairs.
\end{Theorem}
For further details see Section \ref{sec:corrmatrix} and Appendix \ref{app:AIIIproof}. As in the BDI case, this results allows us to conclude that there exists an exact MPS representation of the ground state.

To find the entanglement spectrum requires using the definition of $M(n,\lambda)$ given in Section~\ref{sec:corrmatrix} (with the changes for class AIII given in Appendix~\ref{app:AIIIproof}), as well as finding the Smith canonical form of the relevant matrix. By analogous calculations\footnote{The matrix from the Smith canonical form needed for this calculation is given in Appendix~\ref{app:AIIIproof}.} to those in Example~\ref{examplecorrmatrix},
we find that for $f(z) = \frac{1}{z}(z-b)^2$ with $\lvert b\rvert <1$, $b\in \mathbb{C}$:
\begin{align*}
\det\bigl(\lambda-\hat{A}_N\bigr)&= \bigl( \lambda^2-1\bigr)^{N-1}\bigl(\lambda^2-\lvert b\rvert^{2N}\bigr). \end{align*}
\section{String correlators---analysis}\label{sec:string}
As stated above, the BDI string correlators satisfy
\begin{align*}(-1)^{N(\alpha-1)}\langle\mathcal{O}_\alpha(1) \mathcal{O}_\alpha(N+1)\rangle =D_N[t_1(z)]=D_N[t_1(1/z)], \end{align*}
 where $t_1(z)= \sqrt{{f(z)}/{f(1/z)}}z^{-\alpha}$. Moreover, in AIII the string correlators satisfy
\begin{align*}\big\langle\tilde{\mathcal{O}}_\alpha(1) \tilde{\mathcal{O}}_\alpha(N+1)\big\rangle =\big\lvert D_N[t_2(z)]\big\rvert^2=\big\lvert D_N[t_2(1/z)]\big\rvert^2, \end{align*}
 where \smash{$t_2(z)= \sqrt{{f(z)}/{\overline{f}(1/z)}}z^{-\alpha}$}. We hence see that we can understand both classes by analysing $D_N[t(z)]$ for $t(z)=t_2(z)$. Then for class BDI we reach Theorem \ref{thm:correlators} by noting that $\overline{f}(1/z)=f(1/z)$ for $t_\alpha\in\mathbb{R}$ and, since zeros come in complex-conjugate pairs, we can simplify certain formulae. For class AIII we reach Theorem \ref{thm:correlatorsAIII} by simply taking the absolute value squared of the result.

We will show now that for
\begin{align}
f(z) = \frac{\rme^{\rmi\theta}}{z^{n_P}} \prod_{j=1}^{n_z} \left(z-z_j\right)^2\prod_{k=1}^{n_Z} \left(z-Z_k\right)^2, \label{fzcanon2}
\end{align}
then either $D_N[t(z)]$ or $D_N[t(1/z)]$ can be evaluated using Day's formula.
Some further analysis is then needed to reach our results. First, note that we can simplify by setting $\theta =0$. In the BDI case, $\theta\in\{0,\pi\}$; if $\theta=\pi$ we take the other branch of the square-root and multiply the Toeplitz matrix by $-1$. This sign hence gives an additional factor of $(-1)^N$ when we evaluate the determinant, as mentioned in Remark \ref{remark:nongeneric}. In the AIII case $\theta\in[0,2\pi)$, this would give an additional complex phase when we evaluate the determinant, but since we take the absolute value this drops out.
If $n_z=0$ and $n_Z=0$ then $f(z)=z^{-n_P}$ and the string correlators are trivial. We treat the case $n_z>0$ and $n_Z>0$ first. Then we analyse the remaining cases where either $n_z=0$ or $n_Z=0$.

\subsection[Proof of Theorems 4.2 and 6.2 with zeros inside and outside the unit circle]{Proof of Theorems \ref{thm:correlators} and \ref{thm:correlatorsAIII} with zeros inside\\ and outside the unit circle}
Suppose first that $n_z>0$ and $n_Z>0$. By inserting \eqref{fzcanon2} into $t(z)$ and rearranging, we reach
\begin{align*}
&t(z)= z^{n_z+n_Z-n_P-\alpha}\left(\prod_{j=1}^{n_Z}\bigl(-\overline{Z}_j\bigr)\right)^{-1} \frac{\prod_{j=1}^{n_z} \left(z-z_j\right)\prod_{k=1}^{n_Z} \left(z-Z_k\right)}{\prod_{j'=1}^{n_z} \bigl(1-z/\overline{z}_{j}^{-1}\bigr)\prod_{k=1}^{n_Z} \bigl(z-\overline{Z}_{k'}^{-1}\bigr)},\\
&t(1/z)= {z^{n_P+\alpha-n_z-n_Z}} \left(\prod_{j=1}^{n_z} (-z_j)\right)\left(\prod_{k=1}^{n_Z} \frac{Z_k}{\overline{Z}_k}\right) \frac{\prod_{j=1}^{n_z} \bigl(z-z_j^{-1}\bigr)\prod_{k=1}^{n_Z} \bigl(z-Z_k^{-1}\bigr)}{\prod_{j=1}^{n_z} \bigl(z-\overline{z}_j\bigr)\prod_{k=1}^{n_Z} \bigl(1-z\overline{Z}_k^{-1}\bigr)}.
\end{align*}
Recall that in Theorem \ref{thm:Day}, the canonical form \eqref{daycanon} for the symbol has the degree of the numerator greater than or equal to the degree of the denominator. Hence, for $n_z+n_Z-n_P-\alpha \geq 0$ we evaluate $D_N[t(z)]$ and for $n_z+n_Z-n_P-\alpha \leq 0$ we evaluate $D_N[t(1/z)]$. Note also that in order to apply Day's formula we must have pairwise distinct zeros in the numerator. Hence, for~$n_z+n_Z-n_P-\alpha \geq 0$, we write
\begin{align}
&t(z)= \lim_{\eps \rightarrow 0}\rho\frac{\prod_{j=1}^{2n_z+2n_Z-n_P-\alpha} (z-\tau_j)}{\prod_{j=1}^{n_z} \bigl(1-z/\overline{z}_j^{-1}\bigr)\prod_{j=1}^{n_Z} \bigl(z-\overline{Z}_j^{-1}\bigr)},\nonumber \\
&\tau_j \in \{ z_1,\dots, z_{n_z},Z_1,\dots, Z_{n_Z},\eps x_1, \dots, \eps x_{n_z+n_Z-n_P-\alpha}\},\qquad \rho = \left(\prod_{k=1}^{n_Z}\bigl(-\overline{Z}_k\bigr)\right)^{-1} \label{tcanon}
 \end{align}
and for $n_z+n_Z-n_P-\alpha \leq 0$:
\begin{align*}
&t(1/z)=\lim_{\eps \rightarrow 0}\rho\frac{\prod_{j=1}^{n_P+\alpha} \left(z-\tau_j\right)}{\prod_{j=1}^{n_z} \left(z-\overline{z}_j\right)\prod_{j=1}^{n_Z} \bigl(1-z\overline{Z}_j^{-1}\bigr)},\\
&\tau_j \in \big\{ z_1^{-1},\dots, z_{n_z}^{-1},Z_1^{-1},\dots, Z_{n_Z}^{-1}, \eps x_1, \dots, \eps x_{n_P+\alpha-n_z-n_Z}\big\}, \\
&\rho=\left(\prod_{j=1}^{n_z} (-z_j)\right)\left(\prod_{j=1}^{n_Z} \frac{Z_j}{\overline{Z}_j}\right) .
\end{align*}
In each case, before the limit is taken these are in the canonical form \eqref{daycanon} and we can evaluate the determinants using Theorem \ref{thm:Day}---this has the form $\sum_M C_M^{\vphantom N} r_M^N$ where the sum is over sets~$M$ defined in Theorem \ref{thm:Day}. Then, by taking the limit, we can determine the string correlator. Recall that in Theorems \ref{thm:correlators} and \ref{thm:correlatorsAIII} we assume the appropriate generic case.
This means that, for $\eps\neq0$, $C_M$ is a product of finite, non-zero terms. Coincident zeros require taking a limit, as discussed in Remark \ref{remark:nongeneric}, while if we have \smash{$z_j = \overline{Z}_k^{-1}$}, then $C_M$ can be zero\footnote{In fact, as we discussed in Remark \ref{remark:nongeneric}, the ground state is equivalent to a model with those zeros removed. Hence, terms containing those zeros should not appear in the result, and so if we have a set $M$ where these zeros appear in $r_M$, the corresponding $C_M$ must be zero. We can see this in Example \ref{correxample} by setting $b=1/a$. }.
\par
Now, a difficulty with the limit $\eps\rightarrow 0$ is that when $\lvert n_z+n_Z-n_P-\alpha \rvert >1$, there are choices of $M$ where both $M$ and $M^c$ contain\footnote{We are abusing notation slightly: $M$ corresponds to a set of indices, but it is helpful to also think of $M$ as the set of $\tau_k$ corresponding to those indices.} $\tau_k$ of order $\eps$. This means that
 \[C_M \propto \prod_{k \in M^c,j\in M} (\tau_k - \tau_j)^{-1}=\Theta(\eps^{-m})\]
 for some $m>0$. However, since $M^c$ by assumption contains a $\tau_k$ of order $\eps$, $r_M$ is $\Theta(\eps^n)$ for some $n>0$. Then, since the contribution of this $M$ to the $N\times N$ determinant is $C_M^{\vphantom N} r_M^N$, for $N$ sufficiently large this term will have a positive power of $\eps$, and so will vanish upon taking the limit $\eps \rightarrow 0$. More precisely, we have
\begin{Lemma}\label{lemma1}
For $N\geq N_{\alpha} = \max\{\lvert n_z+n_Z-n_P-\alpha \rvert,1\}$, in evaluating $D_N[t(z)]$ using Day's formula, any set $M^c$ that contains a term of order $\eps$ does not contribute to the determinant in the limit $\eps \rightarrow 0$.
\end{Lemma}
Let us first consider $D_N[t(z)]$ in the case $n_z+n_Z-n_P-\alpha \geq 0.$
By Lemma \ref{lemma1}, the only sets $M$ that contribute are those where $M^c$ contains no terms of order $\eps$.
By comparing \eqref{tcanon} and~\eqref{daycanon} we see that $p=n_Z$, and thus the sum over $M$ gives all subsets of $\{\tau_j\}$ of size $n_Z$. Since~$\lvert M \rvert = n_Z$ such sets exist when $n_Z \geq n_z+n_Z-n_P- \alpha$, which is equivalent to $\alpha\geq n_z-n_P$. Hence, for~$\alpha< n_z-n_P$ and $N \geq N_{\alpha}$ the determinant is zero. For the other cases under consideration, $ n_z-n_P \leq \alpha \leq n_z+n_Z-n_P$, by applying Day's formula we reach the result given in Theorem \ref{thm:correlators}. Since all $M$ that have a non-zero $r_M$ in the limit $\eps \rightarrow 0$ contain all terms of order $\eps$, in stating the theorem we simplify the formulae by removing this fixed number of terms of order $\eps$ from the definition of $M$, and adjusting the definition of the $C_M$ to account for this.
Note that here, and in the other cases below, the sign in the definition of $r_M$ in Day's formula does not depend on $M$ and so we combine it with the oscillatory factor of the correlator. For $1\leq N < N_{\alpha}$ we can still evaluate the determinant using Day's formula, but the analysis and simplifications just discussed will not apply and so one must
take the limit after applying Theorem \ref{thm:Day} directly for the symbol \eqref{tcanon}. \par

An analogous discussion applies to the case $n_z+n_Z-n_P-\alpha \leq 0$. In this case, $\lvert M \rvert = n_z$ and so choices of $M$ where $M^c$ contains no terms of order $\eps$ exist when $\alpha \leq 2n_z +n_Z -n_P$. Hence, the correlator is zero for $\alpha>\omega+n_Z$. By applying Day's formula in the remaining cases, and simplifying to only include the $M$ that contribute, this completes the proof of Theorem \ref{thm:correlators} with $n_z>0$ and $n_Z>0$.

\begin{proof}[Proof of Lemma \ref{lemma1}]
Consider the case $n_z+n_Z-n_P-\alpha > 0$. Let us suppose that $M$ is such that $M^c$ contains $n>0$,~$\tau_k$ of order $\eps$. This means that $r_M = \Theta(\eps^n)$, and that $M$ contains $(n_z+n_Z-n_P-\alpha-n)$,~$\tau_k$ of order $\eps$. Then, we have that \begin{align*}
C_Mr_M^N = \Theta\bigl(\eps^{-n(n_z+n_Z-n_P-\alpha-n)+nN}\bigr) = \Theta\bigl(\eps^{n^2+n(N-n_z-n_Z+n_P+\alpha)}\bigr),
\end{align*}
which is a positive power of $\eps$ for $N\geq n_z+n_Z-n_P-\alpha$. The analogous proof goes through for the case that $n_z+n_Z-n_P-\alpha < 0$. In the case $\alpha = n_z+n_Z-n_P$ all $\tau_k = \Theta(1)$, then Day's formula applies for $N\geq 1$.\renewcommand{\qed}{}
\end{proof}

\subsection[Proof of Theorems 4.2 and 6.2 for n\_z=0 or n\_Z=0]{Proof of Theorems \ref{thm:correlators} and \ref{thm:correlatorsAIII} for \texorpdfstring{$\boldsymbol{n_z=0}$}{nz=0} or \texorpdfstring{$\boldsymbol{n_Z=0}$}{nZ=0}}
\subsubsection{No zeros outside the unit circle}
Let us now consider the case $n_Z =0$, and $n_z-n_P-\alpha < 0$. Note that we have $n_z>0$ as we do not include the trivial case. The strict inequality allows us to apply Day's formula. In particular,
\begin{align*}
&t(1/z)= {z^{n_P+\alpha-n_z}} \left(\prod_{j=1}^{n_z} (-z_j)\right) \frac{\prod_{j=1}^{n_z} \bigl(z-z_j^{-1}\bigr)}{\prod_{j=1}^{n_z} \left(z-\overline{z}_j\right)}= \lim_{\eps \rightarrow 0}\lim_{Z_1 \rightarrow \infty} \rho\frac{\prod_{j=1}^{n_P+\alpha} \left(z-\tau_j\right)}{\bigl(1-z\overline{Z}_1^{-1}\bigr)\prod_{j=1}^{n_z} \left(z-\overline{z}_j\right)},\\
&\tau_j\in \big\{ z_1^{-1},\dots, z_{n_z}^{-1},Z_1^{-1}, \eps x_1, \dots, \eps x_{n_P+\alpha-n_z-1}\big\}, \qquad \rho=\left(\prod_{j=1}^{n_z} (-z_j)\right).
\end{align*}
Thus, the result we want follows from the formula for $n_Z = 1$ and then taking the limit $Z_1 \rightarrow \infty$. Taking the limit by setting $Z_1^{-1} = \eps x_0$, we derive that contributions from sets $M^c$ that contains a $\tau_j = \Theta(\eps)$ are zero for~$N \geq n_P+\alpha-n_z$. Moreover, note that the correlator is exactly zero for~$\alpha>\omega$ and $N\geq n_P+\alpha-n_z$, since in that case all sets $M^c$ will contain a $\tau_j$ that goes to zero.

 In the case $n_z-n_P-\alpha \geq 0$, it is easier to analyse the determinant directly. Since $n_Z=0$, the Fourier coefficients of the symbol, as calculated in Appendix \ref{app:fourier}, can be seen to be one-sided. In particular $t_n =0$ for $n<n_z-n_P-\alpha$. This means that if $n_z-n_P-\alpha >0$, the determinant is zero. For the only remaining case, $n_z-n_P-\alpha=0$, we have that
\begin{align*}
D_N[t(z)] = t_0^N = \prod_{j=1}^{n_z} (-z_j)^N, \qquad N\geq 1.
\end{align*}

\subsubsection{No zeros inside the unit circle}
Finally, let us consider the case $n_z =0$, $n_Z>0$ and $n_Z-n_P-\alpha > 0$. The strict inequality allows us to apply Day's formula. In particular,
\begin{align*}
&t(z)= z^{n_Z-n_P-\alpha} \left(\prod_{k=1}^{n_Z}\bigl(-\overline{Z}_k\bigr)\right)^{-1} \frac{\prod_{k=1}^{n_Z} \left(z-Z_k\right)}{\prod_{k=1}^{n_Z} \bigl(z-\overline{Z}_k^{-1}\bigr)}
 \\
 &\phantom{t(z)}{}= \lim_{\eps \rightarrow 0}\lim_{z_1 \rightarrow 0} \left(\prod_{k=1}^{n_Z}\bigl(-\overline{Z}_k\bigr)\right)^{-1} \frac{\prod_{j=1}^{2n_z+2n_Z-n_P-(\alpha+2)} \left(z-\tau_j\right)}{\bigl(1-z/\overline{z}_1^{-1}\bigr)\prod_{k=1}^{n_Z} \bigl(z-\overline{Z}_k^{-1}\bigr)},\\
&\tau_j \in \{ z_1,Z_1,\dots, Z_{n_Z},\eps x_1, \dots, \eps x_{n_z+n_Z-n_P-(\alpha+2)}\},\qquad \rho = \left(\prod_{k=1}^{n_Z}\bigl(-\overline{Z}_k\bigr)\right)^{-1} .\end{align*}
As before, we can then apply the result for $n_z=1$ and take the limit $z_1 \rightarrow 0$, while also replacing~$\alpha$ by $\alpha+2$. Taking the limit by setting $z_1 = \eps x_0$, we derive that contributions from sets $M^c$ that contains a $\tau_j = \Theta(\eps)$ are zero for $N \geq n_Z-n_P-\alpha$. Moreover, for $\alpha <-n_P = \omega$ all sets $M^c$ contain a $\tau_j = \Theta(\eps)$, and so for $N \geq n_Z-n_P-\alpha$ these correlators are exactly zero. Finally, as in the case $n_z=0$, one can argue we have one-sided Fourier coefficients that imply the correlator is zero for $n_Z-n_P-\alpha <0$ and for $\alpha = n_Z-n_P$, we have
\begin{align*}
D_N[t(z)] = t_0^N =\prod_{k=1}^{n_Z} \bigl(-\overline{Z}_k\bigr)^{-N}, \qquad N\geq 1.
\end{align*}
We have considered all cases and thus, by using that the zeros come in conjugate pairs, we have proved Theorem \ref{thm:correlators} for the BDI class, and by taking the absolute value squared we have proved
Theorem \ref{thm:correlatorsAIII} for the AIII class.

\subsection{Emptiness formation probability}\label{sec:emptiness}
Here we prove that the emptiness formation probability, $P(N)$, introduced in Remark \ref{remark:emptiness}, can be evaluated using Day's formula. Following \cite{Franchini2005}, for a translation-invariant BDI Hamiltonian as in \eqref{HBDI}, we have that
\begin{align*}
P(N) = \left \lvert D_N\left[\frac{1}{2}\left(1-\frac{f(z)}{\lvert f(z) \rvert} \right)\right] \right\rvert.
\end{align*}
Then, for a generic BDI model of the form given in \eqref{fzcanon}, $P(N) = \left \lvert D_N\left[t(z)\right] \right\rvert$, where
\begin{align*}
t(z) &= \frac{1}{2}\left({\frac{\prod\limits_{j=1}^{n_z} \bigl(1-z/z_j^{-1}\bigr)\prod\limits_{k=1}^{n_Z} \bigl(z-Z_k^{-1}\bigr)-\sigma z^{n_z+n_Z-n_P}\prod\limits_{j=1}^{n_z} \bigl(z-z_j\bigr)\prod\limits_{k=1}^{n_Z} \bigl(1-zZ_k^{-1}\bigr)}{\prod\limits_{j=1}^{n_z} \bigl(1-z/z_j^{-1}\bigr)\prod\limits_{k=1}^{n_Z} \bigl(z-Z_k^{-1}\bigr)}} \right).
\end{align*}
This is a rational symbol, where one can check that the numerator has degree at least the degree of the denominator. We can allow for degenerate zeros by taking an appropriate limit, and, assuming that $n_z>0$ and $n_Z>0$, we can evaluate $P(N)$ using Theorem \ref{thm:Day}.

If $n_z=0$ and $n_Z=0$, then $t(z)=\frac{1}{2}(1-\sigma z^{n_P})$---if $n_P\neq 0$, then $P(N)=2^{-N}$, while if~$n_P=0$ then $P(N)=0$ for $\sigma=1$ and $P(N)=1$ for $\sigma=-1$. If $n_Z=0$ and $n_z>0$, then one can write~${t}(z) = \lim_{\eps\rightarrow 0}\frac{z-\eps}{z-\eps}t(z)$, and then evaluate with Day's formula before taking the limit $\eps \rightarrow 0$. Similarly, if $n_z=0$ and $n_Z>0$, one can take ${t}(z) = \lim_{\eps\rightarrow 0}\frac{1-\eps z}{1-\eps z}t(z)$.

To derive Example \ref{example:emptiness}, recall that $f(z) = z^{-2} (z-a)^2(z-b)^2$, with $\lvert a\rvert<1$ and $\lvert b\rvert>1$. We suppose the generic case that $b\neq 1/a$. Then we can write
\begin{align*}
P(N) = \left \lvert D_N\left[\frac{(1/b-a)}{2}\frac{(z+1)(z-1)}{(1-az)(z-1/b)}\right] \right\rvert,
\end{align*}
which is of the form \eqref{daycanon} with $p=q=1$ and $s=2$. There are thus two contributions to the determinant, leading to the given formula. Note that this formula is strictly positive given~$b\neq 1/a$.

\section{Correlation matrix---analysis}\label{sec:corrmatrix} We can find the characteristic polynomial of the BDI correlation matrix by evaluating the block Toeplitz determinant generated by \eqref{blocksymbol}. Note that $D_N\left[\Phi(z,\lambda)\right]$ is the determinant of a~${2N\times 2N}$ matrix, with $\rmi\lambda$ on the diagonal. This means that ${D_N \left[\Phi(z,\lambda)\right]=(-1)^N\prod_{j=1}^{N} \bigl(\lambda^2- \nu_j^2\bigr)}$. Moreover, since this is a correlation matrix we have that $0\leq \nu_j \leq 1$.
 \par
 \subsection[A canonical form for rational symbols and the definition of M(n,λ)]{A canonical form for rational symbols and the definition of \texorpdfstring{$\boldsymbol{M(n,\lambda)}$}{M(n,λ)}}
  Let us define $\tilde\Phi(z,\lambda)$ to be the symbol $\Phi(z,\lambda)$ with $f(z)$ restricted to be of the form \eqref{fzcanon}.
 Then define \begin{align*}
\tilde{g}(z) &= \prod_{j=1}^{n_z} (z-z_j)\prod_{k=1}^{n_Z} \bigl(z-Z_k^{-1}\bigr), \qquad
\tilde{h}(z)= \prod_{j=1}^{n_z} (1-zz_j)\prod_{k=1}^{n_Z} \bigl(1-zZ_k^{-1}\bigr).
\end{align*}
We then have
\begin{align*}
\tilde\Phi(z,\lambda) &= \frac{a(z)}{\tilde{g}(z)\tilde{h}(z)}= \frac{1}{\tilde{g}(z)\tilde{h}(z)} \left(\begin{matrix} a_{11}(z)& a_{12}(z)\\ a_{21}(z)& a_{22}(z)\end{matrix}\right),
\end{align*}
where
\begin{align}
a_{11}(z)&= a_{22}(z) =\rmi \lambda \frac{\bigl(\prod_{j=1}^{n_z} (-z_j)\bigr)}{\bigl(\prod_{k=1}^{n_Z} (-Z_k)\bigr)} \prod_{j=1}^{n_z}(z-z_j)\bigl(z-z_j^{-1}\bigr) \prod_{k=1}^{n_Z}(z-Z_k)\bigl(z-Z_k^{-1}\bigr), \label{a11}\\
a_{12}(z)&=z^{n_z+n_Z-n_P} \frac{1}{\prod_{k=1}^{n_Z} Z_k^2} \prod_{j=1}^{n_z}(z-z_j)^2 \prod_{k=1}^{n_Z}(z-Z_k)^2, \label{a12}\\
a_{21}(z)&=-z^{-n_z-n_Z+n_P} \prod_{j=1}^{n_z} z_j^2\prod_{j=1}^{n_z}\bigl(z-z_j^{-1}\bigr)^2 \prod_{k=1}^{n_Z}\bigl(z-Z_k^{-1}\bigr)^2 \label{a21}.
\end{align}

Now, in order to apply Theorem \ref{thm:Gorodetsky}, $a(z)$ must be a matrix polynomial of the form $\sum_{j=0}^s a_j z^j$. This is the case if $n_z+n_Z=n_P$, and in fact if this condition is not satisfied we cannot apply Gorodetsky's formula\footnote{If $n_z+n_Z \neq n_P$, then there is a pole in either \eqref{a12} or \eqref{a21}---we could try to pull this pole out and redefine~$\tilde g(z)$, however, in that case we have that $a_s$ has only one non-zero entry and one of the conditions of Gorodetsky's formula is that $a_s$ is invertible. As explained in Section \ref{sec:corrmatrixresults}, we can analyse $n_P \neq n_z+n_Z$ by taking an appropriate limit of a case with $n_P=n_z+n_Z$.}: we thus fix $n_z+n_Z=n_P$.
We then have that all $a_{ij}(z)$ are polynomials of degree $2(n_z+n_Z)$, and
\begin{align*}
\det(a_s)=\bigl(1- \lambda^2\bigr) \frac{\prod_{j=1}^{n_z} z_j^2}{\prod_{k=1}^{n_Z} Z_k^2}.
\end{align*}
Define the set of zeros and inverse zeros by
\begin{align*}
\{\tau_i\}_{i=1,\dots, 2(n_z+n_Z)} \in \big\{z_{j_1}^{\vphantom {-1}},z_{j_2}^{-1},Z_{k_1}^{\vphantom {-1}},Z_{k_2}^{-1}\big\}_{j=1,\dots, n_z; k=1,\dots, n_Z}.
\end{align*}
Recall that we do not assume that the $\{\tau_i\}$ are all pairwise distinct, however, in the statement of the theorem we assume\footnote{In fact, if this is the case one can still apply Gorodetsky's formula as given in \cite{Bottcher}, but the given formulae will be slightly altered. Alternatively one can note that, as discussed in Remark \ref{remark:nongeneric}, this assumption on our model is without loss of generality.} that $z_j \neq Z_k^{-1}$ for any $j,k$.
Then there exist $2\times 2$ matrix polynomials~$y(z)$ and $w(z)$ such that $\det{y(z})$ and $\det{w(z)}$ are non-zero and independent of~$z$~and
\begin{align}
y(z) a(z) w(z) = \left(\begin{matrix}1 & 0 \\0 & \displaystyle \prod_{j=1}^{n_z}(z-z_j)^2\bigl(z-z_j^{-1}\bigr)^2 \prod_{k=1}^{n_Z}(z-Z_k)^2\bigl(z-Z_k^{-1}\bigr)^2\end{matrix}\right) \label{smith};
\end{align}
this is the Smith canonical form of $a(z)$ \cite{Bottcher,Gohberg}. Define
\begin{align*}
&G(z)=\tilde g(z)\bigl( 1,z, \dots, z^{n_z+n_Z-1}\bigr),\\
 &H(z) =\tilde h(z)\bigl( 1,z, \dots, z^{n_z+n_Z-1}\bigr),\\
&m^{(n)}(z,\lambda)= \underbrace{\bigl(y_{21}(z,\lambda)H(z), y_{22}(z,\lambda)H(z), y_{21}(z,\lambda)z^nG(z),y_{22}(z,\lambda)z^nG(z) \bigr)}_{4(n_z+n_Z)},
\end{align*}
where the second row of $y(z)$ from the Smith canonical form appears in the definition of $m^{(n)}$, and we make the $\lambda$ dependence explicit.
Then we define a $4(n_z+n_Z)\times 4(n_z+n_Z)$ matrix
\begin{align*}
M(n,\lambda) = \left(\begin{matrix}m^{(n)}(\tau_1,\lambda)  \\  \partial_z m^{(n)}(\tau_1,\lambda) \\  m^{(n)}(\tau_2,\lambda)  \\  \partial_z m^{(n)}(\tau_2,\lambda) \\  \vdots  \\  m^{(n)}(\tau_{2(n_z+n_Z)},\lambda) \\  \partial_z m^{(n)}(\tau_{2(n_z+n_Z)},\lambda)  \end{matrix}\right).
\end{align*}
This matrix is, in our generic case, the matrix $\mathcal{M}\big[n,\tilde{\Phi}(z)\big]$ appearing in Theorem \ref{thm:Gorodetsky}.

\subsection{Proof of Theorem \ref{thm:corrmatrix}}
Given the canonical form of $\tilde\Phi(z,\lambda)$ and the definitions of the relevant functions above, we simply apply Theorem \ref{thm:Gorodetsky} as given in \cite{Bottcher}, leading to
\begin{align*}
D_N \big[\tilde\Phi(z,\lambda)\big]= \bigl(1- \lambda^2\bigr)^N \left(\frac{\prod_{j=1}^{n_z} z_j^2}{\prod_{k=1}^{n_Z} Z_k^2}\right)^N \frac{\det{M(N,\lambda)}}{\det{M(0,\lambda)}}. \end{align*}
This is the first part of our result, we now show that there are only a finite number of zeros of the characteristic polynomial $D_N \big[\tilde\Phi(z,\lambda)\big]$ that are not equal to $\pm 1$.
Firstly, note that $\lambda$ dependence in $\det{M(n,\lambda)}$ comes from $y_{21}(z,\lambda)$ and $y_{22}(z,\lambda)$. Importantly, the definition \eqref{smith} of $y$ does not depend on $N$. One can determine $y$ and $w$ through a finite sequence of elementary transformations of $a(z)$. In particular, the elementary transformations with $\lambda$ dependence reduce the order (as a polynomial in $z$) of the matrix elements of $a_{11}$, $a_{12}$, $a_{21}$ by taking linear combinations of rows or columns multiplied by coefficients from these polynomials at each step. These coefficients will be polynomials in $\lambda$, since initially all matrix elements of $a(z)$ are polynomials in $z$ and $\lambda$ \eqref{a11}--\eqref{a21}. Hence, there exists a $d_0 \in \mathbb{N}$ such that we can find\footnote{Note that after applying this sequence of transformations, the matrix we have will differ from the right-hand side of \eqref{smith} by a polynomial in $\lambda$ in each diagonal entry. We get the Smith canonical form by dividing each entry of $w(z,\lambda)$ by one of these polynomials as appropriate. Hence, we can find $y(z,\lambda)$, $w(z,\lambda)$ such that $y$ is a~polynomial in $\lambda$ while $w$ is a rational function of $\lambda$. Examples are given in Appendix \ref{app:smith}.} a $y(z,\lambda)$ with entries that are polynomial in $\lambda$ of degree at most $d_0$. This means that $\det{M(n,\lambda)}$ is a~polynomial in~$\lambda$ of degree $2d'+d''\leq 4d_0(n_z+n_Z)$, where this degree does not depend on $n$.

Now, recalling that $D_N \big[\tilde\Phi(z,\lambda)\big]=(-1)^N\prod_{j=1}^{N} \bigl(\lambda^2- \nu_j^2\bigr)$, we have
\begin{align*}
(-1)^N\prod_{j=1}^{N} \bigl(\lambda^2- \nu_j^2\bigr)&= \bigl(1- \lambda^2\bigr)^N \left(\frac{\prod_{j=1}^{n_z} z_j^2}{\prod_{k=1}^{n_Z} Z_k^2}\right)^N \frac{\det{M(N,\lambda)}}{\det{M(0,\lambda)}}\\
&=\bigl(1- \lambda^2\bigr)^N \left(\frac{\prod_{j=1}^{n_z} z_j^2}{\prod_{k=1}^{n_Z} Z_k^2}\right)^N \frac{\alpha_N \prod_{j=1}^{d'}\bigl(\lambda^2-\tilde \nu_j(N)^2\bigr)\prod_{j=1}^{d''}(\lambda-c_j)}{\alpha_0 \prod_{j=1}^{d'}\bigl(\lambda^2-\tilde \nu_j(0)^2\bigr)\prod_{j=1}^{d''}(\lambda-c_j)}.
\end{align*}
By comparing the two sides of this equation, we have that \begin{align*}
\det{M(n,\lambda)} = \mu \left(\frac{\prod_{j=1}^{n_z} z_j^2}{\prod_{k=1}^{n_Z} Z_k^2}\right)^{-n} \prod_{j=1}^{d'}\bigl(\lambda^2-\tilde \nu_j(n)^2\bigr)\prod_{j=1}^{d''}(\lambda-c_j),
\end{align*}
where $\mu$ and $c_j$ are independent of $n$ and~$\tilde \nu_j(n)$ satisfies $\tilde \nu_j(0)=1$, $\tilde \nu_j(n)\leq 1$ for $n\geq 1$ and~$\tilde \nu_j(n)<1$ for some $n>0$. Thus we have $d'\in\mathbb{N}$
non-trivial eigenvalues of the correlation matrix, even in the limit $N\rightarrow \infty$. Theorem \ref{thm:corrmatrixAIII} is proved similarly, with the relevant changes pointed out in Appendix \ref{app:AIIIproof}.

\section{The MPS-transfer matrix} \label{sec:transfer}
As discussed in Remark \ref{MPSremark}, Theorems \ref{thm:corrmatrix} and \ref{thm:corrmatrixAIII} allow us to deduce the existence of an exact MPS representation for the ground state of BDI and AIII models with $f(z)$ of the form given in \eqref{fzcanon} and \eqref{fzcanonAIII} respectively. Moreover, an exact MPS representation is constructed for the BDI class in \cite{Jones21}. Given an MPS, it can be put into a canonical form \cite{Perez2007}---we will henceforth suppose that the MPS under consideration is in this form. From the MPS, we can define the MPS-transfer matrix (hereafter referred to as the transfer matrix) and in this section we show how Theorem \ref{thm:correlators} allows us to deduce properties of this transfer matrix for the BDI case. As an application of this, we can find a lower bound on the bond dimension of an MPS representation of the ground state---this complements the upper bound found in \cite{Jones21}. In certain cases, the upper and lower bounds coincide and allow us to give the optimal bond dimension of such an MPS. Throughout this section we will work with the spin chain model (see Appendix \ref{app:spin}), and use standard graphical notation to give an intuitive illustration of some formulae \cite{Cirac2020,Pollmann2017}. While we focus on the BDI case, an analogous discussion based on Theorem \ref{thm:correlatorsAIII} would allow us to deduce properties of the transfer matrix in the AIII class.

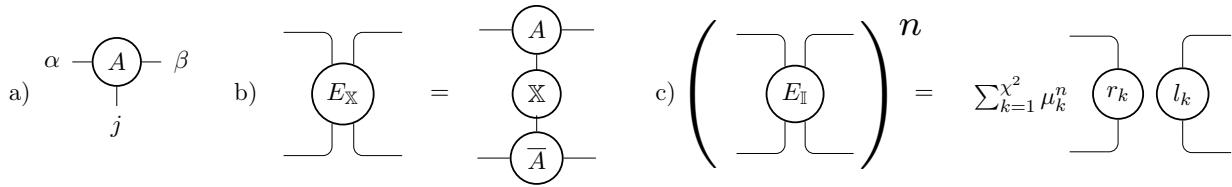
\begin{figure}[t]\centering \scalebox{0.83}{
\begin{tikzpicture}
 \begin{scope}[shift={(0,.5)}]
 \begin{scope}[shift={(2,0)}]
\node at (-4.5,0) {a)};

\node (nodeA) at (-3,0.5) [thick,minimum size=.15cm,circle,draw] {$A$};
\node at (-3,-.5) {$j$};
\node at (-4,0.5) {$\alpha$};
\node at (-2,0.5) {$\beta$};
\draw[rounded corners] (node cs:name=nodeA, angle = 180)--+(-.3,0.);
\draw[rounded corners] (node cs:name=nodeA, angle = 0) --+(.3,0.);
\draw[rounded corners] (node cs:name=nodeA, angle = -90) --+(0,-0.3);
\end{scope}
\node at (1,0) {b)};

\node (nodeE) at (2.5,0) [thick,minimum size=.15cm,circle,draw] {$E_\mathbb{X}$};
\draw[rounded corners] (node cs:name=nodeE, angle = 70) -- +(0,.5)--+(.75,0.5);
\draw[rounded corners] (node cs:name=nodeE, angle = -70) -- +(0,-.5)--+(.75,-0.5);
\draw[rounded corners] (node cs:name=nodeE, angle = 110) -- +(0,.5)--+(-.75,0.5);
\draw[rounded corners] (node cs:name=nodeE, angle = -110) -- +(0,-.5)--+(-.75,-0.5);

\node at (4,0) {$=$};

\node (nodeA1) at (5.5,1) [thick,minimum size=.15cm,circle,draw] {$A$};
\draw[rounded corners] (node cs:name=nodeA1, angle = 180)--+(-.5,0.);
\draw[rounded corners] (node cs:name=nodeA1, angle = 0) --+(.5,0.);
\node (nodeX) at (5.5,0) [thick,minimum size=.15cm,circle,draw] {$\mathbb{X}$};
\node (nodeA2) at (5.5,-1) [thick,minimum size=.15cm,circle,draw] {$\overline{A}$};
\draw[rounded corners] (node cs:name=nodeA2, angle = 180)--+(-.5,0.);
\draw[rounded corners] (node cs:name=nodeA2, angle = 0) --+(.5,0.);
\draw[rounded corners] (node cs:name=nodeA1, angle = -90) --+(0,-0.25);
\draw[rounded corners] (node cs:name=nodeA2, angle = 90) --+(0,0.25);
\begin{scope}[shift={(6.5,0)}]

\node at (1,0) {c)};
\node[scale=2] at (1.5,0) {$\Bigg($};

\node (nodeE) at (3,0) [thick,minimum size=.15cm,circle,draw] {$E_\mathbb{I}$};
\draw[rounded corners] (node cs:name=nodeE, angle = 70) -- +(0,.5)--+(.75,0.5);
\draw[rounded corners] (node cs:name=nodeE, angle = -70) -- +(0,-.5)--+(.75,-0.5);
\draw[rounded corners] (node cs:name=nodeE, angle = 110) -- +(0,.5)--+(-.75,0.5);
\draw[rounded corners] (node cs:name=nodeE, angle = -110) -- +(0,-.5)--+(-.75,-0.5);
\node[scale=2] at (4.5,0) {$\Bigg)^n$};

\node at (5,0) {$=$};
\node[scale=1] at (6.5,0.) {$\sum_{k=1}^{\chi^2} \mu_k^n$};

\node (noder) at (8,0) [thick,minimum size=.15cm,circle,draw] {$r_k$};
\draw[rounded corners] (node cs:name=noder, angle = 90) -- +(0,.5)--+(-.75,0.5);
\draw[rounded corners] (node cs:name=noder, angle = -90) -- +(0,-.5)--+(-.75,-0.5);
\node (nodel) at (9,0) [thick,minimum size=.15cm,circle,draw] {$l_k$};
\draw[rounded corners] (node cs:name=nodel, angle = 90) -- +(0,.5)--+(.75,0.5);
\draw[rounded corners] (node cs:name=nodel, angle = -90) -- +(0,-.5)--+(.75,-0.5);

\end{scope}
\end{scope}
\end{tikzpicture}}
\caption{Graphical notation: a) is the MPS tensor with bond indices $\alpha, \beta$; b) is the generalised transfer matrix; c) is a formula for $E_\mathbb{I}^n$ when the transfer matrix is diagonalisable.}\label{MPSfigure}
\end{figure}

\subsection{Overview}
For a periodic spin-\sfrac{1}{2} chain with $L$ sites, a translation-invariant MPS representation of a state is of the form
\begin{align*}
\ket{\psi} = \sum_{j_1, \dots, j_n = \pm 1/2} \tr(A_{j_1}\dots A_{j_L})\ket{j_1,\dots,j_L} .
\end{align*}
We will take $\ket{\psi}$ to be the ground state of a model defined in \eqref{fzcanon}.
For a fixed value of $j$, $A_{j}$ is a $\chi \times \chi$ matrix, and one can think of $A_{j}$ as a $\chi\times\chi\times2$ tensor. Given $A_{j}$, one can compute various quantities for a system of size $L$ and then take the limit $L\rightarrow \infty$. A useful operator is the (generalised) transfer matrix
\begin{align*}
E_\mathbb{X} = \sum_{j,k=\pm 1/2} \mathbb{X}_{kj} A_{j} \otimes \overline{A}_{k} \qquad \textrm{for} \quad \mathbb{X}_{kj} = \bra{k}\mathbb{X}\ket{j}.
\end{align*}
We can view this as a $\chi^2 \times \chi^2$ matrix, where we group indices connecting sites to the left and sites to the right\footnote{If we put the indices in we have $E^{(\alpha\alpha';\beta\beta')}_\mathbb{X} = \sum_{j,k=\pm 1/2} \mathbb{X}_{kj} A^{(\alpha;\beta)}_{j} \overline{A}^{(\alpha';\beta')}_{k}$.}
(see Figure \ref{MPSfigure}). The transfer matrix, $E_\mathbb{I}$, can be used to calculate the norm of the state, through
\begin{align*}
\braket{\psi}{\psi}&=\lim_{L\rightarrow \infty} \tr\bigl(E_\mathbb{I}^L\bigr). \end{align*}
Note that $A_j$ can be rescaled so that the state is normalised in the limit, the canonical form fixes a rescaling \cite{Perez2007}. Moreover, correlations of single-site operators $\mathbb{X}_1$ and $\mathbb{X}_{N+1}$ take the form
\begin{align}
\bra{\psi}\mathbb{X}_1\mathbb{X}_{N+1}\ket{\psi}&=\lim_{L\rightarrow \infty} \tr\bigl(E_\mathbb{X}^{\vphantom{N-1}}E_\mathbb{I}^{N-1}E_\mathbb{X}^{\vphantom{N-1}}E_\mathbb{I}^{L-(N+1)}\bigr).\label{MPScorr}\end{align}
It is simple to extend this to more general correlators, for example, \begin{align*}
\bra{\psi}\mathbb{X}_1\mathbb{X}'_2\mathbb{X}_{N+1}\ket{\psi}&=\lim_{L\rightarrow \infty} \tr\bigl(E_\mathbb{X}^{\vphantom{N-1}}E_\mathbb{X'}^{\vphantom{N-1}}E_\mathbb{I}^{N-2}E_\mathbb{X}^{\vphantom{N-1}}E_\mathbb{I}^{L-(N+1)}\bigr).\end{align*}
 Thus, the transfer matrix $E_\mathbb{I}= \sum_{j,k=\pm 1/2} A_{j} \otimes \overline{A}_{j}$ is intimately related to correlations---indeed in this limit our MPS is a finitely correlated state \cite{Fannes1992,Perez2007,Perez2008}. We now discuss how the results of Theorem \ref{thm:correlators} relate to this framework.
\subsection{General discussion}
Viewing the transfer matrix $E_\mathbb{I}$ as a $\chi^2 \times \chi^2$ matrix, then it has a Jordan normal form with~$\chi^2$ eigenvalues (counting multiplicity).
Let us first consider the case where $E_\mathbb{I}$ is diagonalisable. For~$A$ in canonical form, the largest eigenvalue of $E_\mathbb{I}$ (in absolute value) equals one, and moreover, for the purposes of this discussion, let us suppose this largest eigenvalue is unique (see Remark~\ref{remark:injective}).
Then \eqref{MPScorr} simplifies to
\begin{align}
\bra{\psi}\mathbb{X}_1\mathbb{X}_{N+1}\ket{\psi}= \sum_{k=1}^{\chi^2} \mu_k^{N-1} \bra{l_1} E_\mathbb{X} \ket{r_k}\bra{l_k}E_\mathbb{X} \ket{r_1},\label{MPScorr2}
\end{align}
where $\mu_1=1$, $\lvert \mu_k \rvert < 1$ for $k>1$ and $\{\ket{r_k},\bra{l_j}\}$ are the left and right eigenvectors of $E_\mathbb{I}$, satisfying $\braket{l_j}{r_k}=\delta_{jk}$.

Now, consider the operators $\mathcal{O}_\alpha$ defined in \eqref{Odef}. For $\alpha=2k+1$ these are correlators of local operators, for example,
 \begin{align*}
&\langle \mathcal{O}_{1}(1) \mathcal{O}_{1}(N+1) \rangle= \sum_{k=1}^{\chi^2} \mu_k^{N-1} \bra{l_1} E_{X} \ket{r_k}\bra{l_k}E_{X} \ket{r_1},\\
&\langle \mathcal{O}_{-1}(1) \mathcal{O}_{-1}(N+1) \rangle= \sum_{k=1}^{\chi^2} \mu_k^{N-1} \bra{l_1} E_{Y} \ket{r_k}\bra{l_k}E_{Y} \ket{r_1},\\
&\langle \mathcal{O}_{3}(1) \mathcal{O}_{3}(N+1) \rangle= \sum_{k=1}^{\chi^2} \mu_k^{N-3} \bra{l_1} E_{X}E_{Y}E_{X} \ket{r_k}\bra{l_k}E_{X}E_{Y}E_{X} \ket{r_1}.
\end{align*}
Thus, using the results of Theorem \ref{thm:correlators}, the terms proportional to $r_M^N$ that appear mean we can identify eigenvalues of the transfer matrix: $\big\{(-1)^{n_P+1}r_M\big\} \subseteq \{\mu_k \}$. Moreover, the terms $C_M$ will correspond to the overlaps such as $\bra{l_1} E_{X} \ket{r_k}$.
Note that if $E_\mathbb{I}$ had off-diagonal terms in the Jordan normal form, then these would appear in correlators as, e.g., $n \mu_k^{n-1}$. Such correlations do not appear in Theorem \ref{thm:correlators} in the generic case, although they can appear for non-generic cases, for example in \eqref{eq:nongenericcorrelator}. In any case, since non-trivial Jordan blocks correspond to degenerate eigenvalues, the dimension of the transfer matrix is always lower bounded by the number of distinct $r_M$.

Now, for $\mathcal{O}_\alpha$ with $\alpha =2k$ we have a non-local operator. However, we can do a similar analysis, using the generalised transfer matrix $E_Z$. For example, we have
\begin{align*}
\langle \mathcal{O}_{2}(1) \mathcal{O}_{2}(N+1) \rangle&=\lim_{L\rightarrow \infty} \tr\bigl(E_X^{\vphantom{N-1}}E_Y^{\vphantom{N-1}}E_{Z}^{N-2}E_Y^{\vphantom{N-1}}E_X^{\vphantom{N-1}}E_\mathbb{I}^{L-(N+2)}\bigr)\\&= \sum_{k=1}^{\chi^2} \tilde\mu_k^{N-2} \bigl\langle l_1\vphantom{\tilde l}\bigr|  E_{X}E_{Y} \bigl| \tilde{r\vphantom{l}}_k\bigr\rangle \bigl \langle\tilde{l}_k\bigr| E_{Y}E_{X} \bigl| \vphantom{\tilde{l}}r_1\bigr\rangle ,
\end{align*}
where we suppose for simplicity that we can diagonalise \begin{align*}E_Z= \sum_{k=1}^{\chi^2} \tilde{\mu}_k \bigl|\tilde{r\vphantom{l}}_k\bigr\rangle \bigl\langle \tilde{l}_k\bigr|.\end{align*}

\begin{figure}[t]\centering\scalebox{.9}{
\begin{tikzpicture}
 \begin{scope}[shift={(-5,.5)}]

\node (nodeE) at (2.5,0) [thick,minimum size=1cm,circle,draw] {$E_{Z}$};
\draw[rounded corners] (node cs:name=nodeE, angle = 70) -- +(0,.5)--+(.75,0.5);
\draw[rounded corners] (node cs:name=nodeE, angle = -70) -- +(0,-.5)--+(.75,-0.5);
\draw[rounded corners] (node cs:name=nodeE, angle = 110) -- +(0,.5)--+(-.75,0.5);
\draw[rounded corners] (node cs:name=nodeE, angle = -110) -- +(0,-.5)--+(-.75,-0.5);

\node at (4,0) {$=$};

\node (nodeA1) at (5.5,1) [thick,minimum size=.15cm,circle,draw] {$A$};
\draw[rounded corners] (node cs:name=nodeA1, angle = 180)--+(-.5,0.);
\draw[rounded corners] (node cs:name=nodeA1, angle = 0) --+(.5,0.);
\node (nodeX) at (5.5,0) [thick,minimum size=.15cm,circle,draw] {$Z$};
\node (nodeA2) at (5.5,-1) [thick,minimum size=.15cm,circle,draw] {$\overline{A}$};
\draw[rounded corners] (node cs:name=nodeA2, angle = 180)--+(-.5,0.);
\draw[rounded corners] (node cs:name=nodeA2, angle = 0) --+(.5,0.);
\draw[rounded corners] (node cs:name=nodeA1, angle = -90) --+(0,-0.25);
\draw[rounded corners] (node cs:name=nodeA2, angle = 90) --+(0,0.2);
\node at (7,0) {$=$};
\begin{scope}[shift={(5,0)}]
\node (nodeU1) at (6.5,1) [thick,minimum size=.15cm,circle,draw] {$U^\dagger$};
\node (nodeU2) at (4.5,1) [thick,minimum size=.15cm,circle,draw] {$U^{\vphantom{\dagger}}$};
\node (nodeU3) at (3.5,1) [thick,minimum size=.15cm,circle,draw] {$W$};
\node (nodeA1) at (5.5,1) [thick,minimum size=.15cm,circle,draw] {$A$};
\draw[rounded corners] (node cs:name=nodeA1, angle = 180)--+(-.22,0.);
\draw[rounded corners] (node cs:name=nodeA1, angle = 0) --+(.17,0.);
\draw[rounded corners] (node cs:name=nodeU3, angle = 0) --+(.15,0.);
\draw[rounded corners] (node cs:name=nodeU3, angle = 180) --+(-.25,0.);
\draw[rounded corners] (node cs:name=nodeU1, angle = 0) --+(.25,0.);
\node (nodeA2) at (5.5,-1) [thick,minimum size=.15cm,circle,draw] {$\overline{A}$};
\draw[rounded corners] (node cs:name=nodeA2, angle = 180)--+(-2.3,0.);
\draw[rounded corners] (node cs:name=nodeA2, angle = 0) --+(1.3,0.);
\draw[rounded corners] (node cs:name=nodeA1, angle = -90) -- (node cs:name=nodeA2, angle = 90);

\node at (8,0) {$=$};
\begin{scope}[shift={(9,0)}]
\node (nodeU1) at (3.5,1) [thick,minimum size=.15cm,circle,draw] {$U^\dagger$};
\node (nodeU2) at (1.5,1) [thick,minimum size=.15cm,circle,draw] {$U^{\vphantom{\dagger}}$};
\node (nodeU3) at (0.5,1) [thick,minimum size=.15cm,circle,draw] {$W$};
\draw[rounded corners] (node cs:name=nodeU3, angle = 0) --+(.15,0.);
\draw[rounded corners] (node cs:name=nodeU1, angle = 0) --+(.25,0.);
\draw[rounded corners] (node cs:name=nodeU3, angle = 180) --+(-.25,0.);
\node (nodeE) at (2.5,0) [thick,minimum size=1cm,circle,draw] {$E_\mathbb{I}$};
\draw[rounded corners] (node cs:name=nodeE, angle = 70) -- +(0,.52)--(node cs:name=nodeU1, angle = 180);
\draw[rounded corners] (node cs:name=nodeE, angle = -70) -- +(0,-.5)--+(1.55,-0.5);
\draw[rounded corners] (node cs:name=nodeE, angle = 110) -- +(0,.52)--(node cs:name=nodeU2, angle = 0);
\draw[rounded corners] (node cs:name=nodeE, angle = -110) -- +(0,-.5)--+(-2.5,-0.5);
\end{scope}

\end{scope}
\end{scope}
\end{tikzpicture}}
\caption{Graphical representation of the relationship between $E_Z$ and $E_\mathbb{I}$. $U$ is a unitary matrix, and~$W$ is a diagonal matrix of phase factors. This follows from Theorem 5 of \cite{Sanz2009}. }\label{MPSsymmetry}
\end{figure}
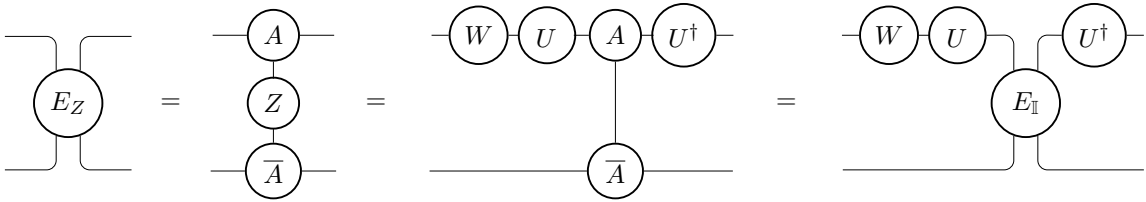

Because $\prod_j Z_j$ is a symmetry of our system, it can be shown that $E_Z$ and $E_\mathbb{I}$ have the same eigenvalues up to phase factors \cite{Sanz2009}. This follows from the result that a symmetry operator on the physical index corresponds to a transformation on the bond indices of the form $A\rightarrow WUAU^\dagger$ where $U$ is unitary and $W$ is a diagonal matrix of phase factors $\rme^{\rmi\theta_j}$. This is illustrated in Figure~\ref{MPSsymmetry}.
As before, we can use Theorem \ref{thm:correlators} to identify eigenvalues $\tilde \mu_k$ of $E_Z$. Moreover, we know that for each $k$, $\lvert \tilde \mu_k \rvert = \lvert \mu_j \rvert$ for some $j$. Hence, we can lower bound $\chi^2$ by the number of $r_M$ that give us distinct $\mu_j$ and distinct $\tilde{\mu}_j$ where none of these $r_M$ have the same absolute value.
\begin{Remark}\label{remark:injective}
An MPS is called injective if the transfer matrix has a unique largest eigenvalue.\footnote{In some literature this is called a pure MPS \cite{Pollmann2017}, and injective has a slightly different definition \cite{Perez2007,Sanz2009}.} In the models considered in this paper, the MPS ground state is not necessarily injective. Indeed, suppose that $\omega=2k+1$ for $k\in \mathbb{Z}$. Then we have a local order parameter $\mathcal{O}_{2k+1}$. Note that~$\mathcal{O}_{2k+1}$ is odd under the symmetry $\prod_j Z_j$. Hence, given a unique largest eigenvalue, we must have
\begin{align}
  \langle \mathcal{O}_{\omega}(1)\rangle = \bra{l_1} E_{X}E_{Y}\dots E_{X} \ket{r_1}= 0 , \label{eq:mixed}
\end{align}
where for definiteness we fix $\omega>0$. From \eqref{MPScorr2}, the two-point correlation function will behave like:
\begin{align}
\lim_{N\rightarrow\infty}\langle \mathcal{O}_{\omega}(1) \mathcal{O}_{\omega}(N+1) \rangle=\bra{l_1} E_{X}E_{Y}\dots E_{X} \ket{r_1}\bra{l_1}E_{X}E_{Y}\dots E_{X} \ket{r_1},\label{eq:mixed2}
\end{align}
 where the left-hand side has a non-zero limit given in Remark \ref{remark:orderparameter}. Equations \eqref{eq:mixed} and \eqref{eq:mixed2} are inconsistent, and so for $\omega =2k+1$ we must have further eigenvalues of absolute value one. This means that if the MPS is injective, we must have $\omega =2k$.

It is simple to generalise \eqref{MPScorr} to the non-injective case, and the relation between $E_Z$ and~$E_\mathbb{I}$ illustrated in Figure \ref{MPSsymmetry} does not rely on the MPS being injective. However, if we do have an injective MPS, this means that $W$ is a matrix of the form $\rme^{\rmi \theta} \mathbb{I}$. By applying the symmetry transformation on the physical index twice we have $A \rightarrow \rme^{2\rmi\theta} U AU^\dagger$. Since the transfer matrix~$E_{Z^2}=E_\mathbb{I}$, we can conclude that $\rme^{\rmi \theta}=\pm1$. Hence, in the injective case, either the spectrum of~$E_Z$ is the same as the spectrum of $E_\mathbb{I}$ or it is the same as the spectrum of $-E_\mathbb{I}$.
\end{Remark}

\begin{Remark}\label{genericremark}
Throughout the paper we have considered the generic case to be where the zeros and inverse zeros $\big\{z_{j_1}^{\vphantom {-1}},z_{j_2}^{-1},Z_{k_1}^{\vphantom {-1}},Z_{k_2}^{-1}\big\}_{j_1,j_2=1,\dots, n_z; k_1,k_2=1,\dots, n_Z}$ are pairwise distinct.
For the purposes of finding a lower bound on the bond dimension, we will now suppose the following `strongly generic' condition.
For $m_z\subseteq\{1,\dots,n_z\}$ and $m_Z\subseteq\{1,\dots,n_Z\}$
define the products
\begin{align*}
R(m_z,m_Z)=\prod_{j \in m_z} z_j \prod_{k \in m_Z} Z_k^{-1}.
\end{align*}
The `strongly generic' condition is that $\lvert R(m_z,m_Z) \rvert =\lvert R(m'_z,m'_Z) \rvert$ if and only if $m_z$ and $m_{z}'$ (similarly, $m_Z$ and $m_Z'$) differ only by replacing the index of any zero by the index of its complex-conjugate.\footnote{Recall that since $f(z)$ has real coefficients, zeros are either real or come in complex-conjugate pairs.} This is a natural condition given the form of $f(z)$, and moreover implies the usual generic case assumed above.
\end{Remark}

\subsection{Lower bound on the bond dimension}
We will now consider the number of different $r_M$ that appear in Theorem \ref{thm:correlators}, and use this to find a lower bound on the bond dimension. For ease of notation, we analyse the case $n_P=n_z+n_Z$, where, for $n_P$ even, we can find a lower bound that coincides with the upper bound in \cite{Jones21}. The general lower bound then follows easily. We remark that the $N_{\alpha}$ that appears in Theorem \ref{thm:correlators} has a natural explanation for this value of $n_P$. Recall that for $N\geq N_{\alpha}$ the given formula for the correlator applies, where $N_{\alpha} =\max\{\lvert \alpha \rvert,1\}$ (for $N<N_{\alpha}$ we can take a limit to evaluate the correlator). This is exactly the number of sites that the local endpoint operator of $\mathcal{O}_\alpha$ occupies---in particular, this is the value of $N$ for which the product of $E_\mathbb{X}$ at each endpoint is fixed, and as $N$ increases we simply increase the number of $E_\mathbb{I}$ or $E_Z$ that appear in between the endpoints.
\begin{Corollary}\label{corollary:bound}
Suppose that $f(z)$ satisfies the strongly generic condition given in Remark {\rm \ref{genericremark}}. Then
\begin{align*}
 2^{n_z+n_Z}\leq \chi^2.
\end{align*}
\end{Corollary}
In \cite{Jones21}, the upper bound $ \chi^2 \leq 2^{2 \lceil\textrm{range}(H)/2\rceil}$ is derived, where $\textrm{range}(H)$ is defined in Remark~\ref{MPSremark}. If $n_P=n_z+n_Z$, then $\textrm{range}(H)= n_z+n_Z$. Hence, when $n_P=n_z+n_Z$ is even (implying also that $\omega$ is even),
combining the two inequalities gives us that $2^{n_z+n_Z}= \chi^2$. When~$n_P=n_z+n_Z$ is odd, and for other values of $n_P$, $2 \lceil\textrm{range}(H)/2\rceil \geq n_z+n_Z$. Hence, these two inequalities do not fix the optimal value of $\chi$ in these cases, although we conjecture that the upper bound is optimal;\footnote{More precisely, we make this conjecture given the generic condition used throughout the paper: i.e., that $\big\{z_{j_1}^{\vphantom {-1}},z_{j_2}^{-1},Z_{k_1}^{\vphantom {-1}},Z_{k_2}^{-1}\big\}_{j_1,j_2=1,\dots, n_z; k_1,k_2=1,\dots, n_Z}$ are pairwise distinct. If this does not hold then there are counter examples. For example, $f(z)=z^{-2}(z-2)^2(z-1/2)^2$ has a product state ground state with bond dimension~$\chi=1$, while $\textrm{range}(H)=2$.} i.e., $\log_2(\chi)=\lceil\textrm{range}(H)/2\rceil $.
\subsubsection{Proof of Corollary \ref{corollary:bound}}\label{sec:corollaryproof}
Let us fix $n_P=n_z+n_Z$, Theorem \ref{thm:correlators} gives non-zero correlators for $-n_Z\leq \alpha\leq n_z$. Note that there are $2^{n_z+n_Z}$ subsets of $S=\big\{z_1,\dots, z_{n_z}, Z_1^{-1},\dots, Z_{n_Z}^{-1}\big\}$. Given any subset, one can take the product of the elements contained within it. Clearly all $r_M$ appearing in Theorem \ref{thm:correlators} are of this form. Moreover, we show in Appendix \ref{app:corollaryproof} that for any such product, there is a set $M$ that appears in $\langle \mathcal{O}_{\alpha}(1) \mathcal{O}_{\alpha}(N+1) \rangle =\sum_M C_M^{\vphantom N}r_M^N$ for $-n_Z\leq \alpha\leq n_z$ such that $r_M$ gives this product. The constants $C_M$ are non-zero, so we can identify $r_M$ as eigenvalues $\mu_k$ of $E_\mathbb{I}$ or $\tilde{\mu}_k$ of $E_Z$. The local (non-local) correlation functions are those with $\alpha$ odd (even). Note that in Theorem \ref{thm:correlators}, for a fixed $\alpha$ the products $r_M$ that appear always correspond to either an even-size or an odd-size subset of $S$. If $\omega$ is odd, then
the local order parameter contains only even-size subsets, since there must be an $M$ such that $r_M=1$. If $\omega$ is even, then the non-local string order parameter contains only even-size subsets for the same reason. Moreover, if we shift $\alpha$, we alternate between correlators containing only even- and only odd-size subsets. Suppose $\omega$ is odd, then products of terms in all of the even-size subsets of $S$ (and only these products) appear as $r_M$ in correlations of local operators. Using the strongly generic condition, this gives us~$2^{n_z+n_Z-1}$ different $\mu_k$. Moreover, products of terms in all of the odd-size subsets of $S$ (and only these products) appear as $r_M$ in correlations of non-local operators. This gives us $2^{n_z+n_Z-1}$ different~$\tilde\mu_k$. Since each~$\tilde\mu_k$ is a product of an odd-size subset in $S$, by the strongly generic condition, it cannot have the same absolute value as any $\mu_j$ that appears in a local correlator. This means that there must be additional eigenvalues of $E_\mathbb{I}$ that correspond to each of these $\tilde\mu_k$. If $\omega$ is even the same argument goes through exchanging odd-size and even-size. Hence, we can conclude that there are at least~$2^{n_z+n_Z}$ different eigenvalues of $E_\mathbb{I}$, i.e., $2^{n_z+n_Z}\leq \chi^2$. Moreover, for $n_P= n_z+n_Z+k$, we simply shift the labels of the $\mathcal{O}_\alpha$ by $\alpha \rightarrow \alpha-k$ in each formula. This means that $2^{n_z+n_Z}\leq \chi^2$ holds in general.

\subsection{Spectrum of the transfer matrix}
In the proof of Corollary \ref{corollary:bound}, we identify eigenvalues of the transfer matrix, $E_\mathbb{I}$, up to complex phase factors, in order to find a lower bound on the bond dimension. In the case that we have~$n_P=n_z+n_Z$ is an even integer, and using the upper bound on the bond dimension from~\cite{Jones21}, we can go further and find the spectrum of the transfer matrix. This is summarised as follows:

\begin{Corollary} \label{corollary:spectrum}
Consider a model of the form given in \eqref{fzcanon}, with $\sigma = \pm 1$, that satisfies the strongly generic condition given in Remark {\rm\ref{genericremark}} and has $n_P=n_z+n_Z=2n$ for some $n\in\mathbb{Z}$.
We define $S=\big\{z_1,\dots, z_{n_z}, Z_1^{-1},\dots, Z_{n_Z}^{-1}\big\}$.
Then the MPS is injective and the $2^{n_z+n_Z}$ eigenvalues of the transfer matrix are labelled by subsets $s\subseteq S$ and are given by
\begin{align}
\mu(s) = (-\sigma)^{\lvert s \rvert}\prod_{\tau_j \in s} \tau_j. \label{eq:spectrum}
\end{align}
\end{Corollary}
In Remark \ref{remark:transfermatrixspectrum}, we show that subject to the strongly generic condition, and with $n_P=n_z+n_Z+k=2n$, then the non-zero eigenvalues are given by \eqref{eq:spectrum}, and all further eigenvalues of the transfer matrix are zero. Note that if $n_P$ is even, then so is $\omega$. We do not give results for the spectrum of the transfer matrix in cases where $\omega$ is odd.

\begin{Remark}
Let us define $\rme^{\rmi \theta}=-\sigma$ and\begin{align*}
\eps_j &=\begin{cases} -\log{z_j}, &1\leq j\leq n_z, \\
-\log{Z_{j-n_z}^{-1}},&n_z+1\leq j \leq n_z+n_Z,
\end{cases}
\end{align*}
where $\eps_j \in \mathbb{C}$ in general.
Then given the assumptions in Corollary \ref{corollary:spectrum}, we have that the eigenvalues of the transfer matrix are given by
\begin{align*}
\mu_{n_1,\dots, n_{n_z+n_Z}} =\rme^{-\sum_j n_j (\eps_j+\rmi \theta)}.
\end{align*}
We can view these eigenvalues as coming from an free-fermion effective Hamiltonian (in general non-hermitian)
\begin{align}
H_{\mathrm{eff}}&= \sum_{j=1}^{n_z+n_Z}\left(\eps_j+\rmi\theta\right) a^\dagger_j a_j \label{eq:h_eff}
\end{align}
for some fermionic modes $a_j$. Then the spectrum of $E_\mathbb{I}$ coincides with the spectrum of $\rme^{-H_{\mathrm{eff}}}$. It would be interesting to determine whether there exist choices of $a_j$ such that $E_\mathbb{I}$ \emph{is equal to}
$\rme^{-H_{\mathrm{eff}}}$
 and, if so, whether the form of the $a_j$ can be determined straightforwardly from properties of the function $f(z)$. Such a free-fermion form of $H_{\mathrm{eff}}$ in the XY model\footnote{Except on the disorder line $a=b$, this model is outside the subclass analysed here. The MPS and transfer matrix that correspond to $H_\mathrm{eff}$ are exact only in the limit of infinite bond dimension. } with $f(z) = \frac{1}{z}(z-a)(z-b)$ is analysed analytically for the case of $a,b\in\mathbb{R}$ in \cite{Rams2015}. The zeros are restricted to be real so that the quantum to classical mapping can be used \cite{Suzuki71}. For the case of complex zeros, the eigenvalues of the transfer matrix were studied numerically in \cite{Zauner2015}, but no formula for $H_\mathrm{eff}$ is given.
\end{Remark}
\begin{Remark}
As shown in the proof of Corollary \ref{corollary:spectrum}, the phase factor $\rme^{\rmi\theta} = \pm 1$ is the phase difference between $E_\mathbb{I}$ and $E_Z$. In particular, whereas the dominant eigenvalue of $E_\mathbb{I}$ is always $1$ (by normalisation), the dominant eigenvalue of $E_Z$ is $\rme^{\rmi\theta}$. This is a $\mathbb Z_2$ invariant of the gapped phase of matter \cite{Jones2019}. Its value is independent of the topological invariant $\omega$; indeed, even when~$\omega = 0$, one can have $\theta = 0$ or $\theta = \pi$. These are known as two distinct \emph{symmetry-protected trivial} (SPt) phases of matter \cite{Fuji15}: whilst they are not accompanied by protected edge modes or degeneracies in the entanglement spectrum, they are nevertheless separated by a quantum phase transition.

Physically, the SPt phases associated to $\theta=0$ or $\theta=\pi$ are distinguished by noting whether or not the string order parameter with long-range order oscillates, i.e., $\langle \mathcal O_\omega(1) \mathcal O_\omega(N+1) \rangle = \textrm{constant} \times \rme^{\rmi\theta N} (1+o(1)) $. Hence, one can interpret $\theta$ as the momentum of the string order parameter. Intuitively, the reason this is quantized (i.e., that it has to take on a discrete value~$\theta \in \{0,\pi\}$) is due to the string order parameter being related to a $\mathbb Z_2$ symmetry, implying that $\rme^{2\rmi\theta} = 1$. As discussed in Remark \ref{remark:injective}, this can be made precise in the MPS framework by noting that $E_{Z^2} = E_\mathbb{I}$.
\end{Remark}
\begin{Remark}\label{remark:transfermatrixspectrum}
As derived in Remark \ref{limitremark}, all models with $n_P = n_z +n_Z-k$ for $k>0$ are limiting cases of models with $n_P = n_z' +n_Z$, where $n_z' = n_z+k$ and we have $k$ additional zeros $\eps x_j$ with $x_j$ pairwise distinct. Suppose also that the zeros of these models (including our chosen $x_j$) satisfy the strongly generic condition, and moreover that $n_P$ (and therefore $\omega$) is even along this path. Then all models with $\eps>0$ have a transfer matrix that can be analysed using Corollary \ref{corollary:spectrum}; this means that the MPS is injective and that we can identify the spectrum of the transfer matrix as $2^{n_z+n_Z+k}$ eigenvalues that are products of the zeros and inverse zeros. In the limit $\eps=0$, the limiting MPS is the ground state of a model with $n_P = n_z +n_Z-k$.
 The corresponding limiting transfer matrix has $2^{n_z+n_Z}$ known non-zero eigenvalues, and has $2^{n_z+n_Z+k}-2^{n_z+n_Z}$ zero eigenvalues (arising from products of the form \eqref{eq:spectrum} for sets that contained any $\eps x_j$).
Note that a priori this limiting MPS need not be in canonical form; bringing it to canonical form can potentially reduce the bond dimension, thereby removing zero eigenvalues of the transfer matrix.\footnote{However, our conjecture for generic models implies that the limiting MPS does have optimal bond dimension---see the discussion below Corollary \ref{corollary:bound}.}
An analogous argument can be made for $k<0$ by considering $f(1/z)$ and noting that the models defined by $f(1/z)$ and $f(z)$ differ by an on-site change of basis and so have the same transfer matrix.
\end{Remark}

\subsubsection{Proof of Corollary \ref{corollary:spectrum}}
The proof essentially follows from the proof of Corollary \ref{corollary:bound}. In particular, given our assumptions we have that $\omega$ is even and so, as explained above, we have $2^{n_z+n_Z-1}$ distinct eigenvalues of $E_\mathbb{I}$ labelled by odd-size sets $s\subseteq S$:
\begin{align*}
\mu(s) = \sigma (-1)^{n_P+1}\prod_{\tau_j \in s} \tau_j = -\sigma \prod_{\tau_j \in s} \tau_j.
\end{align*}
(Recall that if $\sigma=-1$ in \eqref{fzcanon}, we have an additional factor of $(-1)^N$ in all correlators given in Theorem \ref{thm:correlators}.) Furthermore, we also have $2^{n_z+n_Z-1}$ distinct eigenvalues of $E_Z$, labelled by the even-size subsets
\begin{align*}
\tilde\mu(s) = -\sigma \prod_{\tau_j \in s} \tau_j.
\end{align*}
 Moreover, by the strongly generic condition, $\lvert \tilde\mu(s)\rvert \neq \lvert \mu(s')\rvert$ for any $s$, $s'$. Now, the upper bound on the bond dimension means that we have identified all eigenvalues of $E_\mathbb{I}$ up to phase factors, and since all $\lvert \tau_j \rvert<1$ we have a unique largest eigenvalue: this is $ \tilde\mu(s) =-\sigma$, where $s$ is the empty set. Hence, the MPS is injective.
As discussed in Remark \ref{remark:injective}, having an injective MPS implies that the spectrum of $E_Z$ is either the same as the spectrum of $E_\mathbb{I}$ or the same as the spectrum of $-E_{\mathbb{I}}$. We can then conclude that $\pm \tilde\mu(s)$ is an eigenvalue of $E_\mathbb{I}$ for all $s$ (where the sign does not depend on $s$). We can identify this sign since we know that $\mu=1$ is an eigenvalue of $E_\mathbb{I}$. Hence, if $\sigma=-1$, we have that $\tilde\mu(s)$ is an eigenvalue of $E_\mathbb{I}$ for all $s$, while if $\sigma=1$, then $-\tilde\mu(s)$ is an eigenvalue of $E_\mathbb{I}$ for all $s$. We have hence identified $2^{n_z+n_Z}$ eigenvalues of $E_\mathbb{I}$, and by the upper bound on the bond dimension this means we have the full spectrum.

\begin{figure}[t]\centering\scalebox{.9}{
\begin{tikzpicture}
 \begin{scope}[shift={(-5,.5)}]

\node (nodeA1) at (5.5,1) [thick,minimum size=.15cm,circle,draw] {$A$};
\draw[rounded corners] (node cs:name=nodeA1, angle = 180)--+(-.5,0.);
\draw[rounded corners] (node cs:name=nodeA1, angle = 0) --+(.5,0.);
\node (nodeA2) at (5.5,-1) [thick,minimum size=.15cm,circle,draw] {$\overline{A}$};
\draw[rounded corners] (node cs:name=nodeA2, angle = 180)--+(-.5,0.);
\draw[rounded corners] (node cs:name=nodeA2, angle = 0) --+(.5,0.);
\draw[rounded corners] (node cs:name=nodeA1, angle = -90) -- (node cs:name=nodeA2, angle = 90);

\node at (7,0) {$=$};

\begin{scope}[shift={(.5,0)}]

\node (noder) at (8,0) [thick,minimum size=.15cm,circle,draw] {$r_1$};
\draw[rounded corners] (node cs:name=noder, angle = 90) -- +(0,.5)--+(-.75,0.5);
\draw[rounded corners] (node cs:name=noder, angle = -90) -- +(0,-.5)--+(-.75,-0.5);
\node (nodel) at (9,0) [thick,minimum size=.15cm,circle,draw] {$l_1$};
\draw[rounded corners] (node cs:name=nodel, angle = 90) -- +(0,.5)--+(.75,0.5);
\draw[rounded corners] (node cs:name=nodel, angle = -90) -- +(0,-.5)--+(.75,-0.5);

\end{scope}
\begin{scope}[shift={(10.5,0)}]
\node at (1,0) {$-\frac{1}{b} \times \frac{1}{C_X}$};

\node (nodeE) at (2.5,0) [thick,minimum size=.15cm,circle,draw] {$E_X$};
\node (nodev) at (3.5,0) [thick,minimum size=.15cm,circle,draw] {$r_1$};
\draw[rounded corners] (node cs:name=nodeE, angle = 70) -- +(0,.5)--+(.8,0.5)-- (node cs:name=nodev, angle = 90);
\draw[rounded corners] (node cs:name=nodeE, angle = -70) -- +(0,-.5)--+(.8,-0.5)-- (node cs:name=nodev, angle = -90);
\draw[rounded corners] (node cs:name=nodeE, angle = 110) -- +(0,.5)--+(-.75,0.5);
\draw[rounded corners] (node cs:name=nodeE, angle = -110) -- +(0,-.5)--+(-.75,-0.5);
\node (nodeE) at (5.5,0) [thick,minimum size=.15cm,circle,draw] {$E_X$};
\node (nodev) at (4.5,0) [thick,minimum size=.15cm,circle,draw] {$l_1$};
\draw[rounded corners] (node cs:name=nodeE, angle = 110) -- +(0,.5)--+(-.8,0.5)-- (node cs:name=nodev, angle = 90);
\draw[rounded corners] (node cs:name=nodeE, angle = -110) -- +(0,-.5)--+(-.8,-0.5)-- (node cs:name=nodev, angle = -90);
\draw[rounded corners] (node cs:name=nodeE, angle = 90) -- +(0,.5)--+(.75,0.5);
\draw[rounded corners] (node cs:name=nodeE, angle = -90) -- +(0,-.5)--+(.75,-0.5);
\end{scope}
\begin{scope}[shift={(7,-3.5)}]
\node at (1,0) {$-a\times\frac{1}{C_Y}$};

\node (nodeE) at (2.5,0) [thick,minimum size=.15cm,circle,draw] {$E_Y$};
\node (nodev) at (3.5,0) [thick,minimum size=.15cm,circle,draw] {$r_1$};
\draw[rounded corners] (node cs:name=nodeE, angle = 70) -- +(0,.5)--+(.8,0.5)-- (node cs:name=nodev, angle = 90);
\draw[rounded corners] (node cs:name=nodeE, angle = -70) -- +(0,-.5)--+(.8,-0.5)-- (node cs:name=nodev, angle = -90);
\draw[rounded corners] (node cs:name=nodeE, angle = 110) -- +(0,.5)--+(-.75,0.5);
\draw[rounded corners] (node cs:name=nodeE, angle = -110) -- +(0,-.5)--+(-.75,-0.5);
\node (nodeE) at (5.5,0) [thick,minimum size=.15cm,circle,draw] {$E_Y$};
\node (nodev) at (4.5,0) [thick,minimum size=.15cm,circle,draw] {$l_1$};
\draw[rounded corners] (node cs:name=nodeE, angle = 110) -- +(0,.5)--+(-.8,0.5)-- (node cs:name=nodev, angle = 90);
\draw[rounded corners] (node cs:name=nodeE, angle = -110) -- +(0,-.5)--+(-.8,-0.5)-- (node cs:name=nodev, angle = -90);
\draw[rounded corners] (node cs:name=nodeE, angle = 90) -- +(0,.5)--+(.75,0.5);
\draw[rounded corners] (node cs:name=nodeE, angle = -90) -- +(0,-.5)--+(.75,-0.5);
\end{scope}
\begin{scope}[shift={(14,-3.5)}]
\node at (1,0) {$+\frac{a}{b}\times\frac{1}{C_{\tilde{Z}}}$};

\node (nodeE) at (2.5,0) [thick,minimum size=.15cm,circle,draw] {$E_{\tilde Z}$};
\node (nodev) at (3.5,0) [thick,minimum size=.15cm,circle,draw] {$r_1$};
\draw[rounded corners] (node cs:name=nodeE, angle = 70) -- +(0,.5)--+(.8,0.5)-- (node cs:name=nodev, angle = 90);
\draw[rounded corners] (node cs:name=nodeE, angle = -70) -- +(0,-.5)--+(.8,-0.5)-- (node cs:name=nodev, angle = -90);
\draw[rounded corners] (node cs:name=nodeE, angle = 110) -- +(0,.5)--+(-.75,0.5);
\draw[rounded corners] (node cs:name=nodeE, angle = -110) -- +(0,-.5)--+(-.75,-0.5);
\node (nodeE) at (5.5,0) [thick,minimum size=.15cm,circle,draw] {$E_{\tilde{Z}}$};
\node (nodev) at (4.5,0) [thick,minimum size=.15cm,circle,draw] {$l_1$};
\draw[rounded corners] (node cs:name=nodeE, angle = 110) -- +(0,.5)--+(-.8,0.5)-- (node cs:name=nodev, angle = 90);
\draw[rounded corners] (node cs:name=nodeE, angle = -110) -- +(0,-.5)--+(-.8,-0.5)-- (node cs:name=nodev, angle = -90);
\draw[rounded corners] (node cs:name=nodeE, angle = 90) -- +(0,.5)--+(.75,0.5);
\draw[rounded corners] (node cs:name=nodeE, angle = -90) -- +(0,-.5)--+(.75,-0.5);
\end{scope}
\end{scope}
\end{tikzpicture}}
\caption{Graphical representation of the transfer matrix for $f(z)=z^{-2}(z-a)^2(z-b)^2$ with $a<1$ and $b>1$. We define $\tilde{Z} = Z - \frac{\left(a^2-1\right) b^2-a b+1}{b (a-b)}$. The normalisation constants $C_X$, $C_Y$ and $C_{\tilde{Z}}=-C_XC_Y$ are given in \eqref{eq:diagonaltransfer}.}\label{MPSfigure2}
\end{figure}
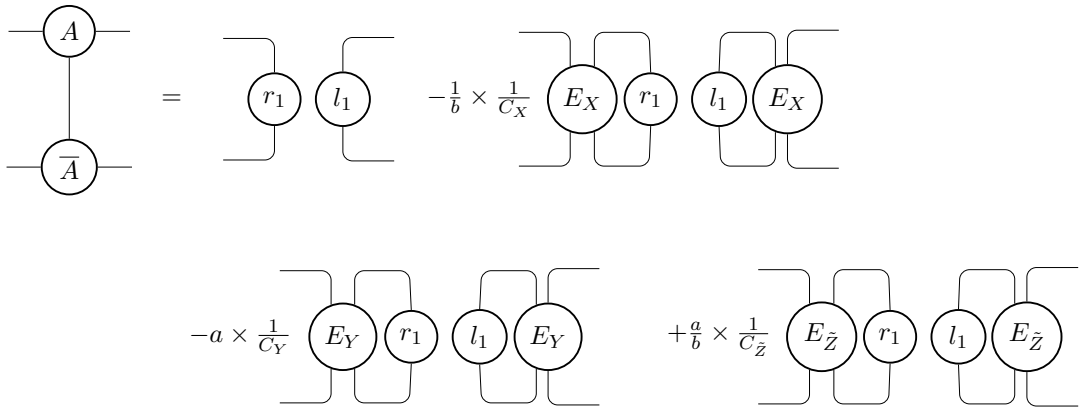

\subsection[Diagonalising the transfer matrix for the case of two zeros of multiplicity two and a pole of order two]{Diagonalising the transfer matrix for $\boldsymbol{f(z) = z^{-2}(z-a)^2(z-b)^2}$}\label{sec:diagonal}
We now explore how our results can allow us to constrain the eigenvectors of the transfer matrix as well as the eigenvalues. Let us consider $f(z) = z^{-2}(z-a)^2(z-b)^2$ with $a<1$ and $b>1$. The correlators are non-zero for $-1\leq \alpha \leq 1$. The values of these correlators are given in Example~\ref{correxample} (after shifting the index appropriately). Let $P_1=\ket{r_1}\bra{l_1}$, then, given the upper bound $\chi^2=4$, in Appendix \ref{app:diagonaltransfer} we prove the following:
\begin{align}
E_\mathbb{I}={}&P_1 -
a \left( \frac{(b-a)}{{\bigl(1-b^{-2}\bigr)(1-ab)}}E_YP_1E_Y \right)-\frac{1}{b} \left(\frac{(a-b)}{{\bigl(1-a^2\bigr)(1-ab)}}E_XP_1E_X\right)\nonumber \\
&+\frac{a}{b} \left( \frac{(a-b)^2}{{\bigl(1-a^2\bigr)\bigl(1-b^{-2}\bigr)(1-ab)^2}}E_{\tilde Z}P_1E_{\tilde Z} \right),\label{eq:diagonaltransfer}
\end{align}
where
\[\tilde{Z} = Z - \frac{\bigl(a^2-1\bigr) b^2-a b+1}{b (a-b)}.\]
 The eigenvalues are $\{1,\!-a,-1/b,a/b\}$ and the bracketed expressions are the relevant ${P_j\!=\!\ket{r_j}\!\bra{l_j}}$. This is given graphically in Figure \ref{MPSfigure2}. Note that we did not need the MPS tensor to derive this result, although it is needed to evaluate this formula. We can furthermore use the formula to identify that, say, $\ket{r_2} \propto E_Y \ket{r_1}$. Given this formula for~$E_\mathbb{I}$, further calculations are required to determine whether we can find expressions for fermionic modes $a_j$ in \eqref{eq:h_eff} such that~$E_\mathbb{I}=\rme^{-H_{\mathrm{eff}}}$.

One can straightforwardly generalise the analysis of Appendix \ref{app:diagonaltransfer} to derive results about the eigenvectors of $E_\mathbb{I}$ in cases where $f(z)$ has more zeros. However, this will not immediately lead to a diagonal form as in \eqref{eq:diagonaltransfer}. In that example, operators such as $E_YP_1E_Y$ corresponded to a unique eigenvalue, while in the general case such operators can correspond to multiple eigenvalues (i.e., we have a sum over more than one set of zeros in the expressions given in Theorem \ref{thm:correlators}).

\section{Recovering generic models via a limiting procedure}\label{sec:generic}

The results in this paper are for the special subclass of BDI and AIII models where all non-zero zeros of $f(z)$ have an even multiplicity. This subclass admits closed results for correlation functions and an exact matrix product state representation of the ground state. In this section, we show how a generic BDI or AIII model can be obtained as a limit\footnote{While it will be a sequence $f_m(z)$ of finite-range models, the range increases linearly with $m$ (i.e., it increases without bound). Nevertheless, the limit $\lim_{m \to \infty} f_m(z)$ is finite-range.} of such special models. This shows that any ground state in these subclasses can be obtained as a limit of matrix product states. Moreover, such a sequence can be used to derive results about generic models by using results derived in the present work.

We note that sequences approximating certain Toeplitz symbols (see Section \ref{sec:toeplitz}) by rational functions are discussed in \cite{Hartwig1969}. The methods of expansion are different in general, although coincide for the case of the quantum Ising model. Moreover, our approach focuses on a sequence of Hamiltonians with MPS ground states, from which one can then derive a sequence of Toeplitz symbols.

\subsection{A sequence of models}
A general gapped BDI or AIII model corresponds, up to normalisation, to a Laurent polynomial of the form
\begin{align}
  f(z) = \sigma \frac{1}{z^{N_P}} \prod_{j=1}^{N_z} \left(z-z_j\right)\prod_{k=1}^{N_Z} \left(z-Z_k\right). \label{fz_generic}
\end{align}
We have $\lvert z_j\rvert <1$, $\lvert Z_k\rvert >1$, and $\sigma \in \{\pm1\}$ for the BDI class \cite{Jones2019}, while for AIII, $\sigma=\rme^{\rmi\theta}$.
For a generic model in this class, we have that all zeros have multiplicity one, and moreover that no zeros coincide with inverse zeros (or inverse conjugate zeros in AIII). We now construct a sequence $f_m(z)$ ($m=1,2,\dots$) of models of the form \eqref{fzcanon} such that in an open region of the complex plane containing the unit circle, we have $\lim_{m \to \infty} f_m(z) = f(z)$. As discussed in Section \ref{sec:model}, the reason that the unit circle is special is because it stores all physical information. To construct the sequence, we first rewrite
\begin{align*}
f(z) = \sigma \left(\prod_{k=1}^{N_Z} \left(-Z_k\right)\right) \frac{1}{z^{N_P-N_z}} g(z)^2 \qquad \textrm{with}\quad g(z) = \prod_{j=1}^{N_z} \sqrt{ 1-\frac{z_j}{z}} \prod_{k=1}^{N_Z} \sqrt{ 1-\frac{z}{Z_k} }.
\end{align*}
By using the series expansion of the square roots, we obtain a well-defined series expansion for $g(z)$ for $\max_j\{|z_j|\} < |z| < \min_k \{|Z_k| \}$ (i.e., an annulus containing the unit circle). More precisely, if we define the partial sums
\begin{align*}
s_m(z) = \sum_{l=0}^m (-1)^l \left( \begin{matrix} 1/2 \\ l \end{matrix} \right) z^l
\end{align*}
then $\sqrt{1-z} = \lim_{m\to \infty} s_m(z)$ (where we take the principal branch of the square root).
We can thus define a sequence
\begin{align}
g_m(z) = \prod_{j=1}^{N_z} s_m(z_j/z) \prod_{k=1}^{N_Z} s_m(z/Z_k), \label{eq:path}
\end{align}
where $g_m(z)$ converges uniformly to $g(z)$ on the annulus given above.
Hence, the functions $f_m(z) \propto z^{N_z-N_P} g_m(z)^2$ define a sequence of polynomials which all belong to the special subclass of models studied in the present work, and where the limit of this sequence is a generic gapped BDI or AIII model.\footnote{One can analyse non-generic models in a similar way, for ease of presentation we focus on the generic case. For example, if $f(z) =(z-z_1)(z-z_2)^2 $, we could take the sequence $g_m(z)=s_m(z_1/z)(z-z_2)$.} We explain below that every truncated model has the same winding number as the limiting $f(z)$.

This path can then be used to extend certain results of the present work to the more general case. We illustrate this now for the order parameter in the BDI class.

\subsection{Order parameter}

In a previous work \cite{Jones2019}, for a general $f(z)$ of the form \eqref{fz_generic}, we derived the following value of the order parameter in class BDI:
\begin{align}
\lim_{N\rightarrow\infty}\lvert\langle\mathcal{O}_\omega(1)\mathcal{O}_\omega(N+1)\rangle \rvert = \left( \frac{ \prod_{j_1,j_2=1}^{N_z} (1-z_{j_1} z_{j_2})
	\prod_{k_1,k_2=1}^{N_Z} \left(1-\dfrac{1}{Z_{k_1} Z_{k_2}}\right)}{\prod_{j=1}^{N_z}\prod_{k=1}^{N_Z} \left (1-\dfrac{z_j}{ Z_k}\right)^2} \right)^{1/4}. \label{eq:orderparameter_generic}
\end{align}
This result is proved using Szeg\H{o}'s theorem \cite{Szego}. To illustrate the usefulness of approximating a~generic $f(z)$ by a sequence of models with degenerate roots, as in \eqref{eq:path}, here we demonstrate how this expression \eqref{eq:orderparameter_generic} is naturally obtained by taking a limit of the formula \eqref{eq:orderparameter} which is derived in this work using Day's formula (see Theorem \ref{thm:Day}).

For any $m$, since $s_m(z)$ is a polynomial with $s_m(0) = 1$, we know there exists a set of complex numbers \smash{$\big\{ \lambda_l^{(m)}\big\}_{l=1,\dots,m}$} such that we can write \[s_m(z) = \prod_{l=1}^m \bigl(1-\lambda_l^{(m)} z \bigr).\] Moreover, since $s_m(z)$ is real-valued on the real-line, the $\lambda_l^{(m)}$ are real or come in complex-conjugate pairs. One can also show\footnote{It is straightforward to see that $\lvert 1-s_m(z) \rvert <1$ for $z=\rme^{\rmi k}$, and the result follows using Rouch\'{e}'s theorem~\cite{Garling14}.} that \smash{$\big\lvert\lambda_l^{(m)}\big\rvert<1$} for all $m$. This means that the winding number of $g_m(z)$ is zero, and so, for all $m$, $f_m(z)$ has winding number $N_z-N_P$.

We thus have
\begin{align}
\sqrt{1-z} = \lim_{m \to \infty} \prod_{l=1}^m \bigl(1-\lambda_l^{(m)} z\bigr), \qquad \lvert z\rvert<1. \label{eq:sqrt_identity}
\end{align}
By taking the square root of both sides of \eqref{eq:sqrt_identity}, we obtain the useful identity
\begin{align}
(1-z)^{1/4} = \lim_{m \to \infty} \prod_{l_1,l_2=1}^m \bigl( 1 - \lambda_{l_1}^{(m)} \lambda_{l_2}^{(m)} z \bigr), \label{eq:sqrt_identity2}
\end{align}
we prove this in Appendix~\ref{app:limit}.

Using the above factorisation of $s_m(z)$, we have that
\begin{align*}
g_m(z) = \prod_{j=1}^{N_z} \prod_{l=1}^m \left(1-\frac{\lambda_l z_j}{z} \right) \times \prod_{k=1}^{N_Z} \prod_{p=1}^m \left(1- \frac{\lambda_p z}{Z_k} \right).
\end{align*}
There are $n_z = N_z m$ roots inside the unit circle given by $\lambda_l z_j$ and $n_Z = N_Z m$ roots outside given by $Z_k/\lambda_p$.

Using the result \eqref{eq:orderparameter} derived in this work, the order parameter for a given $m$ is
\begin{align*}
\lim_{N\rightarrow\infty}\lvert\langle\mathcal{O}&_\omega(1)\mathcal{O}_\omega(N+1)\rangle \rvert \\
&= \frac{ \prod_{j_1,j_2=1}^{N_z} \prod_{l_1,l_2=1}^m (1-\lambda_{l_1} \lambda_{l_2} z_{j_1} z_{j_2}) \prod_{k_1,k_2=1}^{N_Z} \prod_{l_1,l_2=1}^m \left(1-\dfrac{\lambda_{l_1} \lambda_{l_2}}{Z_{k_1} Z_{k_2}}\right)}{\prod_{j=1}^{N_z}\prod_{k=1}^{N_Z} \prod_{ l_1, l_2=1}^m \left (1-\lambda_{ l_1} \lambda_{ l_2} \dfrac{z_j}{ Z_k}\right)^2}.
\end{align*}
Note that, as discussed in Remark \ref{remark:nongeneric}, this formula applies even in the non-generic case. Using~\eqref{eq:sqrt_identity2} to take the limit, we obtain
\begin{align}
&\lim_{m \to \infty} \lim_{N\rightarrow\infty}\lvert\langle\mathcal{O}_\omega(1)\mathcal{O}_\omega(N+1)\rangle \rvert \nonumber\\
&\qquad= \left( \frac{ \prod_{j_1,j_2=1}^{N_z} (1- z_{j_1} z_{j_2} )
\prod_{k_1,k_2=1}^{N_Z} \left(1-\dfrac{1}{Z_{k_1} Z_{k_2}}\right)}{\prod_{j=1}^{N_z}\prod_{k=1}^{N_Z} \left (1- \dfrac{z_j}{ Z_k}\right)^2} \right)^{1/4}. \label{eq:orderparameter_limit}
\end{align}
We see that this coincides with the known formula \eqref{eq:orderparameter_generic}. For this to be a rigorous independent derivation of this formula, one should also prove that the two limits on the left-hand side of~\eqref{eq:orderparameter_limit} commute. This would require bounding the subleading terms which are also given by Theorem~\ref{thm:correlators}, but we will not pursue this further here.

\section{Outlook}\label{sec:outlook}
In this paper we have analysed correlations in a subclass of BDI and a subclass of AIII Hamiltonians. We derived exact formulae for string correlations and for the characteristic polynomial of the correlation matrix in both classes. This allowed us to deduce the existence of an MPS representation, and to give a lower bound on its bond dimension for BDI models. Moreover, for class BDI we showed how our results constrained properties of the transfer matrix, even leading to the full spectrum in certain cases. We furthermore saw how generic models can be recovered as a limit of the models studied in this work. We expect that the analysis of the transfer matrix in class BDI could be straightforwardly generalised to class AIII using the results of Section \ref{sec:AIII}.

There are a number of outstanding questions that emerge from our discussion:
\begin{itemize}\itemsep=0pt
 \item
In Section \ref{sec:transfer}, we saw that the spectrum of the transfer matrix has a free-fermion form, suggesting that there could exist a free-fermion Hamiltonian $H_\textrm{eff}$ (in general, non-hermitian) such that $E_\mathbb{I} = \rme^{-H_\textrm{eff}}$. This is a natural question, with connections to quantum-classical mappings and imaginary time evolution under our class of Hamiltonians \cite{Hutchinson2015,Zauner2015}. Relatedly, in Section \ref{sec:diagonal}, we showed how our results can be used to diagonalise the transfer matrix in a simple case. It would be interesting to see to what extent a similar analysis can be applied to other cases in these classes of models.
\item Our methods, based on Toeplitz determinant theory, allowed us to deduce the existence of an MPS representation of the ground state in both the BDI and AIII classes, but did not give an upper bound on its bond dimension. Based on the analysis in \cite{Jones21}, in the BDI class we have an upper bound of $ \chi^2 \leq 2^{2 \lceil\textrm{range}(H)/2\rceil}$. It would be of interest to derive this using the methods of this paper, in particular to see how this arises through Gorodetsky's formula.
\item In Examples \ref{examplecorrmatrix} and \ref{examplecorrmatrix2}, we gave exact formulae for the correlation eigenvalues $\{\nu_j\}$, both for finite subsystem size $N$, and for $N\rightarrow \infty$. It would be interesting to determine analogous formulae for more general examples in our subclass of models, and to see if there is any simple relationship between zeros of $f(z)$ and these eigenvalues, as is the case with the transfer matrix eigenvalues. Moreover, our results based on Gorodetsky's formula must agree for $N\rightarrow \infty$ with (limiting cases of) the results of Its, Mezzadri and Mo \cite{Its2008}. It would be interesting to clarify the relationship between the two.
\item A further problem is to rigorously prove the degeneracy of the correlation eigenvalues, argued physically in Remark \ref{remark:symmetry}, from the point of view of (block) Toeplitz determinants.
\item In Section~\ref{sec:generic}, we showed how general models with $f(z)$ of the form \eqref{fz_generic} in the BDI or AIII class can be approximated by a sequence of the models considered in this work. We already saw how this gave a new interpretation to the formula for the order parameter obtained in \cite{Jones2019}. This could potentially be used to appropriately generalise results derived in this work to more general Hamiltonians.
\item Finally, it is natural to look for other classes of models where we can find subclasses of models that admit exact closed formulae for correlations. One extension would be to study free-fermion models in other symmetry classes, identifying subclasses where the correlations simplify. The results of \cite{Kraus10,Schuch08} imply that rational symbols for the correlation matrix are a necessary condition for exact Gaussian MPS ground states (or in more than one-dimension, projected entangled pair states (PEPS) \cite{Cirac2020,Kraus10}) in free-fermion models. It would be of interest to see whether any of the Toeplitz determinant methods used in this work would be applicable to such models.
\end{itemize}

\appendix
\section{The corresponding spin chain}
\label{app:spin}
Define the Jordan--Wigner transformation by
\begin{align*}
Z_n = \rmi \tilde\gamma_n \gamma_n,\qquad
X_n = \prod_{m=1}^{n-1} \left(\rmi \tilde\gamma_m \gamma_m\right) \gamma_n, \qquad
Y_n = \prod_{m=1}^{n-1} \left(\rmi \tilde\gamma_m \gamma_m\right) \tilde\gamma_n.
\end{align*}
Then, the Hamiltonian \eqref{HBDI} becomes\footnote{This duality is not quite exact due to the implicit periodic boundary conditions on the spin chain; we will ignore this here since we focus on bulk properties, see \cite{Suzuki71}.}
\begin{align*}
H &= \sum_{n\in\mathrm{sites}} \Bigg(\frac{t_0}{2} Z_n-\sum_{\alpha > 0} \frac{t_\alpha}{2} X_n \left(\prod_{m=n+1}^{n+\alpha-1} Z_m\right)X_{n+\alpha}
-\sum_{\alpha < 0} \frac{t_\alpha}{2}Y_n \left(\prod_{m=n+1}^{n+\lvert\alpha\rvert-1} Z_m\right)Y_{n+\lvert\alpha\rvert}\Bigg).
\end{align*}
The string operators $\mathcal{O}_\alpha$ are local for $\alpha$ odd and non-local for $\alpha$ even. For example,
\begin{align*}
&\mathcal{O}_3(n)=X_nY_{n+1}X_{n+2},\qquad
\mathcal{O}_2(n)=\bigg(\prod_{j< n}Z_j\bigg) Y_nX_{n+1},\\
&\mathcal{O}_1(n)=X_n,\qquad
\mathcal{O}_0(n)=\prod_{j< n}Z_j,\\
&\mathcal{O}_{-1}(n)=Y_n\qquad \mathcal{O}_{-2}(n)=\bigg(\prod_{j< n}Z_j \bigg)X_nY_{n+1},
\end{align*}
and results for their (bulk) correlations carry over. Entanglement properties derived from the correlation matrix also apply to the spin chain. One difference is that phases with odd winding number spontaneously break the symmetry $\prod_j Z_j$. There is no symmetry breaking for the fermionic chain.
\section{Fermionic two-point correlators}\label{app:fourier}

In this section we calculate the Fourier coefficients of
\begin{align*}
\tilde t(z) = \frac{1}{z^{n_z+n_Z}} \frac{\prod_{j=1}^{n_z} \left(z-z_j\right)\prod_{k=1}^{n_Z} \left(z-Z_k\right)}{\prod_{j=1}^{n_z} \left(1/z-\overline{z}_j\right)\prod_{k=1}^{n_Z} \left(1/z-\overline{Z}_k\right)}.
\end{align*}
The Fourier coefficient $\tilde{t}_n$ gives the fermionic two-point correlators in the relevant subclass of both the BDI and AIII classes. For BDI, noting that zeros come in complex-conjugate pairs, $\tilde{t}_n$ gives
\[\langle-\rmi \tilde\gamma_m\gamma_{m+n-k}\rangle\qquad \text{for}\quad \sqrt{\frac{f(z)}{f(1/z)}}=z^k\tilde{t}(z).\] This includes all $f(z)$ in the class considered in the main text. For AIII, $\tilde{t}_n$ gives
\[-2\big\langle c^\dagger_{A,m}c^{\vphantom \dagger}_{B,m-n-k}\big\rangle\qquad \text{for} \quad\sqrt{\frac{f(z)}{\overline{f}(1/z)}}=z^k\tilde{t}(z).\]
Furthermore, these Fourier coefficients are the matrix elements of the Toeplitz determinants that we use to calculate string correlators. These determinants have symbol $\hat{t}(z)=z^k\tilde{t}(z)$ for some appropriate $k$, and we can find the Fourier coefficients using the formulae below along with~${\hat{t}_n=\tilde{t}_{n-k}}$.

We have
\begin{align*}
\tilde{t}_n &= \frac{1}{2\pi \rmi}\frac{1}{\prod_{j=1}^{n_z}(-\overline{z}_j)\prod_{k=1}^{n_Z}\bigl(-\overline{Z}_k\bigr)}\int_{S^1} \frac{\prod_{j=1}^{n_z} (z-z_j)\prod_{k=1}^{n_Z} (z-Z_k)}{\prod_{j=1}^{n_z} \bigl(z-\overline{z}_j^{-1}\bigr)\prod_{k=1}^{n_Z} \bigl(z-\overline{Z}_k^{-1}\bigr)} z^{-n-1} \rmd z ,
\end{align*}
with the integrand in a standard form with simple poles at $z=\overline{z}_j^{-1}, \overline{Z}_k^{-1}$ and, for $n=0$, at~$z=0$. We evaluate by deforming the contour: for $n>0$ we deform the contour out to infinity, and pick up the negative residues at each of the poles $\big\{\overline{z}_j^{-1}\big\}$. For $n <0$ we deform the contour to~$z=0$ and sum over residues at the poles \smash{$\big\{\overline{Z}_k^{-1}\big\}$}. For $n=0$ we deform the contour to $z=0$ and sum over the residues at \smash{$\big\{\overline{Z}_j^{-1}\big\}$} and $z=0$. This gives
\begin{align*}
\tilde{t}_n =- \sum_{k=1}^{n_z} \frac{\prod_{j=1}^{n_z} \left(1-z_j\overline {z}_k\right)\prod_{j=1}^{n_Z} \left(1-Z_j\overline z_k\right)}{\prod_{j=1, j\neq k}^{n_z} \left( \overline z_j-\overline z_k\right)\prod_{j=1}^{n_Z} \bigl(\overline Z_j-\overline z_k\bigr)}\overline z_k^{n-1}, \qquad n>0,\\
\tilde{t}_n = \sum_{k=1}^{n_Z} \frac{\prod_{j=1}^{n_z} \bigl(1-z_j\overline Z_k\bigr)\prod_{j=1}^{n_Z} \bigl(1-Z_j\overline Z_k\bigr)}{\prod_{j=1}^{n_z} \bigl(\overline z_j-\overline Z_k\bigr)\prod_{j=1,j\neq k}^{n_Z} \bigl(\overline Z_j- \overline Z_k\bigr)} \overline Z_k^{n-1}, \qquad n<0
\end{align*}
and for $n=0$,\begin{align*}
\tilde{t}_0 = &\sum_{k=1}^{n_Z} \frac{\prod_{j=1}^{n_z} \bigl(1-z_j\overline Z_k\bigr)\prod_{j=1}^{n_Z} \bigl(1-Z_j\overline Z_k\bigr)}{\prod_{j=1}^{n_z} \bigl(\overline z_j-\overline Z_k\bigr)\prod_{j=1,j\neq k}^{n_Z} \bigl(\overline Z_j- \overline Z_k\bigr)} \overline Z_k^{-1} + \prod_{j=1}^{n_z} (-z_j)\prod_{k=1}^{n_Z} (-Z_k) .\end{align*}
If $n_z=0$ we have $\tilde{t}_0=\prod_{k=1}^{n_Z} \bigl(-\overline{Z}_k^{-1}\bigr)$, one can see this directly by changing variables $z\rightarrow1/z$ before doing the contour integral. In class BDI we can remove all complex-conjugates from the above equations since the zeros appear in complex-conjugate pairs.

\section{Comparison of asymptotics of string correlators} \label{app:asymptotics}
In Remark \ref{remark:asymptotics}, we refer to the analysis of the asymptotics of the correlators found in \cite{Jones2019}. In the notation of that paper, we have
\begin{align*}
l(z) = m(z)^{-1}= \frac{\prod_{j=1}^{n_z} (1-zz_j)\prod_{j=1}^{n_z} (1-z_j/z)}{\prod_{k=1}^{n_Z} \bigl(1-zZ_k^{-1}\bigr)\prod_{k=1}^{n_Z} \bigl(1-Z_k^{-1}/z\bigr)}.
\end{align*}
For the calculation, we need the Fourier coefficients $l_N$ and $m_N$ for large $N$. As in Appendix~\ref{app:fourier}, we can deform the contour and pick up residues from poles outside the unit circle. Then the asymptotics are dominated by the pole(s) nearest to the unit circle. For $l_N$, this is $Z_1$ (and any other $Z_k$ of the same size) and for $m_N$ this is $z_1^{-1}$ \big(and any other $z_j^{-1}$ of the same size\big). This gives us the asymptotics needed in Remark \ref{remark:asymptotics}.
\begin{Example}
Let $f(z) = (z-a)^2(z-b)^2(z-c)^2$, with $\lvert a \rvert <1$, $\lvert b \rvert <1$ and $\lvert c \rvert >1$. We will evaluate the correlator $ \langle \mathcal{O}_{5}(1) \mathcal{O}_{5}(N+1) \rangle$ using the methods of \cite{Jones2019} and compare to the result of Theorem \ref{thm:correlators}.

Firstly, using Theorem \ref{thm:correlators}, we have $\omega = 4$, $n_z=2$, $n_Z=1$, $n_P=0$ and we hence use Case 2. Then $\lvert M \rvert=0$ and so we have one set $M$ such that $r_M = 1/c$. Evaluating $C_M$ gives
\begin{align*}
\langle \mathcal{O}_{5}(1) \mathcal{O}_{5}(N+1) \rangle = (-1)^N c^{-N} \prod_{x \in \{a,b,c\}}\bigl(x^{-1}-a\bigr)\bigl(x^{-1}-b\bigr)\frac{1}{c-a}\frac{1}{c-b}a^2b^2c^4.
\end{align*}

Now, using a result from \cite{Hartwig1969} as in \cite{Jones2019}, we have that
\begin{align}
\langle \mathcal{O}_{5}(1) \mathcal{O}_{5}(N+1) \rangle = (-1)^N \underbrace{\frac{\bigl(1-a^2\bigr)\bigl(1-b^2\bigr)(1-ab)^2\bigl(1-1/c^{2}\bigr)}{(1-a/c)^2(1-b/c)^2}}_{\lim_{M\rightarrow \infty} \lvert\langle \mathcal{O}_{4}(1) \mathcal{O}_{4}(M+1) \rangle \rvert}l_N (1+o(1)),\label{exampleasymptotics}
\end{align}
where \[l_N = c^{-N}\frac{(1-ac)(1-bc)(1-a/c)(1-b/c)}{(1-1/c^2)}\] by the residue theorem. On simplifying the constant, we see that the two approaches agree (and moreover that the $o(1)$ term in \eqref{exampleasymptotics} must be exactly zero).
\end{Example}
\section{Details for Section \ref{sec:corrmatrixresults}}\label{app:smith}
Here we give details of the Smith canonical form needed to derive the results given in Examples~\ref{examplecorrmatrix} and \ref{examplecorrmatrix2}. The results below and the formulae given in the examples follow from details in the main text, where we simplified the relevant expressions using \textsc{Mathematica}~\cite{Mathematica}.

\subsection{Example \ref{examplecorrmatrix} }Recall that $f(z) = \frac{1}{z}(z-b)^2$ with $\vert b\rvert<1$.
Hence, we are interested in the matrix polynomial
\begin{align*}
a(z) = \left(\begin{matrix} -\rmi \lambda b(z-b)(z-1/b) & (z-b)^2\\- b^2(z-1/b)^2&- \rmi \lambda b (z-b)(z-1/b)\end{matrix}\right).
\end{align*}
Then
\begin{align*}
y(z,\lambda) a(z) w(z,\lambda) = \left(\begin{matrix} 1 & 0\\0&(z-b)^2(z-1/b)^2\end{matrix}\right),
\end{align*}
where we define
\begin{align*}
y(z,\lambda)=\left(
\begin{matrix}
 \rmi b \bigl(b^2-bz\bigl(\lambda ^2-1\bigr)+\lambda ^2-2\bigr) & \bigl(b^2-1\bigr) \lambda \\
 \dfrac{\lambda (b z-1) \bigl(b \lambda ^2 (b-z)+b z-1\bigr)}{b^2-1} & \rmi \lambda ^2 (b-z)
\end{matrix}
\right)
\end{align*}
and \begin{align*}
w(z,\lambda) &=
\left(
\begin{matrix}
 -\dfrac{1}{\bigl(b^2-1\bigr)^3 \lambda } & -\dfrac{\rmi (b-z) \bigl(\lambda ^2 (b z-1)^2+b (b-z) (b (b+z)-2)\bigr)}{b^2 \bigl(b^2-1\bigr)^2 \lambda ^2 \bigl(\lambda ^2-1\bigr)} \\
 -\dfrac{\rmi b}{\bigl(b^2-1\bigr)^3} & -\dfrac{{\bigl(b^2-1\bigr)^3}-b (b-z) \bigl(\lambda ^2 (b z-1)^2+b (b-z) (b (b+z)-2)\bigr)}{b^2 \lambda \bigl(\lambda ^2-1\bigr){\bigl(b^2-1\bigr)^2}}
\end{matrix}
\right).
\end{align*}
\subsection{Example \ref{examplecorrmatrix2}} Recall that $f(z) = \frac{1}{z^2}(z-a)^2(z-b)^2$. For the first case $\lvert a\rvert <1$ and $\lvert b\rvert <1$, we are interested in the matrix polynomial
\begin{align*}
a(z) = \left(\begin{matrix} \rmi \lambda a b(z-a)(z-1/a)(z-b)(z-1/b) &(z-a)^2 (z-b)^2\\ -a^2b^2(z-1/a)^2(z-1/b)^2&\rmi \lambda a b (z-a)(z-1/a)(z-b)(z-1/b)\end{matrix}\right).
\end{align*}
The matrix $y(z,\lambda)$ is given by
\begin{align*}
y_{11}(z,\lambda)={} &-ba^2 (b+z)+a^2+a b (1-b z)+b^2,\\
y_{12}(z,\lambda)={} &\rmi \lambda \bigl(a^2-z (a+b)+a b+b^2-1\bigr),\\
y_{21}(z,\lambda)={} &a \bigl(a^2-1\bigr) b \bigl(b^2-1\bigr) (a+b) (a b-1)^2 \bigl(a^2 b^2-a^2-a b-b^2\bigr) (a z-1) (b z-1)\\
&
\times\bigl(a^2 b^2 \lambda ^2 \bigl(a^2+a b+b^2-1\bigr)-a^4 b^2-a^2 b^4-a^3 b+a^2-a b^3+a b+b^2-1\bigr) \\
&\times\bigl(\bigl(1-\lambda ^2\bigr) (a-z)^2 (b-z)^2 \bigl(a^2 b^2 +a^2bz-a^2+a b (b z-1)-b^2\bigr)^2\\
& -\bigl(a^2-1\bigr)^2 \bigl(b^2-1\bigr)^2 (a b-1)^2\bigr),\\
y_{22}(z,\lambda) = &-\rmi a \bigl(a^2-1\bigr) b \bigl(b^2-1\bigr) \lambda (a+b) (a b-1)^2 \bigl(a^2 \bigl(b^2-1\bigr)-a b-b^2\bigr) (a-z) (b-z) \\
&\times\bigl(a^2 b^2 \lambda ^2 \bigl(a^2+a b+b^2-1\bigr)-a^4b^2-a^2 b^4-a^3 b+a^2-a b^3+a b+b^2-1\bigr) \\
&\times\bigl(\bigl(a^2-1\bigr)^2 \bigl(b^2-1\bigr)^2 (a b-1)^2-\bigl(\lambda ^2-1\bigr) (a-z) (a z-1) (b-z) (b z-1)\\
& \times\bigl(a^2-z (a+b)+a b+b^2-1\bigr) \bigl(a^2 (b (b+z)-1)+a b (b z-1)-b^2\bigr)\bigr). \end{align*}

We also have
\begin{align*}
w_{11}(z,\lambda)={}& -\frac{\rmi \bigl(a^2-z (a+b)+a b+b^2-1\bigr)}{{\bigl(a^2-1\bigr)^3 \bigl(b^2-1\bigr)^3 \lambda (a b-1)^3}
} \bigl(b^2 \bigl(\bigl(\lambda ^2-1\bigr) z (b-z)-1\bigr)+1\\
&+a^3 \bigl(b \bigl(-2 b^2+z^2+2\bigr)+\lambda ^2 (b-z) (b (b+z)-1)-z\bigr)\\
&+a^2 \bigl(b^2 \bigl(-\lambda ^2 \bigl(z^2+1\bigr)+z^2+2\bigr)+b \bigl(\lambda ^2-1\bigr) z \bigl(z^2+2\bigr)-\lambda ^2 z^2+z^2-1\bigr)\\&
+a b \bigl(b^2 \bigl(-\lambda ^2 \bigl(z^2+1\bigr)+z^2+2\bigr)+b \bigl(\lambda ^2-1\bigr) z \bigl(z^2+2\bigr)-\lambda ^2 z^2+z^2-1\bigr)\bigr),\\
w_{12}(z,\lambda)={}& -\rmi (a-z) (b-z) \bigl(\lambda ^2 (a z-1) (b z-1) \bigl(a^2-z (a+b)+a b+b^2-1\bigr)\\
&
+(a-z) (b-z) \bigl(a^2 (b (b+z)-1)+a b (b z-1)-b^2\bigr)\bigr)/W,\\
w_{21}(z,\lambda)={}&\frac{-ba^2 (b+z)+a^2+a b (1-b z)+b^2}{\bigl(a^2-1\bigr)^2 \bigl(b^2-1\bigr)^2 (a b-1)^2},\\
w_{22}(z,\lambda)={}&-(a z-1) (b z-1)(a^2-1)(b^2-1)(ab-1)\lambda/W,
\end{align*}
where $W$ is given by
\begin{align*}
W={}&\bigl(a^3 \bigl(a^2-1\bigr)^4 b^3 \bigl(b^2-1\bigr)^4
\lambda \bigl(\lambda ^2-1\bigr) (a+b) (a b-1)^5 \bigl(-a^2\bigl(b^2-1\bigr)+a b+b^2\bigr)\\
&\times \bigl(a^2 b^2 \lambda ^2 \bigl(a^2+a b+b^2-1\bigr)-b^2a^4-a^2 b^4-a^3 b+a^2-a b^3+a b+b^2-1\bigr)\bigr).\end{align*}

The second case $\lvert a\rvert <1$ and $\lvert b\rvert >1$ is similar. Considering equations \eqref{a11}--\eqref{a21} The matrix polynomial is given by replacing $a(z)$ by $a(z)/b^2$ and thus one can use the same $y(z)$ as above. (In the Smith canonical form we should replace $w(z)$ by $w(z)/b^2$, but this does not affect our calculations.)

\section{Details for Section \ref{sec:AIII}}\label{app:AIIIproof}
\subsection{String correlators}
Here we derive the Toeplitz determinant form of string correlators in the AIII class. Define $s_n$ to be the $n$th Fourier coefficient of $s(z) =\sqrt{f(z)/\overline{f}(1/z)}$; then using the definitions of $\tilde{\mathcal{O}}_\alpha(n)$ given in \eqref{stringcorrelatorsAIII}, we have
\begin{align}
&(-1)^N\langle \tilde{\mathcal{O}}_\alpha(1)\tilde{\mathcal{O}}_\alpha(N+1)\rangle = \left\langle \prod_{j=1}^{2N} \left(-\rmi \tilde{\gamma}_{j}\gamma_{j+2\alpha}\right)\right\rangle \label{AIIIcorrelator}\\
&
=\det \! \left(\begin{matrix}\Re(s_{\alpha}) & \Im(s_{\alpha}) & \Re(s_{\alpha-1}) &   \hdots &\Re(s_{\alpha-(N-1)}) & \Im(s_{\alpha-(N-1)}) \\
\Im(s_{\alpha}) & -\Re(s_{\alpha}) & \Im(s_{\alpha-1}) & \hdots &\Im(s_{\alpha-(N-1)}) & -\Re(s_{\alpha-(N-1)}) \\
\Re(s_{\alpha+1}) & \Im(s_{\alpha+1}) & \Re(s_{\alpha}) &   \hdots &\Re(s_{\alpha-(N-2)}) & \Im(s_{\alpha-(N-2)}) \\
\Im(s_{\alpha+1}) & -\Re(s_{\alpha+1}) & \Im(s_{\alpha}) &  \hdots &\Im(s_{\alpha-(N-2)}) & -\Re(s_{\alpha-(N-2)}) \\
\vdots & \vdots & \vdots & \ddots&\vdots&\vdots\\ \Re(s_{N-1}) & \Im(s_{N-1}) & \Re(s_{N-2}) &   \hdots &\Re(s_{0}) & \Im(s_{\alpha}) \\
\!\Im(s_{\alpha+N-1})\! & \!-\Re(s_{\alpha+N-1})\! & \!\Im(s_{\alpha+N-2})\! &  \hdots &\!\Im(s_{\alpha})\! &\! -\Re(s_{\alpha})\! \end{matrix}\right)\!.\nonumber
\end{align}
This uses the following Majorana correlators:
\begin{align*}
&-\rmi\langle \tilde\gamma_{2n-1}\gamma_{2m-1} \rangle = \rmi\langle \tilde\gamma_{2n}\gamma_{2m} \rangle = \Re(s_{m-n}),\\
&-\rmi\langle \tilde\gamma_{2n}\gamma_{2m-1} \rangle =-\rmi\langle \tilde\gamma_{2n-1}\gamma_{2m} \rangle =\Im(s_{m-n})
\end{align*}
that can be derived from \eqref{AIIItwopoint} and, for $\alpha<0$, we used translation invariance in the first equality. We see that \eqref{AIIIcorrelator} has a ($2\times 2$) block-Toeplitz form. To identify the symbol, we conjugate each block by the unitary matrix
\begin{align*}
\frac{1}{2} \left(\begin{matrix} 1+\rmi & -(1-\rmi)\\ 1+\rmi &1-\rmi\end{matrix}\right).
\end{align*}
Then we have that
\begin{align}
\big\langle \tilde{\mathcal{O}}_\alpha(1)\tilde{\mathcal{O}}_\alpha(N+1)\big\rangle &= (-1)^ND_N\left[\left(\begin{matrix} 0 & z^{-\alpha}\sqrt{\frac{f(z)}{\overline{f}(1/z)}}\\ z^{-\alpha}\sqrt{\frac{\overline{f}(z)}{{f}(1/z)}} &0\end{matrix}\right)\right] \label{AIIIstringresult}\\&= D_N\Bigg[\sqrt{\frac{f(z)}{\overline{f}(1/z)}}z^{-\alpha}\Bigg]D_N\Bigg[\sqrt{\frac{\overline{f}(z)}{{f}(1/z)}}z^{-\alpha}\Bigg]
 =\Bigg\vert D_N\left[\sqrt{\frac{f(z)}{\overline{f}(1/z)}}z^{-\alpha}\right] \Bigg\vert^2.\nonumber
\end{align}
If $f(z)$ is of the form \eqref{fzcanonAIII}, then we have
\begin{align*}
\big\langle \tilde{\mathcal{O}}_\alpha(1)\tilde{\mathcal{O}}_\alpha(N+1)\big\rangle
&=\big\lvert D_N\left[t(z)\right] \big\rvert^2,
\end{align*}
where
\begin{align*}
t(z) &= z^{n_z+n_Z-n_P-\alpha}\left(\prod_{k=1}^{n_Z}\bigl(-\overline{Z}_k\bigr)\right)^{-1} \frac{\prod_{j=1}^{n_z} (z-z_j)\prod_{k=1}^{n_Z} (z-Z_k)}{\prod_{j=1}^{n_z} \bigl(1-z/\overline{z}_j^{-1}\bigr)\prod_{k=1}^{n_Z} \bigl(z-\overline{Z}_k^{-1}\bigr)}.
\end{align*}
For $n_z>0$, $n_Z>0$ and $n_z+n_Z-n_P-\alpha\geq0$, this can be analysed using Day's formula as in Section~\ref{sec:string}. For other cases, as above, we can use $D_N[t(z)] = D_N[t(1/z)]$ along with taking appropriate limits to evaluate this determinant with Day's formula.

\subsubsection{Proof of Theorem \ref{thm:AIII}}
To derive the result for the value of the order parameter, we consider the limit of
\[D_N\left[\sqrt{\frac{f(z)}{\overline{f}(1/z)}}z^{-\omega}\right]\qquad \text{as} \quad N\rightarrow \infty\] and then use \eqref{AIIIstringresult} to find the order parameter. The proof follows directly from the analysis in~\cite[Section 6.1]{Jones2019}; we must simply keep track of the complex-conjugate zeros. Suppose that~$f(z)$ is given by \eqref{fz_genericAIII}, then we write
\begin{align*}
\sqrt{\frac{f(z)}{\overline{f}(1/z)}}z^{-\omega} = \rme^{V(z)},\end{align*}
where \begin{align}
V(z) ={}& V_0 + \frac{1}{2}\sum_{j=1}^{N_z}\bigl(\Log(1-z_j/z)-\Log(1-\overline{z_j}z)\bigr)\nonumber\\
&+ \frac{1}{2}\sum_{k=1}^{N_Z}\bigl(\Log\bigl(1-zZ_k^{-1}\bigr)-\Log\bigl(1-\overline{Z_k}^{-1}/z\bigr)\bigr),\label{eq:V(z)}
\end{align}
and $\Log$ is the principal branch of the logarithm. Szeg\H{o}'s theorem expresses the large $N$ asymptotics that we want as follows:
\begin{align*}
D_N\left[\sqrt{\frac{f(z)}{\overline{f}(1/z)}}z^{-\omega}\right] = \rme^{V_0N+\sum_{n=1}^\infty nV_nV_{-n}}(1+o(1)),
\end{align*}
subject to some smoothness conditions that are satisfied by our symbol \cite{Deift2011}. The Fourier coefficients $V_n$ for $n\neq0$ follow simply from \eqref{eq:V(z)}. We also have
\begin{align*}
\rme^{V_0} = \pm \rme^{\rmi \theta} \sqrt{\frac{\prod_k Z_k}{\prod_k \overline{Z}_k}} = \rme^{\rmi \theta'} \qquad \textrm{for}\quad \theta'\in[0,2\pi).
\end{align*}
The $\pm1$ fixes the correct branch of the square-root, so that $\rme^{V(1)}=f(1)/\lvert f(1)\rvert$. Since we will take the absolute value, this oscillatory factor is unimportant and the order parameter is equal to $\big\lvert \exp\bigl({\sum_{n=1}^\infty nV_nV_{-n}}\bigr) \big\rvert^2$. This can be evaluated as in \cite{Jones2019}, leading to the second part of Theorem~\ref{thm:AIII}.

To derive the first part of Theorem \ref{thm:AIII}, firstly note that if we have that \[D_N\left[\sqrt{\frac{f(z)}{\overline{f}(1/z)}}z^{-\alpha}\right]\rightarrow0\qquad \text{as}\quad N\rightarrow \infty,\] then by \eqref{AIIIstringresult} we have that $\big\langle \tilde{\mathcal{O}}_\alpha(1)\tilde{\mathcal{O}}_\alpha(N+1)\big\rangle \rightarrow0 $. If $\alpha \neq \omega$, then
\begin{align*}
\sqrt{\frac{f(z)}{\overline{f}(1/z)}}z^{-\alpha} = z^{\omega-\alpha} \rme^{V(z)},\qquad
\text{where}\quad
\rme^{V(z)}=\rme^{\rmi \theta'} \sqrt{ \frac{\prod_{j=1}^{n_z} (1-z_j/z)\prod_{k=1}^{n_Z} \bigl(1-zZ_k^{-1}\bigr)}{\prod_{j=1}^{n_z} (1-z\overline{z}_j)\prod_{k=1}^{n_Z} \bigl(1-\overline{Z}_k^{-1}/z\bigr)}}.
\end{align*}
Since no zeros are on the unit circle, there exists a $\rho<1$ such that $\rme^{V(z)}$ is analytic on the annulus $\rho < \lvert z \rvert <\rho^{-1}$. Then using \cite[Theorem 4]{Hartwig1969}, the determinant $D_N\big[z^{m} \rme^{V(z)}\big]$ for $m\in\mathbb{Z}$ and $m \neq 0$ will go to zero as $N\rightarrow \infty$. One could go further and use that theorem to find the correlation lengths, as given for the BDI class in \cite{Jones2019}, but we will not pursue this here.

\subsection{Correlation matrix}
\subsubsection{Proof of Theorem \ref{thm:corrmatrixAIII}}
In class AIII we are interested in the eigenvalues of the block-Toeplitz matrix with symbol~$\hat\Phi(z,0)$, where
\begin{align*}
\hat\Phi(z,\lambda) &= \left(\begin{matrix} \lambda & -\sqrt{\dfrac{f(z)}{\overline{f}(1/z)}}\\ -\sqrt{\dfrac{\overline{f}(1/z)}{f(z)}}&\lambda\end{matrix}\right)
\end{align*}
generates the characteristic polynomial. Let us suppose then that $f(z)$ is restricted to be of the form \eqref{fzcanonAIII}, but recall that, unlike in class BDI, the zeros do not necessarily come in complex-conjugate pairs. Note also that since we are taking the determinant, we can conjugate $\hat\Phi(z,\lambda)$ by the unitary matrix $U$, with $U_{11}=\rme^{-\rmi\theta}$, $U_{21}=U_{12}=0$ and $U_{22}=1$. This removes any dependence on $\theta$, so we will set $\theta=0$ in the formulae below. Then, as in Section \ref{sec:corrmatrix}, we will use Gorodetsky's formula to establish that there are only a finite number of non-trivial eigenvalues.

Let us define \begin{align*}
\tilde{g}(z) &= \prod_{j=1}^{n_z} (z-z_j)\prod_{k=1}^{n_Z} \bigl(z-\overline{Z}_k^{-1}\bigr),\qquad
\tilde{h}(z)= \prod_{j=1}^{n_z} (1-z\overline{z_j})\prod_{k=1}^{n_Z} \bigl(1-zZ_k^{-1}\bigr).
\end{align*}
We then have
\begin{align*}
\tilde\Phi(z,\lambda) &= \frac{a(z)}{\tilde{g}(z)\tilde{h}(z)}= \frac{1}{\tilde{g}(z)\tilde{h}(z)} \left(\begin{matrix} a_{11}(z)& a_{12}(z)\\ a_{21}(z)& a_{22}(z)\end{matrix}\right),
\end{align*}
where
\begin{align}
a_{11}(z)={}& a_{22}(z)= \lambda \frac{\bigl(\prod_{j=1}^{n_z} (-\overline{z}_j)\prod_{k=1}^{n_Z} \bigl(-\overline{Z}_k\bigr)\bigr)}{\prod_{k=1}^{n_Z} \lvert Z_k\rvert^2} \nonumber\\
&\times\prod_{j=1}^{n_z}(z-z_j)\bigl(z-\overline{z_j}^{-1}\bigr) \prod_{k=1}^{n_Z}(z-Z_k)\bigl(z-\overline{Z_k}^{-1}\bigr), \label{a11AIII}\\
a_{12}(z)={}&-z^{n_z+n_Z-n_P} \frac{1}{\prod_{k=1}^{n_Z} \lvert Z_k\rvert^2} \prod_{j=1}^{n_z}(z-z_j)^2 \prod_{k=1}^{n_Z}(z-Z_k)^2,\nonumber \\
a_{21}(z)={}&-z^{-n_z-n_Z+n_P} \frac{\bigl(\prod_{j=1}^{n_z} \overline{z}_j^2\prod_{k=1}^{n_Z} \overline{Z}_k^2\bigr)}{\prod_{k=1}^{n_Z} \lvert Z_k\rvert^2}\prod_{j=1}^{n_z}\bigl(z-\overline{z}_j^{-1}\bigr)^2 \prod_{k=1}^{n_Z}\bigl(z-\overline{Z}_k^{-1}\bigr)^2 \label{a21AIII}.
\end{align}
As above we fix $n_z+n_Z=n_P$ so that we have a matrix polynomial $\sum_{j=0}^{2(n_z+n_Z)} a_j z^j$. We then have
\begin{align*}
\det(a_{2(n_z+n_Z)})=\bigl(\lambda^2-1\bigr) \frac{\prod_{j=1}^{n_z} \overline{z}_j^2\prod_{k=1}^{n_Z} \overline{Z}_k^2}{\bigl(\prod_{k=1}^{n_Z} \lvert Z_k\rvert^2\bigr)^2}.
\end{align*}
Define the set of zeros and inverse conjugate zeros by
\begin{align*}
\{\tau_i\}_{i=1,\dots, 2(n_z+n_Z)} \in \big\{z_{j_1}^{\vphantom {-1}},\overline{z}_{j_2}^{-1},{Z}_{k_1}^{\vphantom {-1}},\overline{Z}_{k_2}^{-1}\big\}_{j=1,\dots, n_z; k=1,\dots, n_Z}.
\end{align*}
We assume without loss of generality that $z_j \neq \overline{Z}_k^{-1}$ for any $j$, $k$.
Then we have the Smith canonical form:
\begin{align*}
y(z) a(z) w(z) = \left(\begin{matrix}1 & 0 \\0 & \displaystyle \prod_{j=1}^{n_z}(z-z_j)^2\bigl(z-\overline{z}_j^{-1}\bigr)^2 \prod_{k=1}^{n_Z}(z-Z_k)^2\bigl(z-\overline{Z}_j^{-1}\bigr)^2\end{matrix}\right) .
\end{align*}
The remaining definitions given in Section \ref{sec:corrmatrix} are unchanged. By following the proof of Theorem~\ref{thm:corrmatrix} in Section \ref{sec:corrmatrix}, we reach Theorem \ref{thm:corrmatrixAIII} for class AIII given in Section~\ref{sec:AIII}.

\subsubsection{Smith canonical form for the example}
In Section \ref{sec:AIII}, we give the correlation matrix eigenvalues for $f(z) =z^{-1}(z-b)^2$. We need to calculate the Smith canonical form for $a(z)$ defined in equations \eqref{a11AIII}--\eqref{a21AIII}. The following expression for $y(z)$ follows from the discussion in the main text and was simplified using \textsc{Mathematica}~\cite{Mathematica}:
\begin{align*}
y_{11}(z)={}&\bigl(\lambda ^2-2\bigr) \lvert b\rvert^2+1,
\\
y_{21}(z)={}&-\lambda y_{11}(z)/\overline{b},\\
y_{21}(z)={}&\bigl(\lvert b\rvert ^2-1\bigr) \bigl(\overline{b} \bigl(\bigl(1-\lambda ^2\bigr) z \bigl(\lvert b\rvert ^4-\overline{b} (2 b-z) \bigl(z \overline{b}-1\bigr)\bigr)-2 b\bigr)\\
&-z \bigl(\lvert b\rvert ^2-1\bigr)^2 \overline{b}+\lambda ^2 \lvert b\rvert ^4+1\bigr),\\
y_{22}(z)={}&\frac{1}{\overline{b}}\bigl(\lambda \bigl(\lvert b\rvert ^2-1\bigr) \bigl(\lvert b\rvert ^4 \bigl(\overline{b} \bigl(\lambda ^2 z-b\bigr)-\lambda ^2+1\bigr)+2 \lvert b\rvert ^2 \bigl(\overline{b} (b-z)-1\bigr)\\&
+\overline{b} \bigl(\bigl(\lambda ^2-1\bigr) z^2 \overline{b}^2 ((z-2b))+\bigl(\lambda ^2-1\bigr) z \overline{b} (2 b-z)+b+z\bigr)\bigr)\bigr).
\end{align*}

\section{Details for Section \ref{sec:transfer}}\label{app:transfer}
\subsection[Proof that all subsets of S, correspond to some r\_M]{Proof that all subsets of $\boldsymbol{S=\{z_1,\dots, z_{n_z}, Z_1^{-1},\dots, Z_{n_Z}^{-1}\}}$ \\ correspond to some $\boldsymbol{r_M}$.}\label{app:corollaryproof}

Here we prove a result needed in Section \ref{sec:corollaryproof}.
Suppose that $n_P=n_z+n_Z$, and let us first consider the case $n_z>0$ and $n_Z>0$. Then in calculating correlators using Theorem~\ref{thm:correlators}, we have Case~1 for $-n_Z\leq \alpha \leq 0$ and Case 2 for $0\leq \alpha \leq n_z$.

For $-n_Z\leq \alpha \leq 0$, we have
\begin{align}
r_M&= \frac{\prod_{k \in M^c} \tau_k} {\prod_{j=1}^{n_Z}Z_j}, \label{appendixcase1}
\end{align}
where $\tau_k$ are zeros and $n_z\leq\lvert M^c \rvert \leq n_z+n_Z$. Note that there is an $r_M$ for any set $M^c$ satisfying the inequality.

For $0\leq \alpha \leq n_z$, we have
\begin{align}
r_M&= \prod_{j=1}^{n_z}z_j{\prod_{k \in M^c} \tau_k}, \label{appendixcase2}
\end{align}
where $\tau_k$ are inverse zeros and $n_Z\leq\lvert M^c \rvert \leq n_z+n_Z$. Again, there is an $r_M$ for any set $M^c$ satisfying the inequality.

Now let us take any subsets $m_z\subseteq M_z = \{1,\dots, n_z\}$ and $m_Z\subseteq M_Z = \{1,\dots, n_Z\}$ and take the product:
\begin{align}
  R(m_z,m_Z)&=\prod_{j \in m_z} z_j \prod_{k \in m_Z} Z_k^{-1}= \frac{\prod_{k \in m_z} z_k\prod_{k \in M_Z\setminus m_Z} Z_k} {\prod_{j=1}^{n_Z}Z_j}\nonumber\\
  &= \frac{\prod_{j=1}^{n_z}z_j}{\prod_{k \in M_z\setminus m_z} z_k\prod_{k \in m_Z} Z_k} \label{appendixcases}.
\end{align}
We want to show that $R(m_z,m_Z)$ always appears in a correlator for some $\alpha$, i.e., it is of the form~\eqref{appendixcase1} or \eqref{appendixcase2}. If $n_z\leq \lvert m_z \rvert + n_Z - \lvert m_Z \rvert$ then by the second equality of \eqref{appendixcases} we see that this~$r_M$ appears in Case 1. If $n_z> \lvert m_z \rvert + n_Z - \lvert m_Z \rvert$ then $n_Z< n_z-\lvert m_z \rvert + \lvert m_Z \rvert$ and so by the third equality of \eqref{appendixcases} we see that this $r_M$ appears in Case 2. Hence, for any choice of~$m_z$ and~$m_Z$ the corresponding $r_M$ appears in a correlator (with a non-zero coefficient $C_M$) for some~$\alpha$.

The same analysis can be applied in the case that $n_Z=0$ or $n_z=0$. The only complication is that the expansions $\sum_M C_M^{\vphantom N}r_M^N$ are, in general, some limit of a case already considered. For~$n_Z=0$ and for $\alpha=0$ we see that $\prod_{j=1}^{n_z}z_j$ is an eigenvalue of $E_Z$. For the other non-zero correlators, we put $n_Z=1$ and then take $Z_1\rightarrow\infty$. For any subset $m_z \subset \{z_1,\dots,z_{n_z}\}$ we have that $r_M=\prod_{j\in m_z}z_j$ appears in some correlator, with coefficient $C_M$. It is straightforward to confirm that for each of these coefficients, $\lim_{Z_1\rightarrow\infty} C_M$ is non-zero.
Analogous statements hold for $n_z=0$. Finally, for $n_z=n_Z=0$, we have $f(z)=1$ and the string correlators give one eigenvalue, $\mu =1$, for the transfer matrix.

\subsection{Diagonalising the transfer matrix}\label{app:diagonaltransfer}
We are interested in the model $f(z) = z^{-2}(z-a)^2(z-b)^2$ for $a<1$ and $b>1$. We expect that~$\chi^2=4$, based on the results of \cite{Jones21}. Now, using Example \ref{correxample}, we have
\begin{align*}
&\sum_{k=1}^{4} \mu_k^{N-1} \bra{l_1} E_{Y} \ket{r_k}\bra{l_k}E_{Y} \ket{r_1}=\langle Y_1 Y_{N+1}\rangle = \frac{\bigl(b^2-1\bigr) (a b-1) }{ b^2 (a-b)}(-a)^{N-1}, \qquad N\geq 1,\\
&\sum_{k=1}^{4} \mu_k^{N-1} \bra{l_1} E_{X} \ket{r_k}\bra{l_k}E_{X} \ket{r_1}=\langle X_1 X_{N+1}\rangle = \frac{\bigl(1-a^2\bigr) (1-a b) }{(a-b)}(-b)^{-(N-1)}, \qquad N\geq 1.
\end{align*}
If we define \[\tilde{Z} = Z - \frac{(a^2-1) b^2-a b+1}{b (a-b)},\] we can also compute the correlator
\begin{align*}\nonumber
\sum_{k=1}^{4} \mu_k^{N-1} \bra{l_1} E_{\tilde{Z}} \ket{r_k}\bra{l_k}E_{\tilde{Z}} \ket{r_1}&=\big\langle \tilde{Z}_1\tilde{Z}_{N+1}\big\rangle \\
&= \langle \rmi \tilde\gamma_1\gamma_1\rmi \tilde\gamma_{N+1}\gamma_{N+1} \rangle-\left(\frac{\bigl(a^2-1\bigr) b^2-a b+1}{b (a-b)}\right)^2\\
&=\frac{\bigl(1-a^2\bigr) \bigl(1-b^{-2}\bigr) (a b-1)^2 }{(a-b)^2}\left(\frac{a}{b}\right)^{N-1}.
\end{align*}
this follows from Wick's theorem. The fermionic two-point function is just the Fourier coefficient given in Appendix \ref{app:fourier}.
We can thus identify $\mu_1=1$, $\mu_2=-a$, $\mu_3=-1/b$ and $\mu_4=a/b$.

Note also that the correlators $\langle \mathcal{O}_{\alpha}(1) \mathcal{O}_{\beta}(N+1) \rangle = 0$ for $\alpha \neq \beta$ as a simple consequence of Wick's theorem \cite{Jones2019}.
This means that
\begin{align*}
\sum_{k=1}^{4} \mu_k^{N-1} \bra{l_1} E_{X} \ket{r_k}\bra{l_k}E_{Y} \ket{r_1}=\sum_{k=1}^{4} \mu_k^{N-1} \bra{l_1} E_{Y} \ket{r_k}\bra{l_k}E_{X} \ket{r_1}=0.
\end{align*}
Similarly, all further two-point correlators of $\tilde{Z}$, $X$ and $Y$ with different operators at the different points all vanish due to Wick's theorem. Moreover, $\langle X_n \rangle =\langle Y_n \rangle =\langle \tilde{Z}_n \rangle=0$. This allows us to deduce, for example, that
\begin{align*}
\ket{r_2}\bra{l_2}=\frac{ b^2 (a-b)}{\left(b^2-1\right) (a b-1) } E_Y\ket{r_1}\bra{r_1}E_Y,
\end{align*}
as well as the other results claimed in Section \ref{sec:diagonal}.

\section{Details for Section \ref{sec:generic}} \label{app:limit}
Here we prove that \begin{align*}
 (1-z)^{1/4} = \lim_{m\to \infty} \left[ \prod_{a=1}^m \prod_{b=1}^m \bigl( 1 - \lambda_a^{(m)} \lambda_b^{(m)} z \bigr) \right].
\end{align*}

Firstly, for any $m \in \mathbb N$ and $|z|<1$, we have
\begin{align}
\sqrt{1-z} = s_m(z) + r_m(z) = s_m(z) \left( 1 + \frac{r_m(z)}{s_m(z)} \right), \label{eq:sqrt}
\end{align}
where
\begin{align*}
s_m(z) = \sum_{a=1}^m \left( \begin{matrix} 1/2 \\ a \end{matrix} \right) (-z)^a \qquad \textrm{and} \qquad r_m(z) = \sum_{a=m+1}^\infty \left( \begin{matrix} 1/2 \\ a \end{matrix} \right) (-z)^a .
\end{align*}
Note that $|r_m(z)| \leq \sum_{a=m+1}^\infty |z|^a = \frac{|z|^{m+1}}{1-|z|}$.

We now write
\begin{align*}
s_m(z) = \prod_{a=1}^m \bigl( 1 - \lambda_a^{(m)}z \bigr).
\end{align*}
As discussed in Section~\ref{sec:generic}, we have that $\big|\lambda_a^{(m)}\big|< 1$. Then, by taking the square root of \eqref{eq:sqrt}, we have
\begin{align}
(1-z)^{1/4}={}& \sqrt{s_m(z)} \sqrt{ 1 + \frac{r_m(z)}{s_m(z)} } \nonumber\\
={}& \left( \prod_{a=1}^m \sqrt{ 1 - \lambda_a^{(m)}z } \right) \times \sqrt{ 1 + \frac{r_m(z)}{s_m(z)} } \nonumber\\
={}& \left( \prod_{a=1}^m \left[ s_m \bigl( \lambda_a^{(m)}z \bigr) \times \left( 1 + \frac{r_m \bigl(\lambda_a^{(m)}z \bigr)}{s_m \bigl( \lambda_a^{(m)}z \bigr)} \right) \right] \right) \times \sqrt{ 1 + \frac{r_m(z)}{s_m(z)} } \nonumber\\
={}& \left[ \prod_{a=1}^m s_m \bigl( \lambda_a^{(m)}z \bigr) \right]\times \sqrt{ 1 + \frac{r_m(z)}{s_m(z)} }\times\Bigg( 1\nonumber\\&
+\underbrace{\sum_{a_1=1}^m \frac{r_m \bigl(\lambda_{a_1}^{(m)}z \bigr)}{s_m \bigl( \lambda_{a_1}^{(m)}z \bigr)} +
\sum_{a_1\neq a_2=1}^m \prod_{i=1}^2 \frac{r_m \bigl(\lambda_{a_i}^{(m)}z \bigr)}{s_m \bigl( \lambda_{a_i}^{(m)}z \bigr)}
+ \cdots + \prod_{i=1}^m \frac{r_m \bigl(\lambda_{a_i}^{(m)}z \bigr)}{s_m \bigl( \lambda_{a_i}^{(m)}z \bigr)}}_S
 \Bigg) . \label{eq:temp}
\end{align}

Consider $|z| \leq R < 1$, then $|r_m(z)| \leq \frac{R^{m+1}}{1-R}$ and \begin{align*}
|s_m(z)| = \left| \sqrt{1-z} - r_m(z) \right| \geq \left|\sqrt{1-z}\right| - |r_m(z)| \geq \sqrt{1-R} - \frac{R^{m+1}}{1-R}
\end{align*}
and thus
\begin{align*}
\left| \frac{r_m(z)}{s_m(z)} \right| \leq \frac{R^{m+1}}{(1-R)^{3/2} - R^{m+1}} =: \rho_m .
\end{align*}
In particular, this shows that $\lim_{m \to \infty} \frac{r_m(z)}{s_m(z)} = 0$. Let us now analyse the term $S$ appearing in~\eqref{eq:temp}. Note that since \smash{$\big|\lambda_a^{(m)}\big|< 1$}, the arguments of $r_m$ and $s_m$ have modulus bounded by $R$. Then
\begin{align*}
&\left| \sum_{a_1=1}^m \frac{r_m \bigl(\lambda_{a_1}^{(m)}z \bigr)}{s_m \bigl( \lambda_{a_1}^{(m)}z \bigr)} +
\sum_{a_1\neq a_2=1}^m \prod_{i=1}^2 \frac{r_m \bigl(\lambda_{a_i}^{(m)}z \bigr)}{s_m \bigl( \lambda_{a_i}^{(m)}z \bigr)}
+ \cdots + \prod_{i=1}^m \frac{r_m \bigl(\lambda_{a_i}^{(m)}z \bigr)}{s_m \bigl( \lambda_{a_i}^{(m)}z \bigr)} \right| \nonumber\\
&\leq \left| \sum_{a_1=1}^m \rho_m +
\sum_{a_1\neq a_2=1}^m \rho_m^2
+ \cdots + \rho_m^m \right| = (1+\rho_m)^m-1 = \rme^{m\ln(1+\rho_m)} - 1 \leq \rme^{m \rho_m} - 1.
\end{align*}
Since $\lim_{m \to \infty} m \rho_m = 0$, we thus see that taking the limit of equation \eqref{eq:temp} simplifies to
\begin{align*}
(1+z)^{1/4} = \lim_{m \to \infty} \left[ \prod_{a=1}^m s_m \bigl( \lambda_a^{(m)}z \bigr) \right] = \lim_{m \to \infty} \left[ \prod_{a=1}^m \prod_{b=1}^m \bigl( 1 + \lambda_a^{(m)} \lambda_b^{(m)} z \bigr) \right] .
\end{align*}

\subsection*{Acknowledgements} We thank J. Bibo, B. Jobst, F. Pollmann and A. Smith for many inspiring discussions on this subject and for collaboration on related work \cite{Jones21}. We are also grateful to A.~B\"{o}ttcher, A.~Its, J.~Keating, F.~Mezzadri, N.~Schuch and S.~Stevens for helpful correspondence and to A.~Smith and the anonymous referees for valuable comments on the manuscript. This work was completed while N.G.J.\ held a Heilbronn Research Fellowship at the Mathematical Institute, University of Oxford, and the Heilbronn Institute for Mathematical Research, Bristol, UK. R.V. was supported by the Harvard Quantum Initiative Postdoctoral Fellowship in Science and Engineering and by the Simons Collaboration on Ultra-Quantum Matter, which is a grant from the Simons Foundation (651440, Ashvin Vishwanath).

\pdfbookmark[1]{References}{ref}
\LastPageEnding

\end{document}